\documentclass[12pt]{iopart}
\usepackage{iopams}                                      
\usepackage{mathrsfs}                                    
\usepackage[latin9]{inputenc}                            
\usepackage[english]{babel}                              
\usepackage[T1]{fontenc}                                 
\usepackage{amsopn}
\usepackage{setspace}                                    
\usepackage{enumitem}                                    
\usepackage{color}                                       
\usepackage{graphicx}                                    
\definecolor{bleupsl}{RGB}{61,55,133}
\definecolor{ddlion}{RGB}{255,245,219}
\usepackage{subfigure}                                   
\usepackage{cite}                                        
\usepackage{multirow}                                    
\usepackage{thmtools}                                    
\usepackage{fourier-orns}                                
\usepackage{adforn}                                      
\usepackage{rotating}                                    
\usepackage{url}

\newcommand{\tn}{\textnormal}

\declaretheorem{theorem}
\declaretheorem{conjecture}

\newenvironment{myfig}
{
   \begin{figure}[t]
      \centering
}
{
   \end{figure}
}

\newenvironment{mystab}
{
   \begin{sidewaystable}[p]
}
{
   \end{sidewaystable}
}

\newenvironment{myitem}
{
   \begin{itemize}[label=$\blacktriangleright$,font=\color{bleupsl}]
}
{
   \end{itemize}
}

\newenvironment{myenum}
{
   \begin{enumerate}[label=\protect\ding{\value*},start = 182,font=\large\color{bleupsl}]
}
{
   \end{enumerate}
}

\def\jnl@style{\it}
\def\aaref@jnl#1{{\jnl@style#1}}

\def\aaref@jnl#1{{\jnl@style#1}}

\def\aj{\aaref@jnl{AJ}}                   
\def\araa{\aaref@jnl{ARA\&A}}             
\def\apj{\aaref@jnl{ApJ}}                 
\def\apjl{\aaref@jnl{ApJ}}                
\def\apjs{\aaref@jnl{ApJS}}               
\def\ao{\aaref@jnl{Appl.~Opt.}}           
\def\apss{\aaref@jnl{Ap\&SS}}             
\def\aap{\aaref@jnl{A\&A}}                
\def\aapr{\aaref@jnl{A\&A~Rev.}}          
\def\aaps{\aaref@jnl{A\&AS}}              
\def\azh{\aaref@jnl{AZh}}                 
\def\baas{\aaref@jnl{BAAS}}               
\def\jrasc{\aaref@jnl{JRASC}}             
\def\memras{\aaref@jnl{MmRAS}}            
\def\mnras{\aaref@jnl{MNRAS}}             
\def\pra{\aaref@jnl{Phys.~Rev.~A}}        
\def\prb{\aaref@jnl{Phys.~Rev.~B}}        
\def\prc{\aaref@jnl{Phys.~Rev.~C}}        
\def\prd{\aaref@jnl{Phys.~Rev.~D}}        
\def\pre{\aaref@jnl{Phys.~Rev.~E}}        
\def\prl{\aaref@jnl{Phys.~Rev.~Lett.}}    
\def\pasp{\aaref@jnl{PASP}}               
\def\pasj{\aaref@jnl{PASJ}}               
\def\qjras{\aaref@jnl{QJRAS}}             
\def\skytel{\aaref@jnl{S\&T}}             
\def\solphys{\aaref@jnl{Sol.~Phys.}}      
\def\sovast{\aaref@jnl{Soviet~Ast.}}      
\def\ssr{\aaref@jnl{Space~Sci.~Rev.}}     
\def\zap{\aaref@jnl{ZAp}}                 
\def\nat{\aaref@jnl{Nature}}              
\def\iaucirc{\aaref@jnl{IAU~Circ.}}       
\def\aplett{\aaref@jnl{Astrophys.~Lett.}} 
\def\apspr{\aaref@jnl{Astrophys.~Space~Phys.~Res.}}
\def\bain{\aaref@jnl{Bull.~Astron.~Inst.~Netherlands}} 
\def\fcp{\aaref@jnl{Fund.~Cosmic~Phys.}}  
\def\gca{\aaref@jnl{Geochim.~Cosmochim.~Acta}}   
\def\grl{\aaref@jnl{Geophys.~Res.~Lett.}} 
\def\jcp{\aaref@jnl{J.~Chem.~Phys.}}      
\def\jgr{\aaref@jnl{J.~Geophys.~Res.}}    
\def\jqsrt{\aaref@jnl{J.~Quant.~Spec.~Radiat.~Transf.}}
\def\memsai{\aaref@jnl{Mem.~Soc.~Astron.~Italiana}}
\def\nphysa{\aaref@jnl{Nucl.~Phys.~A}}   
\def\physrep{\aaref@jnl{Phys.~Rep.}}   
\def\physscr{\aaref@jnl{Phys.~Scr}}   
\def\planss{\aaref@jnl{Planet.~Space~Sci.}}   
\def\procspie{\aaref@jnl{Proc.~SPIE}}   

\graphicspath{{Figures/}}

\begin{document}

\title{The instability of anti-de Sitter space-time}
\author{Gr\'egoire Martinon$^1$}

\address{$^1$ LUTH, Observatoire de Paris, PSL Research University, CNRS, Universit\'e Paris Diderot, Sorbonne Paris Cit\'e, 92190 Meudon, France}

\ead{gregoire.martinon@obspm.fr}

\date{\today}

\begin{abstract}
   In this review, we retrace the recent progress in the anti-de Sitter (AdS) instability problem. By instability we mean that for large classes of initial data, any
   perturbation of AdS space-time, however small, leads to the formation of a black hole. Since the seminal work of
   Bizo\'n and Rostworowski in 2011, many different kinds of numerical experiments were performed in asymptotically AdS
   space-times, unveiling a very intricate structure of the instability. In particular, many efforts were dedicated to the search
   of islands of stability, i.e.\ families of initial data that resist black hole formation. Many analytical and numerical tools
   were deployed to disentangle stable from unstable initial data, and shed new light on the necessary and sufficient
   conditions for collapse. Recently, research beyond spherical symmetry became more and more engaged. This is a very promising
   channel of investigation toward a deeper understanding of the gravitational dynamics in asymptotically AdS space-times.
\end{abstract}

\pacs{04.20.Cv, 04.20.Ex 04.20.Ha, 04.25.D-, 04.25.Nx, 04.40.Nr, 11.25.Tq}
\submitto{\CQG}
\noindent{\it Keywords\/}: anti-de Sitter, instability, general relativity, gauge/gravity duality

\maketitle

\tableofcontents

\section{Introduction}

Einstein's equations in vacuum admit three maximally symmetric solutions: Minkowski (flat), de Sitter (dS) and anti-de Sitter
(AdS) space-times. They correspond respectively to zero, positive and negative curvature and cosmological constant $\Lambda$.
Both Minkowski and dS space-times are non-linearly stable, which means that no arbitrarily small perturbation can grow unbounded. Mathematical
demonstrations can be found respectively in \cite{Christodoulou93} and \cite{Friedrich86}. Both proofs rely on a dispersion
mechanism: any perturbation should decay one way or another, and this is possible when waves can propagate freely toward infinity
without back-reacting substantially on the metric. The decay rate of the perturbation is exponential in dS and the non-linear
stability can be inferred relatively easily. In Minkowski space-time on the other hand, the decay rate is borderline and the
non-linear stability proof is much more subtle.

AdS space-time is much different (we refer the reader to
\cite{Abbott82,Henneaux85,Ashtekar84,Ashtekar00,Wald00,Papadimitriou05,Hollands05a,Ashtekar14} for its precise definition). The
negative cosmological constant acts like a gravitational potential that prevents time-like geodesics from ever reaching infinity.
Instead, test particles, whatever their initial speeds, come back to their starting point in a finite proper time. This is even
true for null geodesics. In this case, a photon can reach infinity in a finite time (as measured by a static observer). Assuming
that the total energy (or mass) of space-time is conserved, this photon has to bounce back on space-like infinity, at the
so-called AdS boundary. In other words, this space-time acts like a reflective confining box: any test particle, be it massive or
massless, follows a straight line in vacuum and comes back at its starting point in a finite time.  It thus oscillates
perpetually, as if it were trapped in a $n$-dimensional billiard table, illustrated in figure \ref{geodesics}.  This confining
mechanism is clearly incompatible with the decay argument needed for establishing a non-linear stability proof. And actually, no
such proof exists so far.

It has long been known that AdS was \textit{linearly} stable, since the seminal works of
\cite{Avis78,Breitenlohner82,Abbott82}. It means that no small perturbation can have a mode that is unbounded in time at first
order in amplitude. Nevertheless, linear stability does not imply non-linear stability. Very curiously, the emergence of the
AdS-Conformal Field theories (CFT) correspondence in 1998 (see \cite{Maldacena98,Maldacena99,Witten98,Aharony00,Hubeny15})
triggered an avalanche of papers, but very few of them were concerned about the non-linear stability of AdS.  The first
questioning of this point appeared in 2006 with the works of \cite{Dafermos06,Anderson06}. Notably in \cite{Anderson06} the
following ``rigidity'' theorem was demonstrated:

\begin{myfig}
   \includegraphics[width=0.49\textwidth]{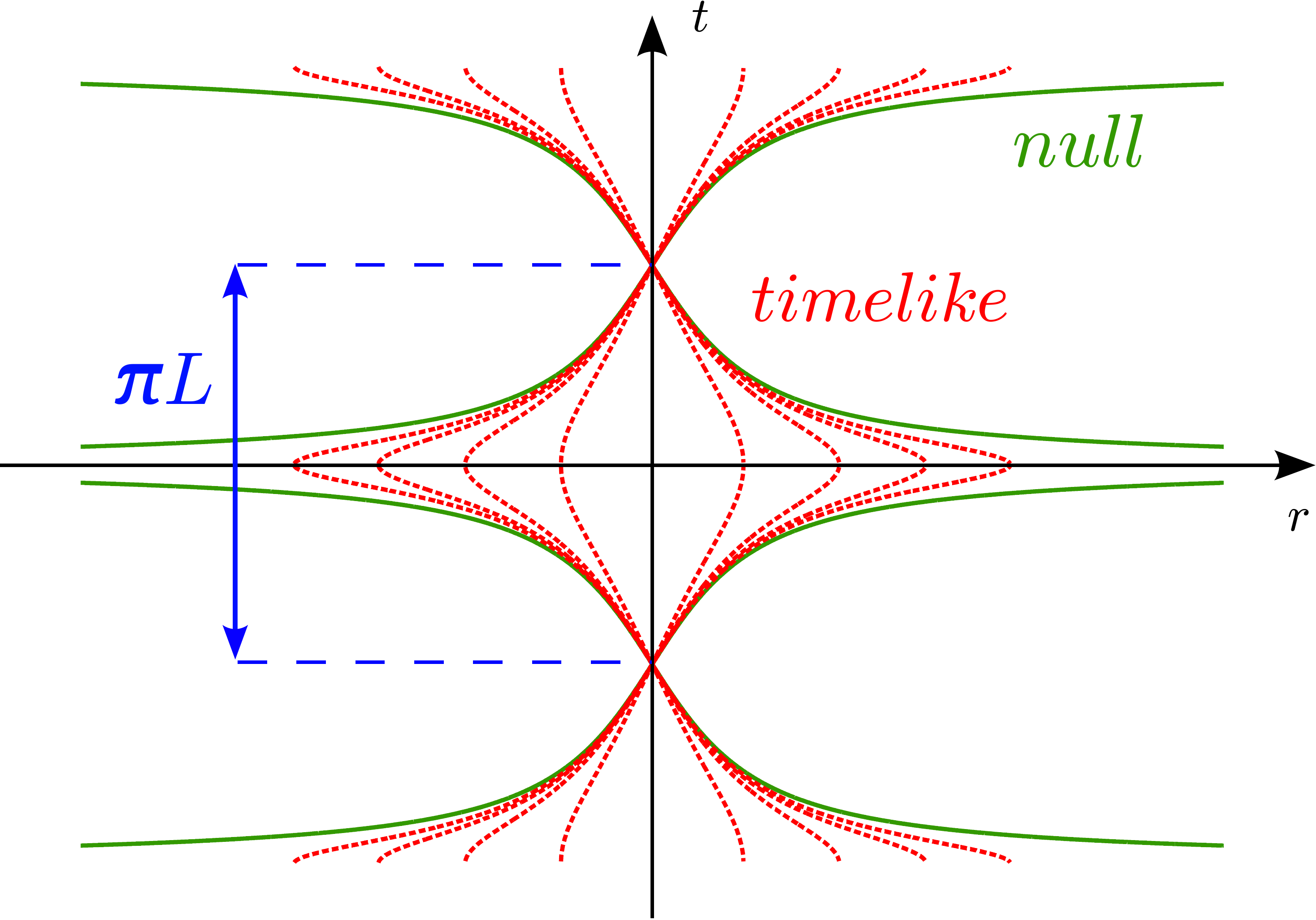}
   \caption{Radial geodesics in vacuum AdS$_n$ space-time in static coordinates $ds^2 = -\left( 1 + \frac{r^2}{L^2} \right)dt^2 + \frac{dr^2}{1+\frac{r^2}{L^2}} + r^2 d\Omega_{n-2}^2$. Time-like geodesics are pictured in dotted red
   lines and null ones in solid green. A bunch of time-like geodesics exhibiting an initial kinetic energy spread comes
   back to its initial position in a unique and finite amount of time, $\pi L$, where $L$ is the AdS radius
   directly linked to the cosmological constant by $\Lambda = -\frac{(n-1)(n-2)}{2 L^2}$.}
   \label{geodesics}
\end{myfig}

\begin{theorem}[Anti-de Sitter rigidity theorem]
The only globally regular asymptotically AdS (AAdS) space-time that tends to AdS at arbitrarily long times is AdS itself.
\end{theorem}

This is clearly the opposite of the decay condition for perturbations and it is quite natural for a space-time that cannot radiate its
energy to infinity. However, the author conjectured that AdS was probably non-linearly stable, and this was the global
consensus at that time, with the notable exception of \cite{Dafermos06}.

The number of researchers working on the AdS non-linear stability problem drastically increased in 2011, after the seminal
paper of Bizo\'n and Rostworowski \cite{Bizon11}. With the help of numerical time evolutions in spherical symmetry, they found that
scalar wave packets were inexorably collapsing to a black hole after several reflections off the AdS boundary. This behaviour
persisted even in the very small amplitude limit, and lead to the so-called AdS instability conjecture. As more and more proofs
were accumulating in favour of a pervasive instability of AdS space-time, other classes of initial
data were discovered, namely islands of stability. These configurations were able to resist black hole collapse for arbitrarily
long times. As more and more such islands came out in the literature, it became clear that the space of initial data (or parameter
space) was chaotically splattered with both stable and unstable configurations in the zero-amplitude limit, a feature at odds with
what happens in Minkowski or de Sitter space-times. A key challenge today is thus to disentangle the structure of the instability
and to unveil necessary and sufficient conditions for collapse.

The more numerous the outcoming papers in the field, the more intricate and subtle the problem seemed to be. This gave rise to
lively debates in the community, but also to an abundance of new numerical experiments and formalisms. In this paper, we give an
overview of the state of the art of the AdS non-linear stability problem and hope to disentangle the information harvested in the
literature. In section \ref{turbulent}, we focus on the properties of unstable initial data, from both a numerical and
perturbative perspective, and introduce the AdS instability conjecture. Section \ref{bhformation} focuses on critical phenomena
associated to black hole formation and singularity theorems in AdS. Islands of stability are discussed in section \ref{quenching},
notably with the two-time framework (TTF) formalism and the analyticity strip method. The general structure of the instability is
addressed in section \ref{structure} and section \ref{conditions} is devoted to the conditions for black hole formation. The
particular case of the instability when no black hole is allowed is considered in section \ref{nobh}. The recent advances beyond
spherical symmetry are discussed in section \ref{beyondspher} and the CFT interpretation is tackled in section \ref{cftinterp}.

Hereafter, we set the speed of light to unity, $c = 1$ and use the signature $(+,-,-,-,\cdots)$.

\section{Weakly turbulent instability of anti-de Sitter}
\label{turbulent}

In 2011, Bizo\'n and Rostworowski studied the collapse of a spherically symmetric scalar wave packet in AdS. The setup of their
experiment was reused so often that we deem useful to reproduce it here.

\subsection{Einstein-Klein-Gordon equations in spherical symmetry}

Consider the Einstein-Klein-Gordon (EKG) system of equations with cosmological constant in four dimensions:
\numparts
\begin{eqnarray}
\label{einsteinscalara}%
   G_{\alpha\beta} + \Lambda g_{\alpha\beta} &= 8\pi G\left(\nabla_\alpha\phi\nabla_\beta \phi - \frac{1}{2}g_{\alpha\beta} \nabla_\mu \phi \nabla^\mu \phi\right),\\
\label{einsteinscalarb}%
   \nabla_\mu \nabla^\mu \phi &= 0,
\end{eqnarray}
\endnumparts
where $g_{\alpha\beta}$ is the metric, $G_{\alpha\beta}$ is Einstein's tensor, $\nabla$ is the connection associated to the
metric, and $\phi$ is a real massless scalar field. In spherical symmetry and in conformal coordinates, the metric can be put into
the form (note that both $t$ and $x$ coordinates have no dimensions, the physical time is thus $Lt$)
\begin{equation}
   ds^2 = \frac{L^2}{\cos^2x}(-Ae^{-2\delta}dt^2 + A^{-1}dx^2 + \sin^2xd\Omega^2), \quad x \in \left[0, \frac{\pi}{2}\right[,
   \label{ansatzscalar}
\end{equation}
where $d\Omega^2 = d\theta^2 + \sin^2\theta d\varphi^2$ is the angular part of the length element. The metric functions $A$, $\delta$ and
scalar field $\phi$ are supposed to depend only on $(t,x)$ and the AdS boundary lies at $x = \pi/2$. Accordingly, we choose notations in
which overdots and primes indicate time and radial derivatives respectively. Defining
\begin{equation}
   \Phi = \phi' \quad \tn{and} \quad \Pi = A^{-1}e^\delta \dot{\phi},
\end{equation}
the system of equations \eref{einsteinscalara}-\eref{einsteinscalarb} boils down to evolution equations for the dynamical variables (in units $4\pi
G = 1$)
\numparts
\begin{eqnarray}
\label{timea}
   \dot{\Phi} &= (Ae^{-\delta}\Pi)',\\
\label{timeb}
   \dot{\Pi} &= \frac{1}{\tan^2x}(\tan^2x A e^{-\delta}\Phi)',
\end{eqnarray}
\endnumparts
and constraint equations
\numparts
\begin{eqnarray}
   \label{time3}
   \dot{A} &= -2\sin x \cos x A^2 e^{-\delta} \Phi\Pi,\\
   \label{cons1}
   A' &= \frac{1 + 2\sin^2x}{\sin x \cos x}(1 - A) - \sin x \cos x A (\Phi^2 + \Pi^2),\\
   \label{cons2}
   \delta' &= -\sin x\cos x(\Phi^2 + \Pi^2).
\end{eqnarray}
\endnumparts
Because the scalar field is massless, the AdS length scale $L$ drops out of the equations. The variables $\Phi$ and
$\Pi$ are evolved in time with \eref{timea}-\eref{timeb} while the constraints \eref{cons1}-\eref{cons2} are used to update the metric
at each time step. The equation \eref{time3} is used as a monitor of code precision. This system of equations supplied with
Dirichlet boundary conditions and compatible initial data is locally well-posed\cite{Holzegel12,Holzegel13a}.

\subsection{Numerical observations}

The initial data is chosen to be a localised wave packet with Gaussian shape:
\begin{equation}
   \Phi(0,x) = 0 \quad \tn{and} \quad \Pi(0,x) = \varepsilon \exp\left( -\frac{\tan^2x}{\sigma^2} \right),
\end{equation}
where $\varepsilon$ denotes the amplitude and $\sigma$ the width of initial data. Apparent horizon formation is signalled by the
vanishing of the blackening factor $A$ in the metric \eref{ansatzscalar}, such that the apparent horizon radius lies at its
largest root. More numerical details can be found in \cite{Maliborski13c}.

Letting the system evolve in time leads to the following observations. If the amplitude is large, the scalar wave packet directly
collapses to a black hole. Lowering the amplitude and repeating the experiment, the horizon radius $x_H$ decreases and eventually
tends to zero when a critical amplitude $\varepsilon_0$ is reached. This would have been the end of the story in asymptotically
flat space-times, as was noted by Choptuik and collaborators in 1993 \cite{Abraham93,Choptuik93}.

If the simulation is run with an amplitude $\varepsilon < \varepsilon_0$, the scalar field starts to contract but does not
form a black hole. It then spreads and reaches infinity in a finite time slightly larger than the null geodesic\footnote{It depends
also on how large the initial data is, i.e.\ how large is the parameter $\sigma$. The wider the initial data, the shorter the
time to reach the boundary.} one $t \gtrsim \pi L/2$, as was noted later in \cite{Garfinkle12}. Because of the reflective
boundary conditions, the field bounces off the AdS boundary. When it comes back to the origin, self-gravitation had the time
to build up a more peaked scalar field profile, so much as to collapse to a black hole when approaching the origin $x = 0$. This leads to a
second branch of collapsing solutions that undergoes one reflection.

When the amplitude is lowered down to a second critical value $\varepsilon_1 < \varepsilon_0$, the resulting black hole has a
horizon radius going to zero. Initial data with amplitudes smaller that $\varepsilon_1$ have to bounce off the AdS
boundary twice before collapsing. And so on and so forth. A sequence of critical amplitude $\varepsilon_n$ can then be
constructed, where $n$ is the number of bounces of the initial data having amplitudes $\varepsilon_n < \varepsilon <
\varepsilon_{n-1}$. This behaviour is illustrated in figure \ref{adsinstability}. Note that on this plot, $x_H$ denotes the horizon radius
right at the point of collapse. In the long term evolution, after several partial absorptions and reflections of the scalar field
on the horizon, all the scalar field falls into the black hole and the metric settles down to the Schwarzschild-AdS solution with a mass
parameter in agreement with the mass of the initial data \cite{Garfinkle12}. 

Furthermore, it was mathematically proved in \cite{Holzegel13a} that the Schwarzschild-AdS solution was non-linearly stable
in spherical symmetry. Soon after black hole formation, the space-time thus settles down to a stable and stationary Schwarzschild
family of solution. The striking feature unveiled by \cite{Bizon11} is that however small the amplitude of the initial data is, a
black hole is formed all the same. This suggests to look at a perturbative approach and see if some indications of collapse can
be inferred.

\begin{myfig}
   \includegraphics[width = 0.49\textwidth]{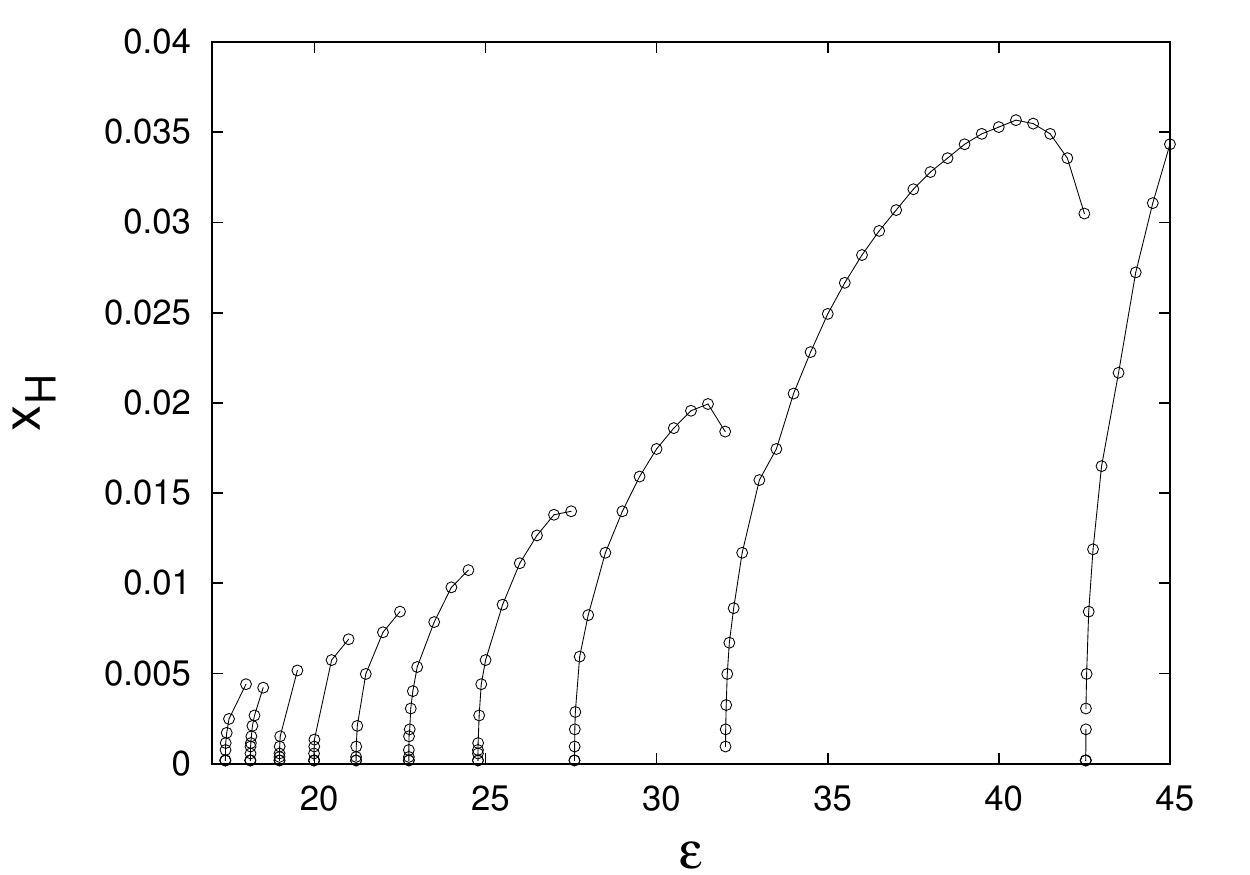}
   \includegraphics[width = 0.49\textwidth]{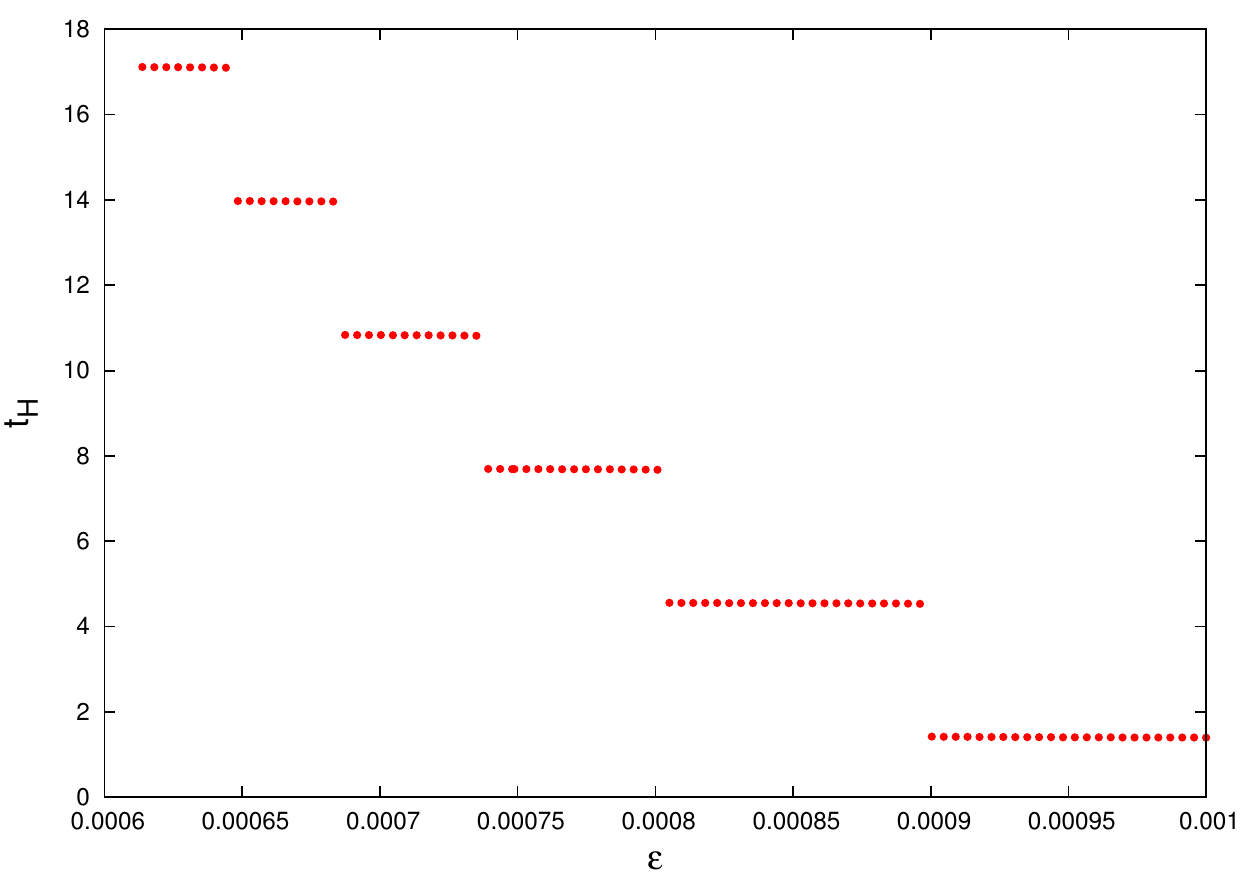}
   \caption{Left: horizon radius $x_H$ as a function of the amplitude $\varepsilon$ of the initial
   wave packet. The far right curve describes the prompt collapse of the scalar wave packet, the other curves on the left are
   respectively for 1 to 9 reflections off the AdS boundary. Right: time of black hole formation $t_H$ as a function of the
   initial amplitude of the scalar wave packet. Each step corresponds to an additional reflection off the AdS boundary. The
   steps are separated by an amount of time $\Delta t \gtrsim \pi L$, i.e.\ slightly larger than the time of a null round
   trip in AdS. Credits: \cite{Bizon11,Jalmuzna11}.}
   \label{adsinstability}
\end{myfig}

\subsection{Perturbative approach}
\label{pertscal}

We denote by $\varepsilon$ the small parameter encoding the initial data amplitude. The EKG system of equations
\eref{einsteinscalara}-\eref{einsteinscalarb} is invariant under the transformation $\phi \to -\phi$. Thus, under the transformation $\varepsilon \to
-\varepsilon$, the scalar field should just change sign and the metric change not at all. This is why the three functions
admits the following even-odd expansion:
\numparts
\begin{eqnarray}
\phi &= \varepsilon\phi_1 + \varepsilon^3 \phi_3 + \ldots,\\
A &= 1 - \varepsilon^2 A_2 - \ldots,\\
\delta &= \varepsilon^2\delta_2 + \ldots.
\end{eqnarray}
\endnumparts
At first order $O(\varepsilon)$, the EKG system boils down to
\begin{equation}
   \ddot{\phi}_1 + \widehat{L}\phi_1 = 0 \quad \tn{with} \quad \widehat{L} = -\frac{1}{\tan^2x}\partial_x(\tan^2x\partial_x).
   \label{pert1}
\end{equation}
This equation can be solved by diagonalising the operator $\widehat{L}$. Defining an inner product on the Hilbert space of solutions as
\begin{equation}
   (f,g) = \int_0^{\pi/2}f(x)g(x)\tan^2x dx,
\end{equation}
an orthonormal basis of solutions is \cite{Maliborski13c}
\numparts
\begin{eqnarray}
\label{eigenmodea}%
   \widehat{L}e_j &= \omega_j^2 e_j \quad \tn{where} \quad \omega_j^2 = (3 + 2j)^2\quad \tn{and} \quad j \in \mathbb{N},\\
\label{eigenmodeb}%
   e_j(x) &= d_j \cos^3 x P_j^{\left(\frac{1}{2},\frac{3}{2}\right)}(\cos(2x)) \quad \tn{with} \quad d_j = \frac{2\sqrt{j!(j+2)!}}{\Gamma(j+3/2)},
\end{eqnarray}
\endnumparts
where $\Gamma$ is Euler's Gamma function and $P_j^{\left(\frac{1}{2},\frac{3}{2}\right)}$ are Jacobi polynomials. All eigenvalues
are real and positive because the operator $\widehat{L}$ is self-adjoint. This means that no eigenmode is unstable,
which is consistent with the linear stability of AdS space-time \cite{Avis78,Breitenlohner82,Abbott82}. In addition, all frequencies
are equidistant: the spectrum is said to be resonant. We can now solve \eref{pert1} through decomposition on the basis
$(e_j)_{j \in \mathbb{N}}$, which gives:
\begin{equation}
   \phi_1(t,x) = \sum_{j=0}^\infty a_j \cos(\omega_j t + \beta_j)e_j(x),
   \label{phi1}
\end{equation}
where $a_j$ and $b_j$ are real constants. In other words, at first order the solution is merely oscillating in time with a spatial
dependence governed by the $(e_j)_{j \in \mathbb{N}}$ functions.

At second order, the equations relative to $A_2$ and $\delta_2$ admit the following solutions:
\numparts
\begin{eqnarray}
   A_2(t,x) &= \frac{\cos^3x}{\sin x}\int_0^x[\dot{\phi_1}^2(t,y) + {\phi_1'}^2(t,y)]\tan^2ydy,\\
   \delta_2(t,x) &= -\int_0^x[\dot{\phi_1}^2(t,y) + {\phi_1'}^2(t,y)]\sin y \cos ydy.
\end{eqnarray}
\endnumparts
At third order, it comes
\begin{equation}
   \ddot{\phi_3} + \widehat{L}\phi_3 = S(\phi_1,A_2,\delta_2) \equiv -2(A_2 + \delta_2)\ddot{\phi}_1 - (\dot{A}_2 + \dot{\delta}_2)\dot{\phi}_1 - (A_2' + \delta_2')\phi_1'.
   \label{phi3}
\end{equation}
Projecting this equation on the basis $(e_j)_{j \in \mathbb{N}}$ yields, denoting $c_j^{(3)} = (\phi_3,e_j)$ and $S_j = (S,e_j)$:
\begin{equation}
   \forall j, \quad \ddot{c}_j^{(3)} + \omega_j^2 c_j^{(3)} = S_j.
   \label{cj3}
\end{equation}
A straightforward (but tedious) look at the right-hand side of \eref{cj3} makes it clear that it contains some resonant terms
$\cos(\omega_jt)$ or $\sin(\omega_j t)$ every time there exists a triplet $(j_1,j_2,j_3)$ such that
\begin{equation}
   a_{j_1} \neq 0, a_{j_2} \neq 0, a_{j_3} \neq 0 \quad \tn{and} \quad \omega_j = \omega_{j_1} + \omega_{j_2} - \omega_{j_3}.
\end{equation}
This is due to the resonant character of the spectrum of the operator $\widehat{L}$ and gives rise to secular resonances, i.e.\ solutions of the form
\begin{equation}
   \phi_3 \sim t\sin(\omega_jt) + \ldots .
\end{equation}
These solutions are thus diverging linearly in time. We infer that when $\phi_3$ and $\phi_1$ are of the same order of magnitude
(namely after a time $t = O(\varepsilon^{-2})$), the perturbative scheme breaks down, as we expect higher orders terms to be smaller than leading terms
in convergent series. Such resonances are quite common in perturbative expansions and they can sometimes be cured by redefining the
expansion parametrisation. For example the Poincar\'e-Lindstedt method consists in expanding the frequencies $\omega_j$ as
\begin{equation}
   \omega_j = \omega_j^{(0)} + \varepsilon^2 \omega_j^{(2)} + \ldots .
\end{equation}
Substituting this expression into \eref{phi1} and \eref{phi3} leads to the suppression of many secular resonances if the
constants $\omega_j^{(2)}$ are chosen astutely. However, in the case under study, if some resonances of \eref{cj3} are indeed
removable, others are not and the expansion is truly diverging on time-scales $t= O(\varepsilon^{-2})$ \cite{Bizon11}.

The perturbative approach thus provides an analytical argument in favour of the black hole formation: any small perturbation
cannot remain small and it takes a time $t= O(\varepsilon^{-2})$ to reach the fully-non linear regime. This is in very good
agreement with numerical results, for which the time of collapse does indeed scales as the inverse square of the amplitude (figure
\ref{adsinstability}). In analogy with quantum mechanics, this statement describes the instability in position-space. In order to get
the momentum-space picture, we can take advantage of the perturbative approach and define the energy per mode of a solution.

\subsection{Energy per mode}

For convenience, we introduce
\begin{equation}
   \quad \Pi_j = (\sqrt{A}\Pi,e_j) \quad \tn{and} \quad \Phi_j = (\sqrt{A}\Phi,e_j'),
\end{equation}
the projections of $\Pi$ and $\Phi$ on the bases $(e_j)_{j \in \mathbb{N}}$ and $(e_j')_{j \in \mathbb{N}}$
respectively (recall that $e_j' = \frac{d e_j}{dx}$). Note that $(e_i,e_j) = \delta_{ij}$ but
$(e_i',e_j') = \omega_j^2 \delta_{ij}$. The total mass of the system can be expressed as\footnote{Recall that the energy density
   $\rho$ measured by a static observer at infinity is $T_{tt}$ where $T_{\alpha\beta} = \left(\nabla_\alpha\phi\nabla_\beta\phi -
   \frac{1}{2}g_{\alpha\beta} \nabla_\mu \phi\nabla^\mu \phi\right)$, so that $\rho = \frac{A^2e^{-\delta}}{2}(\Phi^2 + \Pi^2)$. Due
to spherical symmetry, the mass can be computed by integrating the energy density, but with a rescaled volume form. This is
reminiscent of the famous Tolman-Oppenheimer-Volkoff equations and mass function for spherically symmetric neutron stars.}
\begin{equation}
   M = \frac{1}{2}\int_0^{\pi/2}(A\Phi^2 + A\Pi^2)\tan^2x dx.
\end{equation}
Applying Parseval's identity, it comes
\begin{equation}
   M = \sum_{j=0}^{\infty} E_j(t) \quad \tn{with} \quad E_j = \Pi_j^2 + \frac{\Phi_j^2}{\omega_j^2}.
\end{equation}
We can then reasonably interpret $E_j$ as a proxy for the energy contained in the $j^{th}$ mode (recall however that energy is not local).
It turns out that the signature of the instability is very clear when considering the energy per mode: it is characterised by a
weakly turbulent cascade.

\subsection{Weakly turbulent cascade}
\label{weakturbu}

In order to better understand the mechanism of black hole formation, the authors of \cite{Bizon11} tried to evolve
initial data obtained with a single eigenmode of the linear operator $\widehat{L}$, namely
\begin{equation}
   (\phi,\dot{\phi})_{t=0} = \varepsilon(e_0(x),0).
   \label{singlemode}
\end{equation}
From a perturbative point of view, this single-mode initial data displays a single resonant term in the expression of $S_0$, and
this term can indeed be removed by Poincar\'e-Lindstedt regularisation. It suggests that such initial data should not collapse into a
black hole. This was actually checked via numerical evolution in time: a single-mode excitation is indeed non-linearly stable.

On the other hand, the so-called two-mode initial data (coefficients are defined in equation \eref{eigenmodea}-\eref{eigenmodeb})
\begin{equation}
   (\phi,\dot{\phi})_{t=0} = \varepsilon\left(\frac{e_0(x)}{d_0} + \frac{e_1(x)}{d_1},0\right),
   \label{twomode}
\end{equation}
displays an irremovable secular resonance and does lead to black hole formation in a time $t = O(\varepsilon^{-2})$. It was then
numerically observed in \cite{Bizon11} that the energy of the system was cascading to higher order modes and hence higher spatial frequencies. This
cascade behaviour is illustrated on figure \ref{turbu}. Black hole formation then provides a natural cut-off that eventually
forbids the transfer of energy to smaller and smaller scales indefinitely.

\begin{myfig}
   \includegraphics[width = 0.49\textwidth]{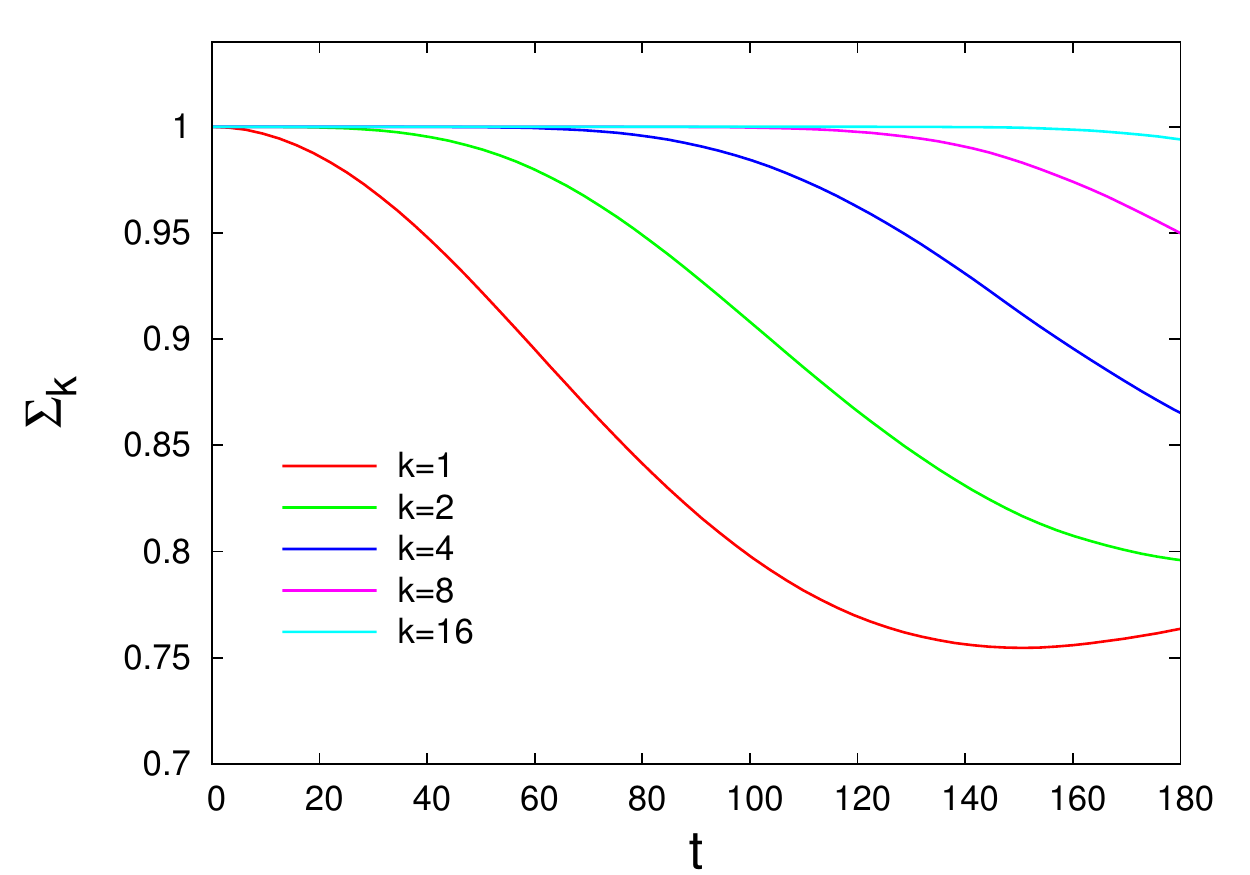}
   \caption{Fraction of the total energy contained in the first $k+1$ modes $\Sigma_k =
      \sum_{j=0}^k E_j$, for an initial excitation of the $j=0$ and $j=1$ modes (equation \eref{twomode}). The energy in the first sixteen modes
      $\Sigma_{16}$ is almost constant while the energy in the first modes $\Sigma_1$, $\Sigma_2$, etc., decreases, which means
      that the energy is flowing to higher modes. Credits: \cite{Bizon11}.}
      \label{turbu}
\end{myfig}

\subsection{The anti-de Sitter instability conjecture}

The important features of the simulations are the following:
\begin{myitem}
   \item a black hole is formed however small the initial amplitude of the perturbation is,
   \item the formation time $t_H$ of the apparent horizon scales like $O(\varepsilon^{-2})$,
   \item energy is flowing from low to high spatial frequencies during evolution.
\end{myitem}

The first and third properties gave rise to the so-called weakly turbulent behaviour. This lead to the following instability
conjecture \cite{Bizon14}:

\begin{conjecture}[Anti-de Sitter instability]
AdS is unstable against black hole formation for a large class of arbitrarily small perturbations.
\end{conjecture}

In \cite{Friedrich14}, it was argued that the reflective boundary conditions were a key ingredient of the
instability precisely because of the absence of a decay condition for perturbations. Let us also mention that a diverging growth
of the frequency of fluctuations was already observed in \cite{Greenwood10} in a AdS space within a big crunch scenario.

\subsection{Other features of the instability}

From the null geodesic knowledge (see figure \ref{geodesics}), it is expected that the crossing time of a massless scalar wave packet
is $\sim \pi L$. However, the authors of \cite{Garfinkle12} observed that the time to form a black
hole after one reflection was slightly larger than the time needed for null geodesics to do one round trip in AdS space-time.
Non-linearities thus tend to slow down the massless scalar field. A snapshot of the time-radial plane of the evolution of the
scalar field is pictured in figure \ref{timeradial} for both direct and delayed collapse.

\begin{myfig}
   \includegraphics[width = 0.49\textwidth]{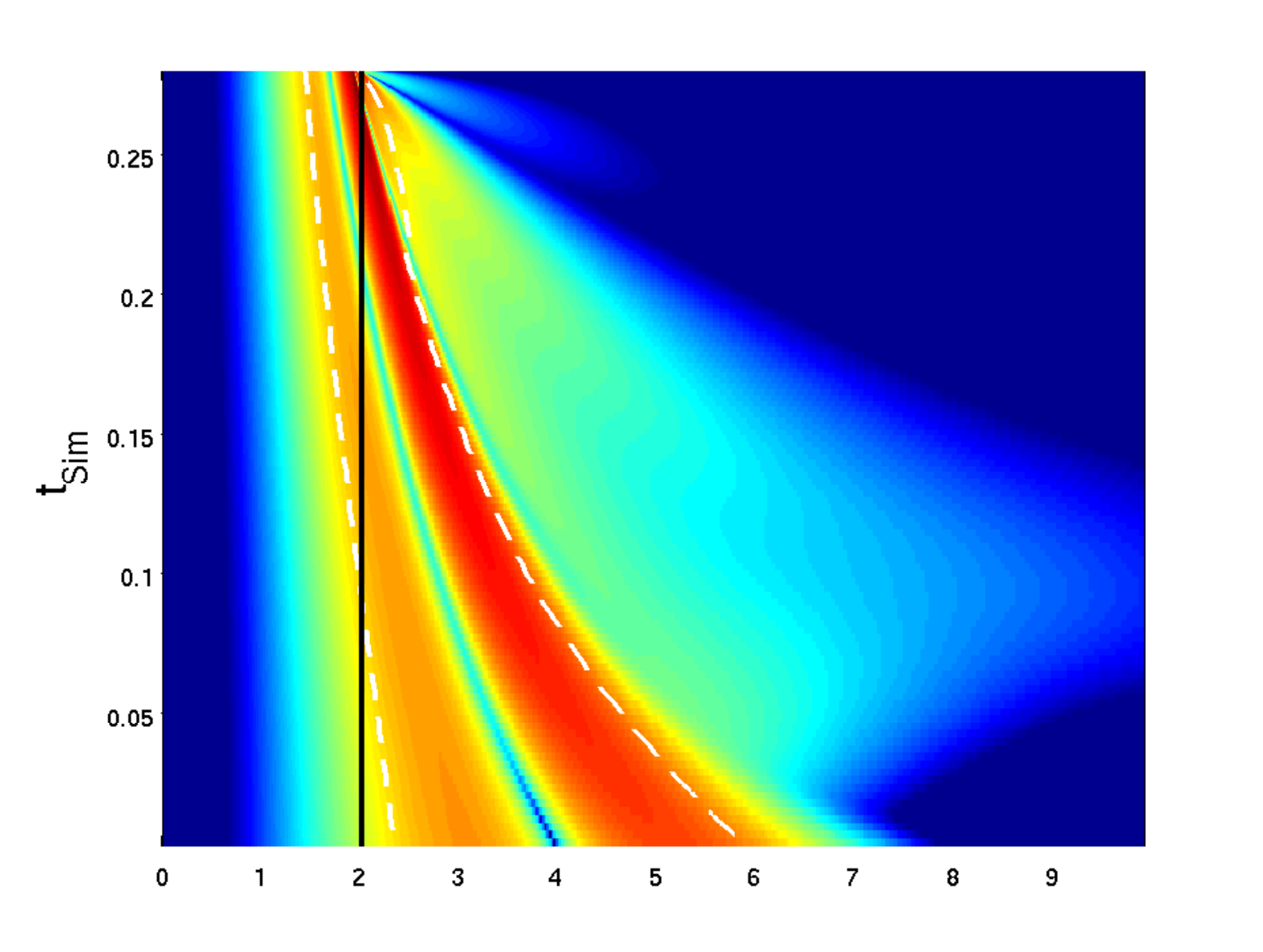}
   \includegraphics[width = 0.49\textwidth]{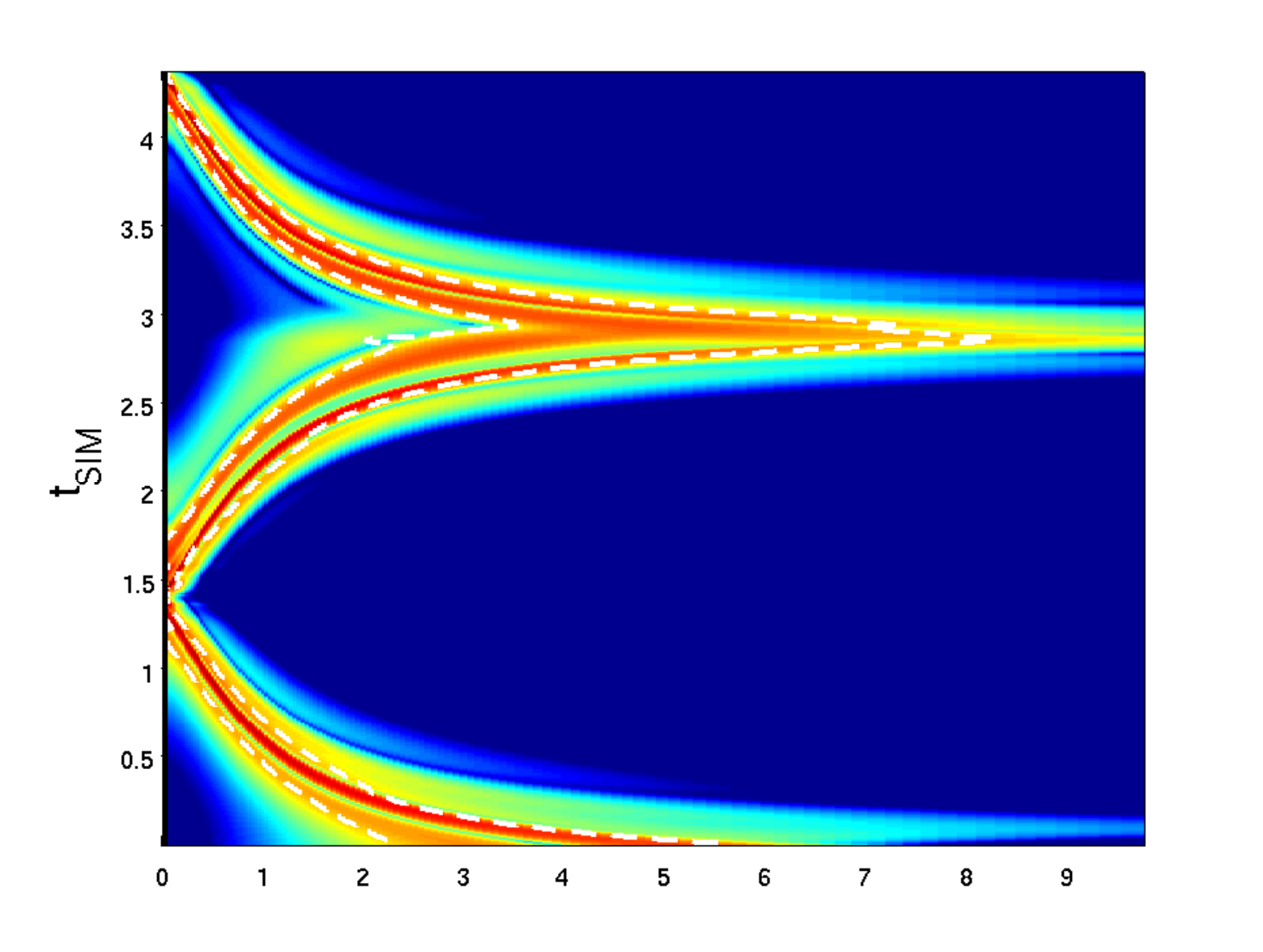}
   \caption{Time-radial planes during the evolution of a
      Gaussian massless scalar field initial data. The radii are in units of $L$, and colour encodes the scalar energy density.
      Vertical black lines indicate the apparent horizon $r_H$ appearing at the end of the simulation. The initial amplitude is
      150 times smaller on the right panel and the corresponding apparent horizon lies at $r_H = 0.0049 L$ after one bounce
      off the AdS boundary. This figure is highly reminiscent of figure \ref{geodesics} for null geodesics. Credits: \cite{Garfinkle12}.}
   \label{timeradial}
\end{myfig}

The instability in five dimensions was first investigated in \cite{Garfinkle11}. In contradiction with \cite{Bizon11}, they did
not observe black hole formation below a certain amplitude threshold. However, this was only due to their spatial resolution (6400
points): the apparent horizon $x_H$ could not be resolved with such a few number of points. This was demonstrated in
\cite{Jalmuzna11} whose authors used $2^{17} \sim 130 000$ spatial points.

The instability is not only present in four and five dimensions, but in all dimensions. Indeed, in \cite{Jalmuzna11}, it was
demonstrated that the spectrum of the linear operator $\widehat{L}$ was resonant in all dimensions, with equally spaced
eigen frequencies. The perturbative approach always gives rise to secular resonant terms at third order, indicating a
breakdown at time $t = O(\varepsilon^{-2})$, independent of the number of dimensions.

More and more arguments were gathered suggesting that the instability was systematic, i.e.\ independent of the initial data. For
example, the instability was recovered for complex scalar fields \cite{Buchel12,Liebling13}. Let us also mention the Vaidya setup
of \cite{Abajo14,Silva15} where the Gaussian wave-packet is replaced by a Gaussian shell with initial data:
\begin{equation}
   \Phi(0,x) = 0 \quad \tn{and} \quad \Pi(0,x) = \varepsilon \exp\left( -\frac{\tan^2\left(\frac{\pi}{2} - x\right)}{\sigma^2} \right),
\end{equation}
where the scalar field is initially concentrated close to the AdS boundary.

The weakly turbulent behaviour is not solely an intrinsic property of AdS, but it was also observed in flat space-time enclosed in a
cavity. The underlying idea is that the reflective boundary conditions of AdS space-times can be mimicked by a flat
space-time with appropriate boundary conditions at a finite radius. However, this analogy holds only for spherically
symmetric distributions. Indeed, the bouncing time is not isotropic in such a cavity if the matter distribution is not spherically
symmetric, whereas AdS space-time is strictly isotropic, whatever is the initial distribution. In \cite{Maliborski12},
Dirichlet boundary conditions were imposed at a finite radius $R$ of Minkowski space-time, with Gaussian initial data. Again,
arbitrarily small amplitudes lead to black hole formation after potentially several bounces on the $r = R$ wall.  Figure
\ref{energy_spec} shows the spectrum $E_j$ versus $j$ of the data at different times of the evolution. The turbulent cascade
toward high-$j$ modes is visible and just before black hole formation, the spectrum approaches a power law of exponent $\alpha
\sim -1.2$. This value seems universal since it is independent of the functional form of the initial data and is also observed in
the 4-dimensional AAdS case.

\begin{myfig}
   \includegraphics[width = 0.49\textwidth]{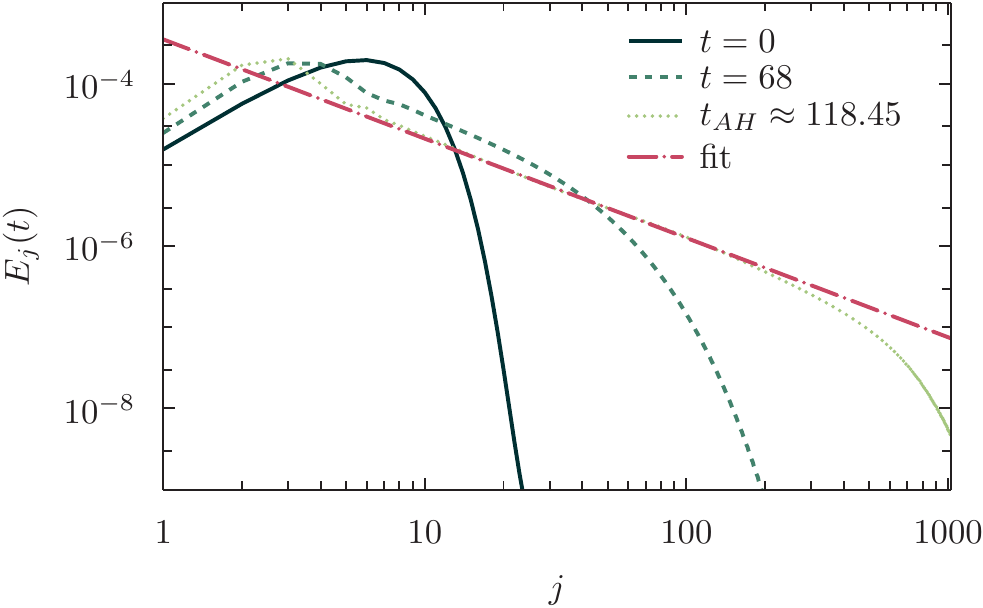}
   \caption{Energy spectrum at different times before collapse of a solution undergoing several
      dozens of reflections in flat space-time enclosed in a cavity. Just before the apparent horizon formation at time $t_{AH}$,
      the spectrum is approximately a power law with a universal exponent $\alpha \sim -1.2$. Credits: \cite{Maliborski12}.}
   \label{energy_spec}
\end{myfig}

Finally, let us observe that the vocabulary of turbulence interfered with the AdS instability field mainly because of the cascade
of energy to higher spatial frequencies. The analogy with fluid turbulence was even pushed further with the study of time frequencies
in \cite{Oliveira13}. In particular, it was shown that the power spectrum of the Ricci scalar at the origin was characterised by a
Kolmogorov-Zakharov power spectrum, i.e.\
\begin{equation}
   P(\omega) = \omega^{-s}\quad \tn{with} \quad s = 1.7 \pm 0.1,
\end{equation}
$\omega$ being the time frequency. The numerical value of $s$ seems to be universal as it holds in both four and five dimensions with
Gaussian or 20-mode initial data.

\section{Black hole formation in AdS space-times}
\label{bhformation}

The details of black hole formation in general relativity (GR) are in striking analogy with phase transitions. This analogy was uncovered by the
seminal work of Choptuik and collaborators \cite{Choptuik93,Abraham93} in asymptotically flat space-times. Since black holes form
more easily in AAdS space-times, and given that there are several branches of black hole formations, it is legitimate to ask how
close to the asymptotically flat case the formation of black hole with a negative cosmological constant is.

\subsection{Critical phenomena in AAdS space-times}
\label{critical}

From an historical perspective, it is important to mention the numerical simulations by Pretorius and Choptuik
in 2000 \cite{Pretorius00} and Husain and collaborators in 2003 \cite{Husain03} in AAdS space-times. In these papers, the
authors evolved in time a spherically symmetric Gaussian wave packet initial data made of a massless scalar field. Like in the asymptotically
flat case, if the scalar wave packet amplitude $\varepsilon$ is larger than a critical value $\varepsilon_\star$, then the scalar
field collapses directly to a black hole with apparent horizon $r_H$. This corresponds to the far right curve of figure
\ref{adsinstability} (left panel). The goal of these papers, however, was to characterise the critical behaviour of black hole
formation. Namely, on the point of collapse, for amplitudes $\varepsilon \gtrsim \varepsilon_\star$, the authors observed that
the apparent horizon radius $r_H$ was governed by
\begin{equation}
   \ln r_H \sim \gamma_r \ln (\varepsilon - \varepsilon_\star) + r_0 + F_r(\ln(\varepsilon - \varepsilon_\star)),
\end{equation}
where $r_0$ is a constant and $F_r$ a sinusoidal function of period (or echoing period) $\Delta_r$. This is a typical feature of
critical phenomena and phase transitions, illustrated on the left panel of figure \ref{critic}. From the numerical simulations, it
was measured that
\numparts
\begin{eqnarray}
   \label{criticscalara}
   \gamma_r &\simeq 1.2 \quad \tn{in 3 dimensions},\\
   \label{criticscalarb}
   \gamma_r &\simeq 0.37 \quad \tn{and} \quad \Delta_r \simeq 3.44 \quad \tn{in 4 dimensions.}
\end{eqnarray}
\endnumparts
These values are universal, in the sense that they are independent of the value of the cosmological constant $\Lambda$ and of the
functional form of the initial scalar wave packet. They also match the values found earlier in asymptotically 4-dimensional flat
space-times. This is to be expected since the physics probed is quite local (and thus independent of the asymptotics) for the formation
of arbitrarily small black holes. Unfortunately, at the time, the authors of \cite{Husain03} were not interested in values of
$\varepsilon$ smaller than $\varepsilon_\star$, so they missed the breakthrough of the so-called AdS weakly turbulent
instability.

The three dimensional case of \cite{Pretorius00} is particular because black holes have a minimum mass below which their formation is strictly
impossible (this is the so-called Ba\~{n}ados-Teitelboim-Zanelli (BTZ) metric \cite{Banados92}). It was thus observed in \cite{Pretorius00} that even if
non-linearities build up in such a setting, the scalar field does not collapse even after several bounces off the AdS
boundary. Still, non-linearities build up and a sub-pulse structure emerges, i.e.\ the initial Gaussian profile breaks off into
several distinct wave packets, as shown on the right panel of figure \ref{critic}.

\begin{myfig}
   \includegraphics[width = 0.49\textwidth]{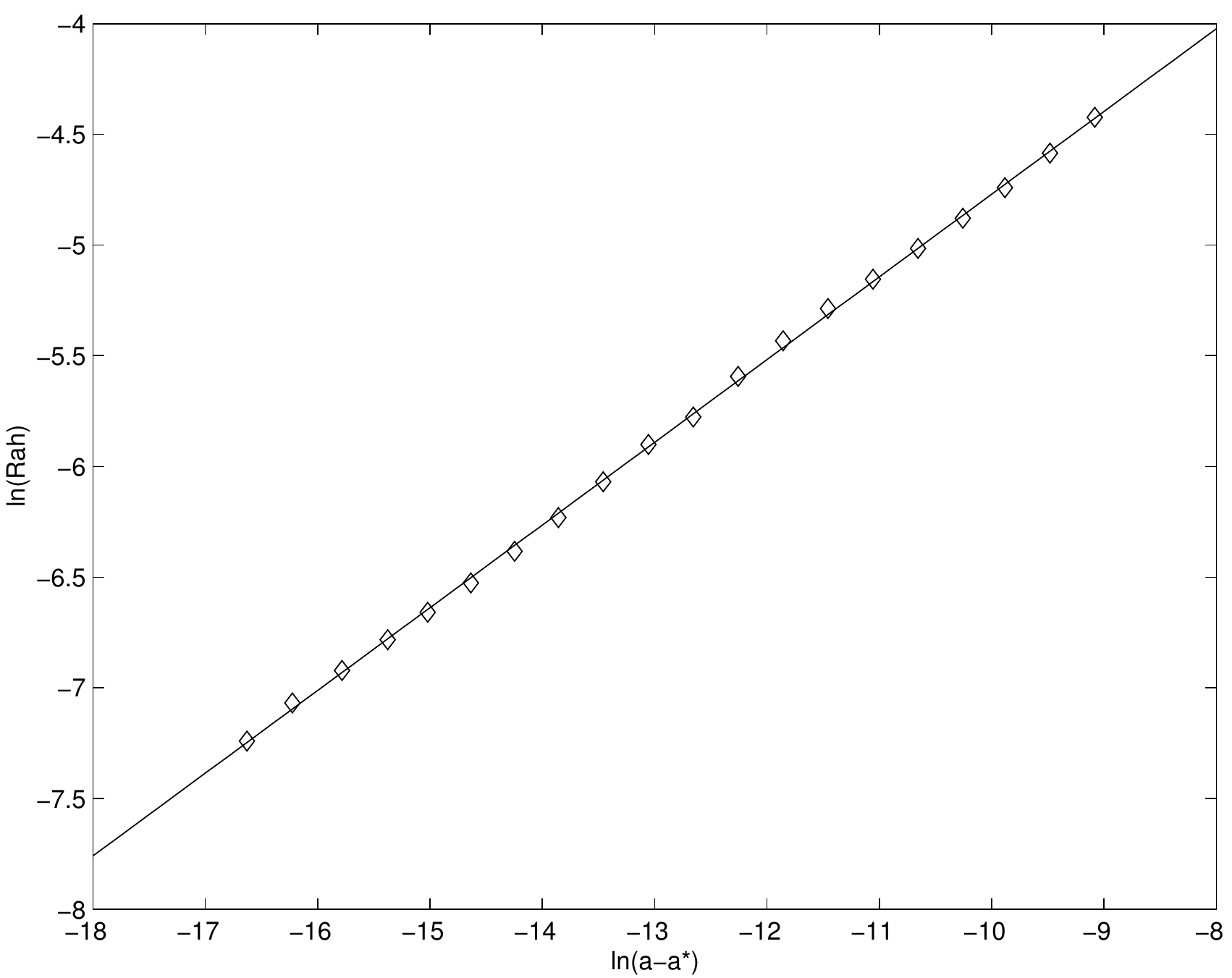}
   \includegraphics[width = 0.26\textwidth]{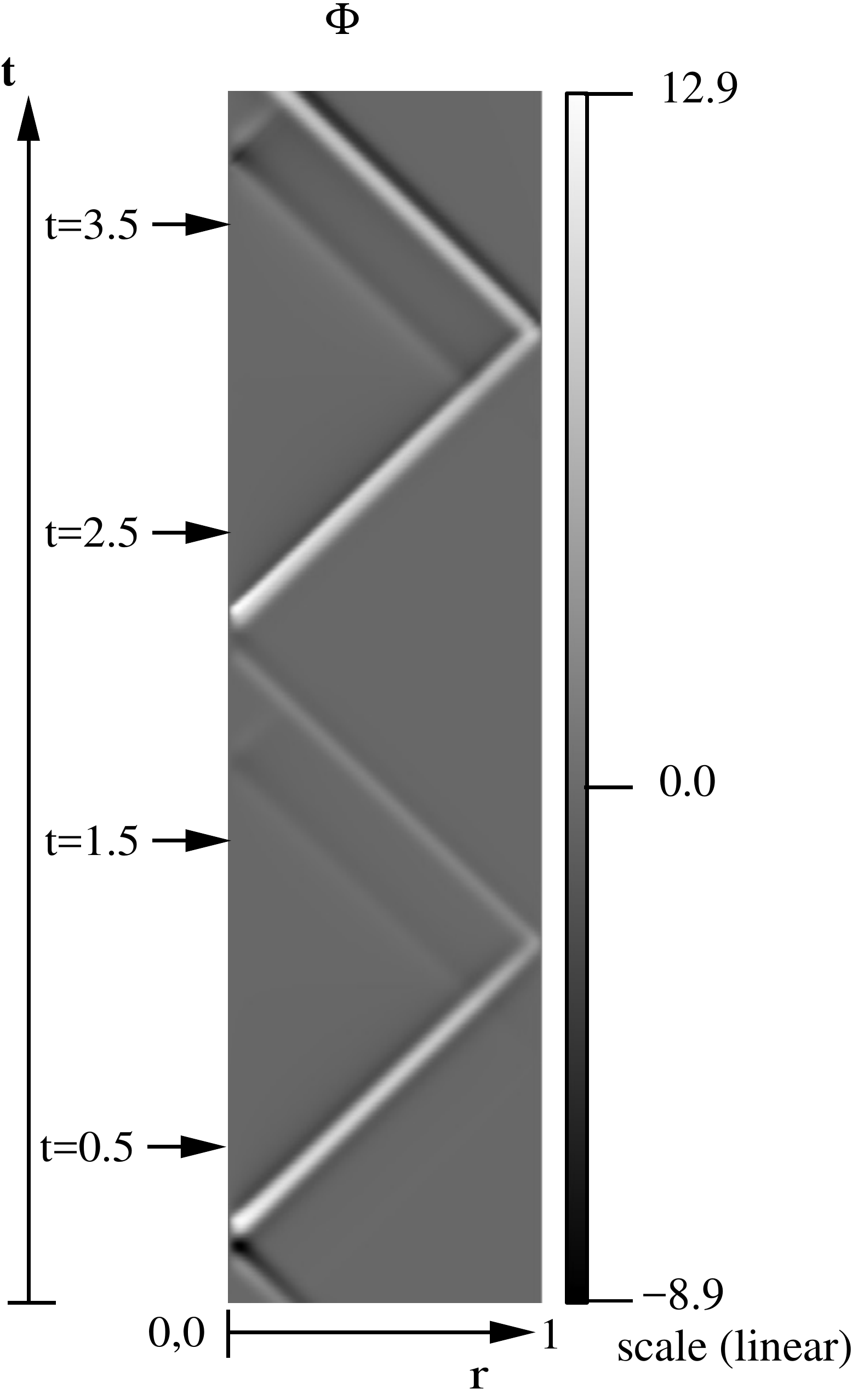}
   \caption{Left: log-log plot of the apparent horizon $r_H$ as a function of the amplitude
      $\varepsilon -\varepsilon_\star$ (denoted by $a$ in the label) of the initial data. Small oscillations of echoing period
      $\Delta$ around the straight line of slope $\gamma$ can be spotted. Right: space-time diagram of the scalar field profile in 3-dimensional
      AdS with Gaussian initial data. The amplitude is smaller than the threshold of black hole formation, so no black hole
      can appear. Credits: \cite{Husain03,Pretorius00}.}
      \label{critic}
\end{myfig}

In the AdS instability context, the authors of \cite{Bizon11} checked that in the right neighbourhood of each critical
amplitude $\varepsilon_n$, i.e.\ for $\varepsilon \gtrsim \varepsilon_n$ (see figure \ref{adsinstability}), the power-law
behaviour of \cite{Choptuik93,Abraham93,Husain03} was recovered, namely for $\varepsilon \gtrsim \varepsilon_n$:
\begin{equation}
   x_H \sim (\varepsilon - \varepsilon_n)^{\gamma_r} \quad \tn{with} \quad \gamma_r \sim 0.37.
\end{equation}

These critical phenomena associated to the AdS instability were refined further in \cite{Olivan16a,Olivan16b}. The authors
were able to resolve precisely the apparent horizon formation and looked at the fine structure of critical collapse. Unlike
previous studies, they focused on the left neighbourhood of critical points that only exist in AdS space-time. Denoting by
$M_g^{n+1}$ the mass gap at which starts the left branch, and $\varepsilon_n$ the corresponding critical amplitude of initial data
undergoing $n$ bounces, they have shown that in the left neighbourhood of critical points ($\varepsilon \lesssim \varepsilon_n$)
the black hole mass $M_H$ was obeying
\begin{equation}
   M_H - M_g^{n+1} \propto (\varepsilon_n - \varepsilon)^{\xi},
   \label{critleft}
\end{equation}
with $\xi \sim 0.7$. This value of $\xi$ is universal, i.e.\ independent of the number of bounces $n$, and of the functional form
of the initial data. Moreover, looking at the maximal value of Ricci scalar at the origin $R_{max}(x=0)$ on the point of collapse,
they observed that in the left neighbourhood of a critical point ($\varepsilon \lesssim \varepsilon_n$)
\begin{equation}
   \ln R_{max}(x=0) = -2\gamma_l \ln (\varepsilon_n - \varepsilon) + b_0 + F_l(\ln(\varepsilon_n - \varepsilon)),
\end{equation}
where again $F_l$ is a sinusoidal function with echoing period equal to $\Delta_l$. The parameters $\gamma_l$ and
$\Delta_l$ are related to their right branch counterparts by (see equations \eref{criticscalara}-\eref{criticscalarb})
\begin{equation}
   \gamma_l = \gamma_r \quad \tn{and} \quad \Delta_l = \frac{\Delta_r}{2\gamma_r}.
   \label{leftright}
\end{equation}
This feature is called the discrete self-similarity near the critical points in AdS space-time and is best illustrated in figure
\ref{selfsim}.

\begin{myfig}
   \includegraphics[width = 0.55\textwidth]{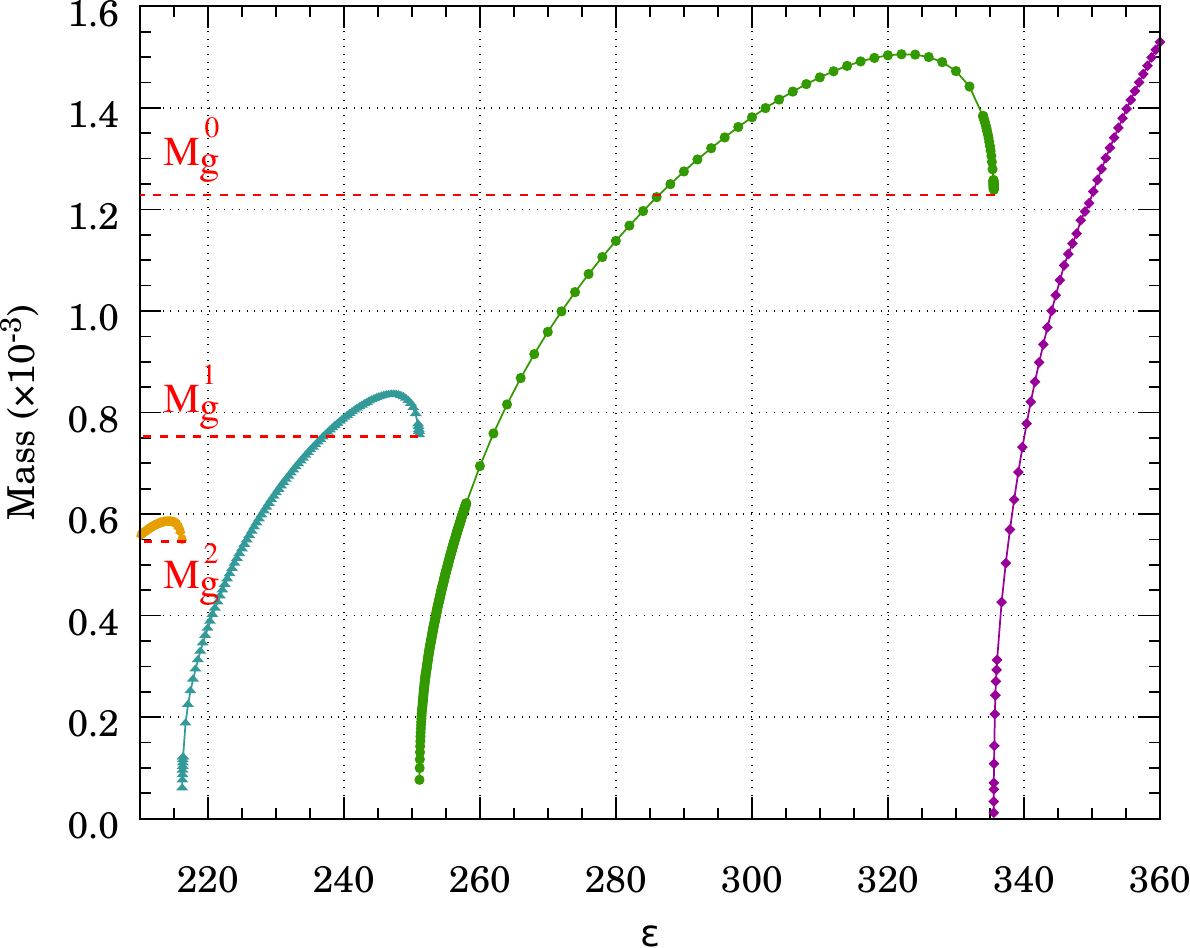}
   \includegraphics[width = 0.44\textwidth]{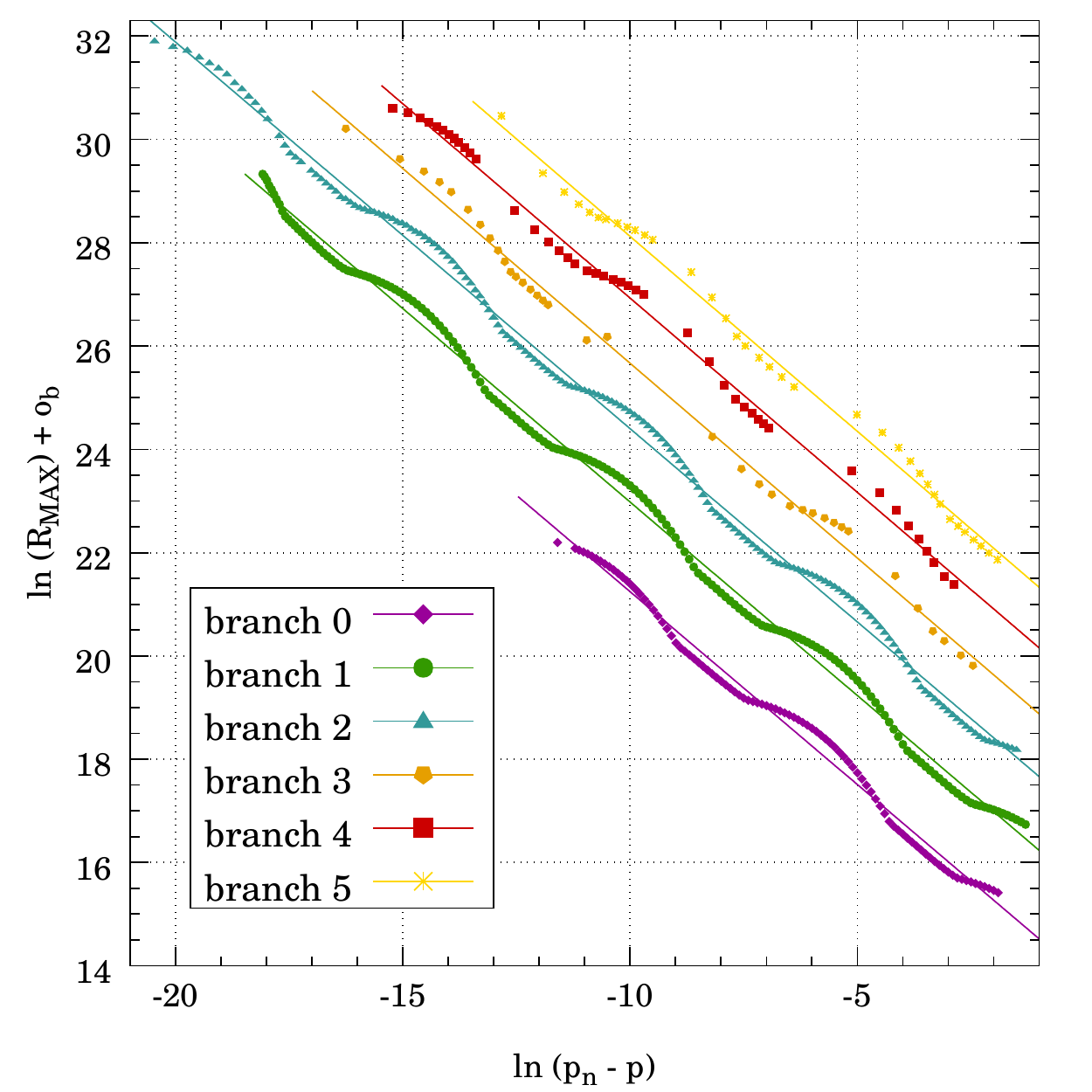}
   \caption{Left panel: mass of the apparent horizon as a
      function of the amplitude of the initial data. In the right neighbourhood of a critical point,
      $M_H \sim (\varepsilon - \varepsilon_n)^{\gamma_r}$ while in the left neighbourhood of a critical point $M_H - M_g^{n+1}
      \sim (\varepsilon_n - \varepsilon)^\xi$. Right panel: critical behaviour of the Ricci scalar at the origin $R_{max}(x=0)$
      for fixed width $\sigma$ of the initial data and for different branches: from direct collapse ($b = 0$) to five-bounce
      collapse ($b=5$). An offset $o_b$ has been added to distinguish between the curves. In the labels, $p$ stands for $\varepsilon$.
      Credits: \cite{Olivan16b}.}
   \label{selfsim}
\end{myfig}

Thus, critical phenomena in AAdS space-times are much richer than in the asymptotically flat case. First, there is an
infinity of black holes formation channels indexed by the number of bounces. Second, each critical point has not only a right
branch but also a left branch (attached to a mass gap), which are related to each other by \eref{leftright}.

\subsection{Singularity theorems in AAdS space-times}

Given the strength of the instability conjecture, a natural question is to know whether we can prove it via a singularity
theorem. In the asymptotically flat case, singularity theorems were proved by Hawking and Penrose \cite{Hawking73}. In a
simplified formulation, the theorems imply that if
\begin{myenum}
   \item the null energy conditions holds, i.e.\ $\forall v, v_{\mu}v^{\mu} \geq 0, R_{\mu\nu}v^\mu v^\nu \geq 0$,
   \item the strong causality or chronology conditions holds, i.e.\ there exists no closed time-like curve,
   \item there exists a region of strong gravity, i.e.\ a closed trapped surface,
\end{myenum}
then the space-time is not time-like nor null geodesically complete, i.e.\ there exist some geodesics that never reach infinity.
Said differently, the space-time is singular and can exhibits a black hole or a naked singularity.

These conditions are discussed in the AAdS case in \cite{Ishibashi12}. In view of the instability conjecture,
condition (c) has to be removed, as it was numerically observed that even weak gravity initial data leads to black hole formation.
However, eliminating this condition rises difficulties that were discussed in details in \cite{Ishibashi12} but not
overcome. Nonetheless, the authors managed to give sufficient conditions for a singularity to form in the simplified case of a
perfect fluid in spherical symmetry, by examining carefully the Raychaudhuri's equation. Notably, the naked singularity formation
was not excluded.

Very recently, the author of \cite{Moschidis17a,Moschidis17b} mathematically and rigorously proved that the spherically symmetric
Einstein-radial massless Vlasov system was non-linearly unstable against black hole formation. This setup, also called the
Einstein-null dust system, is a simplified model of the EKG equations, and can be seen as a high frequency limit of the
latter (some non-linear terms being dropped out). This tour de force can be interpreted as the very first proof of the AdS
instability conjecture in the simplest possible setting, and as such as a specific singularity theorem.

No other attempt of singularity theorem demonstrations in AdS has been attempted to the best of our knowledge. And indeed,
it seems that the formation of black holes is not universal. Several islands of stability\footnote{This denomination was
originally coined in \cite{Dias12b}.} were found in the literature, namely
families of non-linearly stable initial data that never collapse.

\section{Quenching the turbulent cascade}
\label{quenching}

Black holes form for arbitrarily small amplitudes in a very large number of cases (as we have seen in the previous sections),
but is it mandatory? If the instability conjecture was confirmed many times, an even more challenging problem was to find
solutions that resisted black hole formation and circumvented the instability.

\subsection{The hard wall model}

Black hole formation is expected to occur when the energy get concentrated in such a small region that an apparent horizon can form.
The region where such a focus of energy is favoured is obviously the origin in spherical symmetry. What happens then to the
instability if we prevent the scalar field to ever reach the origin? This question was tackled in \cite{Craps14c,Silva16} with
the so-called hard wall implementation. Namely, the authors placed a wall at a radial coordinate $z = z_0$ while the AdS
boundary lied at $z = 0$. The scalar field could only move between these two boundaries, as pictured in figure \ref{hwall}. Furthermore, on
the hard wall, Dirichlet or Neumann boundary conditions were imposed, and both the 3 and 4-dimensional cases were studied.

\begin{myfig}
   \includegraphics[width = 0.49\textwidth]{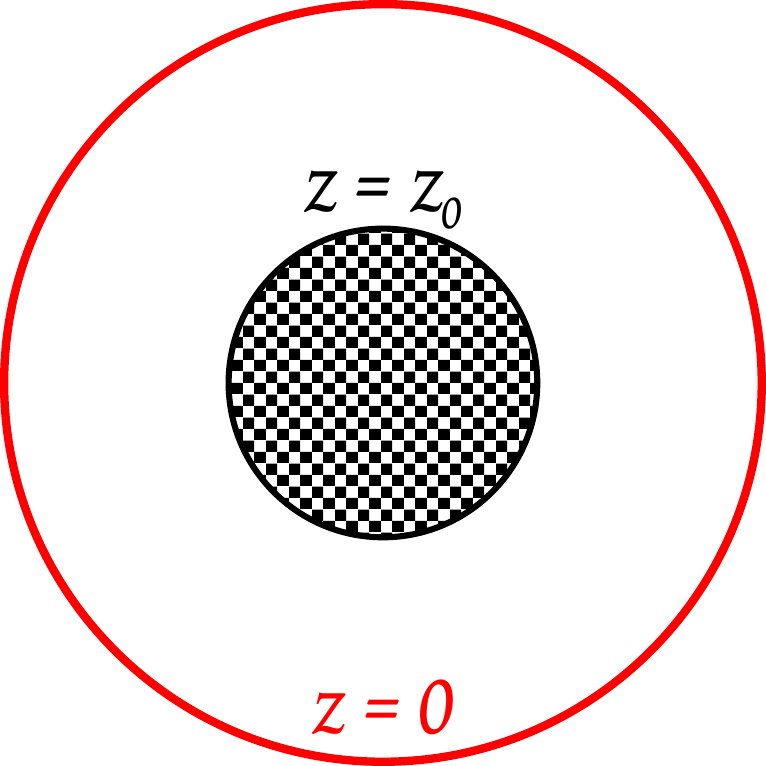}
   \caption{In the hard wall model, a wall, i.e.\ Dirichlet or Neumann boundary conditions, is enforced at a
      coordinate $z_0$. The coordinate $z$ is an inverse radius, so that the AdS boundary lies at $z=0$. The chessboard
      patterned region is forbidden, so that the scalar field is restricted to move only between the wall and the boundary,
   bouncing back and forth between the two.}
   \label{hwall}
\end{myfig}

The only input of data was performed via time-dependent energy injection on the AdS boundary\footnote{For example in
\cite{Krishnan16} the authors advocate that a more natural boundary condition for AdS is to hold the renormalised
boundary stress tensor fixed, instead of the boundary metric.} imposing
\begin{equation}
   \phi(z=0,t) = \varepsilon e^{-\frac{t^2}{\delta t^2}}.
\end{equation}
The authors observed that for small enough amplitudes $\varepsilon$, the scalar pulse generated by the energy injection bounced
forever back and forth between the AdS boundary and the hard wall. For amplitudes larger than some threshold
$\varepsilon_0$, a black hole formed with a horizon smaller than $z_0$ (i.e.\ larger than the radius of the wall). The intuition is
that a black hole is formed if the black hole that would be formed in ordinary AdS space-time (without a hard wall) has
its event horizon outside the hard wall. Otherwise the infalling shell is scattered back by the hard wall before it reaches its
Schwarzschild radius.

This behaviour was observed in all dimensions, with all boundary conditions considered, and is summarised in figure \ref{phasewall}.
These results were sustained by both analytical and numerical arguments. In particular, the frequencies of the linearised modes
did not display obvious resonances, except in the case of Neumann boundary conditions in four dimensions. The hard wall model thus provides a
defocusing mechanism that can suppress the turbulent cascade below a certain amplitude threshold. This is in deep contrast to
unstable initial data that precisely miss the existence of such a threshold.

\begin{myfig}
   \includegraphics[width = \textwidth]{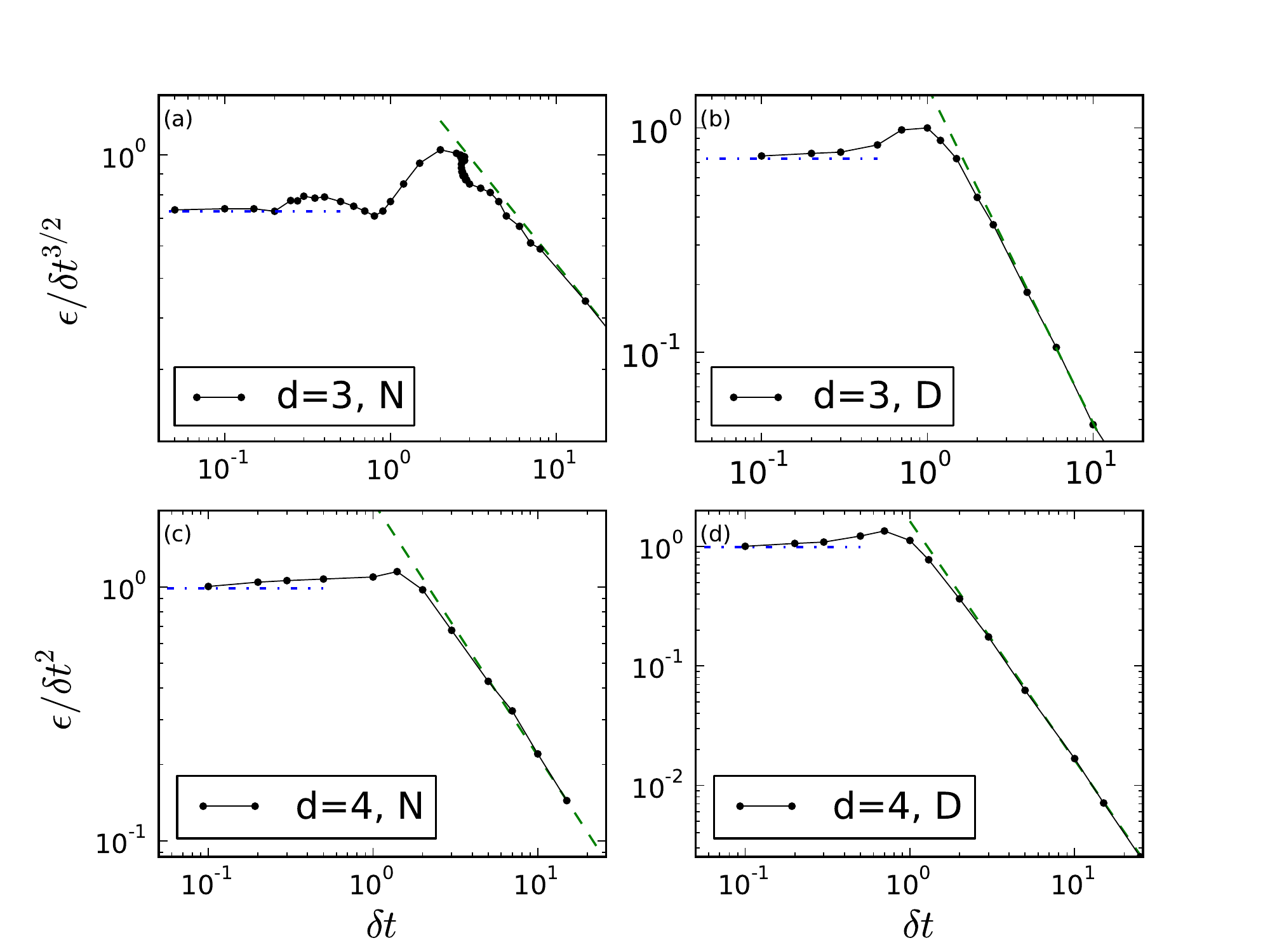}
   \caption{Dynamical phase space diagram for black hole formation in
      the hard wall model. The dimension is $d = 3$ or $4$ and Dirichlet (D) or Neumann (N) boundary conditions are imposed on the
      wall. Black holes are formed for parameters above the dots computed numerically, while the scattering phase occurs below. Straight lines
      correspond to analytical limits. Credits: \cite{Craps14c}.}
   \label{phasewall}
\end{myfig}

\subsection{Time-periodic solutions}
\label{tpsol}

Surprisingly, from a historical perspective, the very first island of stability (non-collapsing solution) was uncovered before the instability conjecture itself.
Fully non-linear spherically symmetric solutions of EKG equation were built as early as 2003, in \cite{Astefanesei03}, where
the authors obtained the first boson stars in AAdS space-times. Their solutions were also proved to be linearly stable against
perturbations and exhibited a maximum mass that was smaller than their asymptotically flat counterparts.

A few years later, the quest for building black holes with scalar hairs in AdS was triggered by \cite{Basu10}. In this work and
its extensions (see e.g.\ \cite{Gentle12,Dias12c,Dias17b}), Reissner-Nordstr\"om black holes surrounded by a spherically symmetric charged scalar
field, in what is called the Abelian-Higgs model, were built either perturbatively or numerically. These configurations have at
least two parameters: one drives the size of the horizon and another drives the amplitude of the scalar cloud. In this formalism, taking the zero-size
limit of the horizon leads naturally to boson stars (when the scalar field is complex and massive) or charged scalar periodic solutions, dubbed solitons. These solutions obtained in
\cite{Astefanesei03,Basu10,Gentle12,Dias12c,Dias17b} were the very first AAdS time-periodic solutions to emerge in the literature,
even before the instability conjecture was formulated. Furthermore, there were clues that these solutions were stable against
linear perturbations. Apart from boson stars (see \cite{Dias12b} and section \ref{geonsec} below), the non-linear stability of
these solitons was never investigated though.

The first numerical evolution of a non-collapsing solution in full AAdS space-time was the single-mode \eref{singlemode} initial
data discussed in section \ref{weakturbu}. Then, other types of stable initial data appeared concomitantly in
\cite{Maliborski13b,Buchel13}, in the spherically symmetric EKG setup. In \cite{Buchel13}, the authors called them ``boson
stars'', but since their scalar field was massless, this denomination was not strictly correct. The authors found initial
conditions that were immune to the non-linear instability below some amplitude threshold $\varepsilon_0$. These solutions
exhibited a power-law spectrum (in terms of pseudo-spectral coefficients) for collapsing solutions while non-collapsing
configurations featured an exponential spectrum. In particular, the authors observed that Gaussian initial data with large width
$\sigma \gtrsim 0.4$ did belong to these non-collapsing solutions, i.e.\ their collapsing time diverged to infinity at a finite
amplitude. This is illustrated in figure \ref{largesigma}. The argument was that widely distributed mass energy prevented the
energy to flow to smaller and smaller scales. Instead, the energy was perpetually exchanged between the first modes. However, this
claim was tempered by \cite{Maliborski13a} whose authors confirmed the results but did reignite instability for even larger values
of the width parameter $\sigma \gtrsim 8$. A similar behaviour was uncovered in the Einstein-Maxwell setup in \cite{Arias16}.

\begin{myfig}
   \includegraphics[width = 0.49\textwidth]{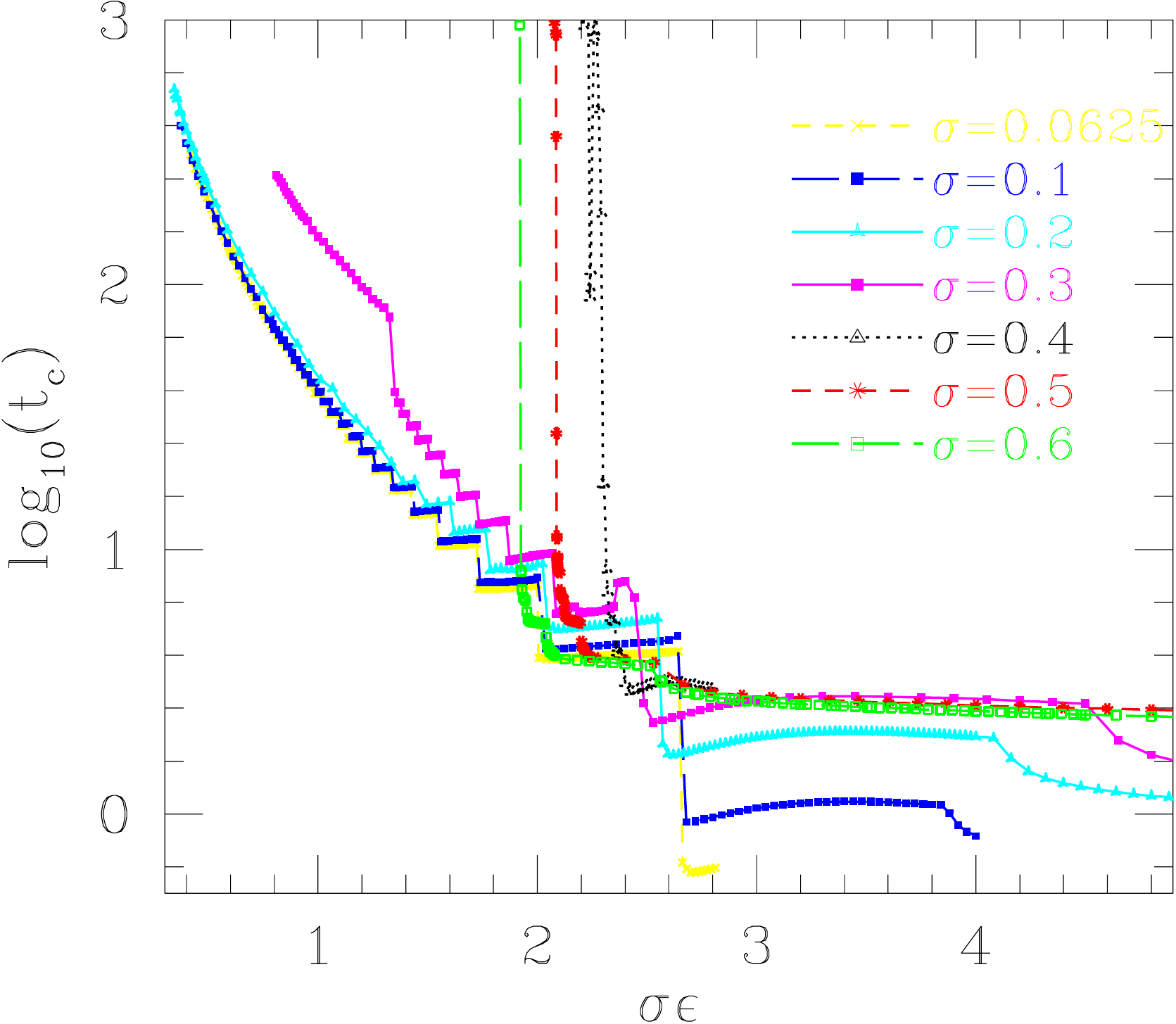}
   \caption{Collapse times for Gaussian initial data of amplitude $\varepsilon$ and
      width $\sigma$. For large $\sigma$, no collapse is observed, i.e.\ the time of collapse $t_c$ diverges to infinity at a
      small but non-zero amplitude. For small $\sigma$, the instability is recovered. In comparison, Bizo\'n and Rostworowski
      \cite{Bizon11} originally used $\sigma = 1/16 = 0.0625$. Credits: \cite{Buchel13}.}
   \label{largesigma}
\end{myfig}

Almost simultaneously in \cite{Maliborski13b}, time-periodic solutions were constructed and shown to be non-linearly stable with
the help of spectral methods. The authors started by constructing perturbatively periodic solutions, that they used as a seed for a
Newton-Raphson solver that could find fully non-linear generalisations. They then plugged the result into an evolution code and
monitored the phase space of spectral coefficients. In particular, they proved that high coefficients remained bounded, as shown
in figure \ref{phasespace}, demonstrating that no turbulent cascade was at play. The perturbed solution was not periodic any more
but quasi-periodic with orbits close to the perturbative periodic solution. This clearly highlighted the existence of a stable periodic
attractor immune to the non-linear stability.

\begin{myfig}
   \includegraphics[width = \textwidth]{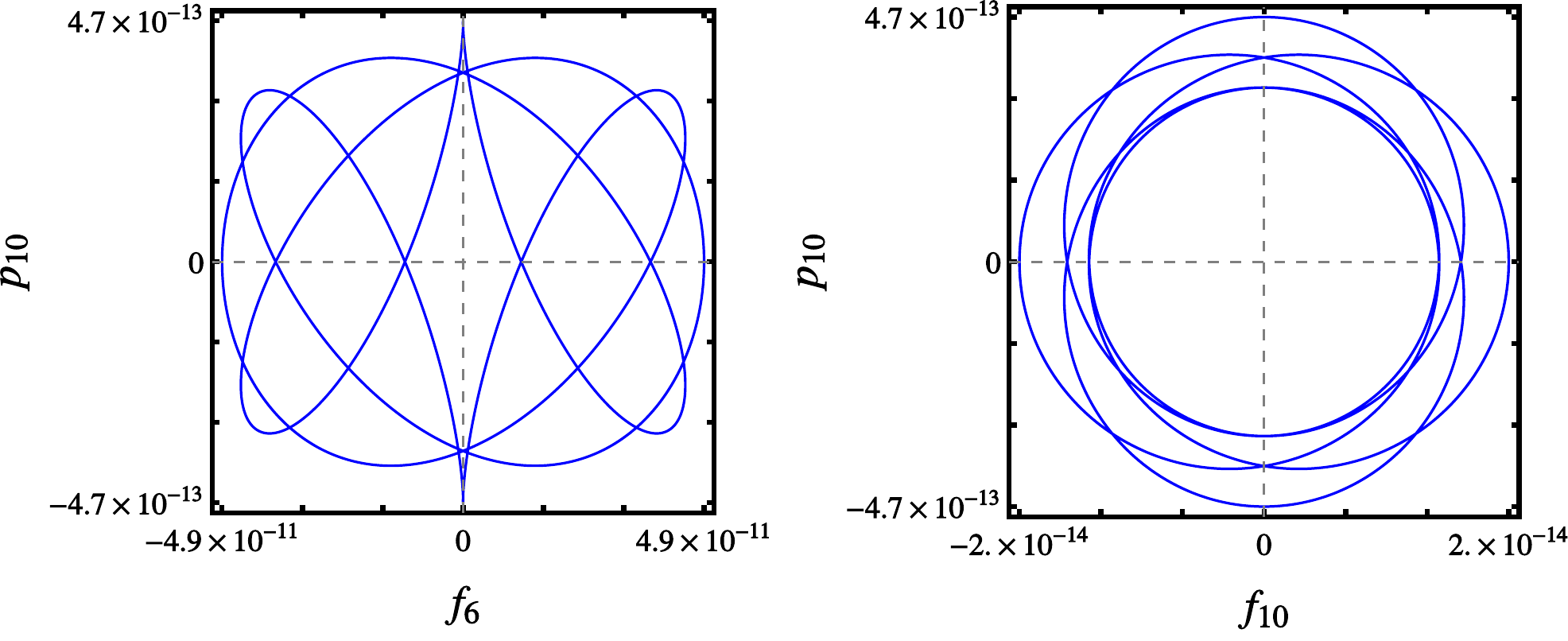}
   \caption{Slices of the coefficients space for a stable and
      near-periodic solution in AdS space-time over 500 periods of time. Not only the coefficients remain bounded and no
      weakly turbulent instability is observed, but the trajectory in phase space turns out to be quasi-periodic. Credits: \cite{Maliborski13b}.}
      \label{phasespace}
\end{myfig}

The perturbative construction of time-periodic solutions consists in choosing an ansatz (see equations
\eref{einsteinscalara}-\eref{einsteinscalarb} and \eref{ansatzscalar})
\begin{equation}
   \phi(t,x) = e^{i\Omega t}f(x), \quad \delta(t,x) = d(x), \quad A(t,x) = \mathcal{A}(x),
\end{equation}
for the three dynamical functions at play. The differential equations for $f$, $d$ and $\mathcal{A}$ are \cite{Maliborski13b}
\numparts
\begin{eqnarray}
   -\Omega^2 \frac{e^d}{\mathcal{A}}f &= \frac{1}{\tan^2 x}(\tan^2x \mathcal{A} e^{-d} f')',\\
   d' &= -\sin x \cos x\left[ {f'}^2 + \left( \frac{\Omega e^d}{\mathcal{A}}f \right)^2 \right],\\
   \mathcal{A}' &= \frac{1 + 2\sin^2 x}{\sin x \cos x} (1 - \mathcal{A}) + \mathcal{A}d'.
\end{eqnarray}
\endnumparts
These equations can be solved order by order by expanding $\Omega$, $f$, $d$ and $\mathcal{A}$ in a small amplitude parameter $\varepsilon$.
Choosing a dominant mode $e_\gamma$ in the zero-amplitude limit, it comes
\numparts
\begin{eqnarray}
   f &= \varepsilon f_1 + \varepsilon^3 f_3 + \ldots \quad \tn{with} \quad f_1(x) \propto e_\gamma(x),\\
   \mathcal{A} &= 1 - \varepsilon^2 \mathcal{A}_2 - \ldots,\\
   d &= \varepsilon^2 d_2 + \ldots,\\
   \Omega &= \omega_\gamma + \varepsilon^2 \Omega_2 + \ldots,
\end{eqnarray}
\endnumparts
where $\Omega$ is expanded around the dominant frequency $\omega_\gamma$ according to the Poincar\'e-Lindstedt method. After projection of the
unknown functions on the eigen basis $(e_j)_{j \in \mathbb{N}}$ all secular resonances are removable by fine-tuning the $\Omega_i$ coefficients. The
perturbative algorithm for the construction of time-periodic solutions was extended in \cite{Kim15} to tachyonic fields and pushed
to $20^{th}$ order in the amplitude.

The formalism of the time-periodic solution \cite{Maliborski13b} was extended to odd spatial dimensions in \cite{Fodor15}. The
authors explored the parameter space further and found bifurcations and resonances in the non-linear solutions that were missed by
\cite{Maliborski13b}. The stability branch of these solutions are shown in figure \ref{fodor}, as well as the agreement between
numerical and perturbative techniques. Stable solutions remain close to the initial data for ever, while unstable ones quickly
collapse to black holes. This is very reminiscent of usual self-gravitating systems that exhibit a maximum mass that is the boundary
between stable and unstable behaviours.

\begin{myfig}
   \includegraphics[width = 0.49\textwidth]{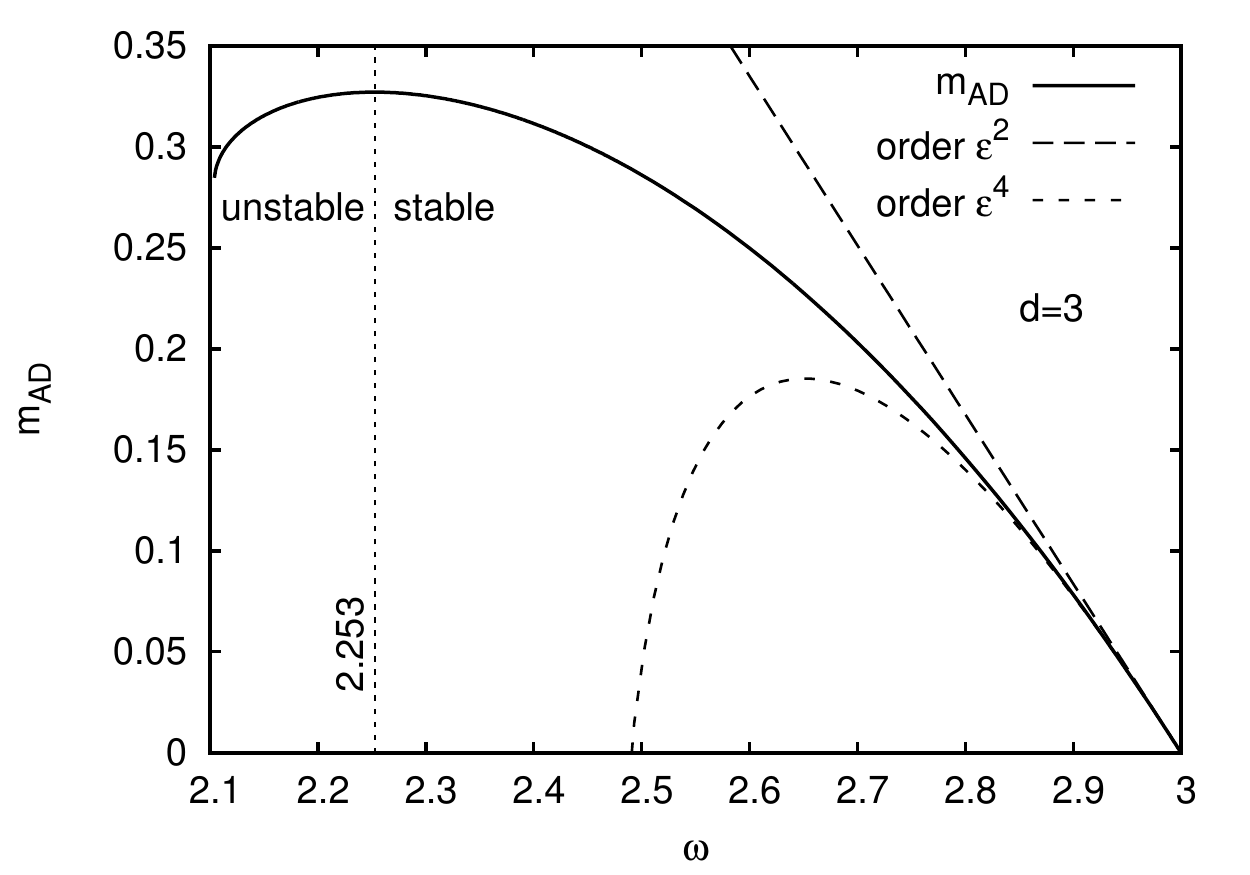}
   \caption{Mass of a time-periodic solution as a function of the
      oscillation frequency $\Omega$. At low amplitudes, $\Omega = 3$. Below a certain critical value $\Omega = 2.253$, the solutions become
      unstable against black hole formation. Perturbative (dashed) and numerical results (solid) agree well each other in the low amplitude limit.
      Credits: \cite{Fodor15}.}
      \label{fodor}
\end{myfig}

Why are time-periodic solutions non-linearly stable? One explanation was given in \cite{Maliborski14} where the authors studied
the spectrum of a linear perturbation superimposed on a time-periodic background. Unlike the vacuum AdS background case,
they observed that the spectrum of the corresponding linear operator $\widehat{L}$ was now only asymptotically resonant:
\begin{equation}
   \omega_j = Cj + D + O\left( \frac{1}{j} \right),
\end{equation}
such that eigen frequencies were equidistant only in the large-$j$ limit. The idea was the following: a resonant spectrum leads to
non-linear instability while an only asymptotically resonant spectrum could lead to an amplitude threshold below which the
instability is suppressed (this point is further discussed in section \ref{roleres} below). The argument gathered
momentum with the time evolution of a Gaussian perturbation around this time-periodic background. This did suppress the turbulent
cascade for sufficiently low amplitudes of the perturbation, highlighting that the time-periodic solution could be an attractor immune
to non-linear instability. This point is illustrated in figure \ref{timepattractor}. For stable solutions, the energy spectrum settles
down to an exponential form at long times, adding weight to the stability and regularity statement.

\begin{myfig}
   \includegraphics[width = 0.49\textwidth]{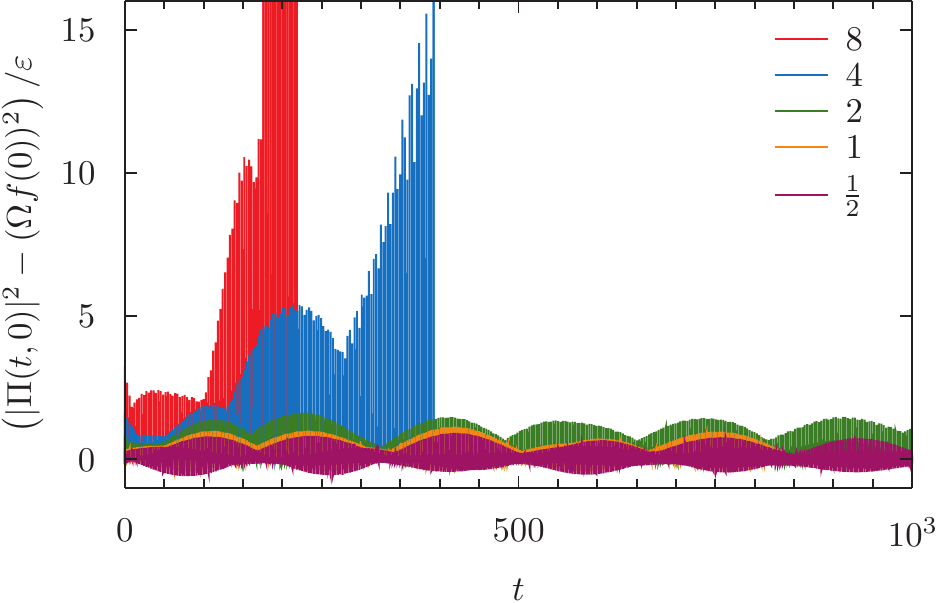}
   \caption{Time-periodic background perturbed by a Gaussian profile. The
      plot shows the time evolution of the scalar field amplitude $\Pi(t,0)$ (with time-periodic background subtracted). The
      amplitude of the perturbation is labelled by different colours. For large amplitudes, the instability is triggered, while
      for low amplitudes, it is suppressed. Credits: \cite{Maliborski14}.}
   \label{timepattractor}
\end{myfig}

The time-periodic and large-width Gaussian solutions were the first non-trivial dynamical examples of non-linearly stable
solutions in AAdS space-times, suggesting that the instability had a much richer structure than was previously thought. It was
noticed in \cite{Abajo14} that the scalar field profiles of large Gaussian initial data of \cite{Buchel13} and of the time-periodic
solutions of \cite{Maliborski13b} were in fact very close to each other and probably belonged to the same island of stability.
Notice that these time-periodic solutions could well be called scalar geons \cite{MartinonPhD}.

\subsection{The two-time framework (TTF)}
\label{ttf}

We have seen in section \ref{pertscal} that the Poincar\'e-Lindstedt method could remove secular resonances arising at third order in the
expansion, at least in some cases. This method works very well and at all orders in the case of time-periodic solutions
(section \ref{tpsol}). However, in the general case, some irremovable resonances appear and are responsible for the
AdS instability. 

The perturbative equations at third order \eref{cj3} are nothing but a system of non-linearly coupled oscillators. Such systems
belong to the class of non-integrable Hamiltonian systems. For a long time, they were believed to obey the ergodic hypothesis. In
1955, in order to test this statement, Fermi and his collaborators performed a numerical simulation of a chain of 64
non-linearly coupled oscillators \cite{Fermi55}. At the time, the authors expected the system would exhibit thermalisation, an
ergodic behaviour in which the system becomes random with all modes excited more or less equally. Instead the system displayed a
very intricate quasi-periodic solution. This is the so-called Fermi-Pasta-Ulam-Tsingou (FPUT) paradox. This result showed that non-integrable
Hamiltonian systems were not always ergodic. In subsequent works (see \cite{Benettin08} for a review), it was shown that the
quasi-periodic behaviour on certain time scales could be studied within the Two-Time Framework (TTF).

Transposed to the AdS instability problem, the ergodicity hypothesis advocates for a systematic instability, since if very
small scales are substantially excited, black hole formation becomes very likely. Thus, the AdS instability would appear as
an ergodic thermalisation process, echoed in the dual CFT (see section \ref{cftinterp}). However, given the similarity between the
FPUT problem and equations \eref{cj3}, it can be reasonably expected that gravitational dynamics in AdS space-time
are not ergodic. There could exist quasi-periodic solutions that do not explore the whole phase space and thus avoid black hole
formation. As for the FPUT problem, TTF might be of great help in finding such solutions.

The TTF aims at providing a systematic way of removing secular resonances and thus building
non-linearly stable solutions. It was first introduced in \cite{Balasubramanian14}, motivated by the FPUT analogy, and refined in
\cite{Craps14a,Craps15a,Buchel15}. Recycling the results of section \ref{pertscal}, we have already seen that at first order, the
solutions could be written
\begin{equation}
   \phi_1(t,x) = \sum_{j=0}^\infty (\alpha_j e^{-i\omega_j t} + \overline{\alpha}_j e^{i\omega_j t})e_j(x),
\end{equation}
where a bar means complex conjugation and $\alpha_j$ are constant complex amplitudes. The idea is the following: given that both
perturbative results and numerical simulations have proved that some initial data becomes singular in a time $t =
O(\varepsilon^{-2})$, let us introduce a new time-scale, or slow-time
\begin{equation}
   \tau \equiv \varepsilon^2 t.
\end{equation}
The intuition is that if the dynamics involves rapid oscillations superimposed on a slow drift behaviour, there should be some
sort of simplified effective description of the slow motion, in which the fast oscillations have been averaged out. This is
at the heart of multiple-scale analysis. We thus expect the small amplitude scalar field to undergo large variations in a time-scale
$\tau = O(1)$, while it oscillates on a much shorter time-scale (i.e.\ it bounces many times before collapsing). This suggests to make
the envelope of oscillations vary slowly, on a time-scale $\tau = O(1)$. We thus write (this is nothing but a
variation of the constants method):
\begin{equation}
   \phi_1(t,\tau,x) = \sum_{j=0}^\infty [\alpha_j(\tau) e^{-i\omega_j t} + \overline{\alpha}_j(\tau) e^{i\omega_j t}]e_j(x).
\end{equation}
Paying attention that now $\partial_t \rightarrow \partial_t + \varepsilon^2 \partial_\tau$, the second order equations for $A_2$
and $\delta_2$ are unchanged but at third order a new term $\partial_t \partial_\tau \phi_1$ appears (compare with
\eref{phi3}), namely
\begin{equation}
   \partial_t^2{\phi_3} + \widehat{L}\phi_3 + 2\partial_t \partial_\tau \phi_1 = S(\phi_1,A_2,\delta_2).
   \label{phi3slow}
\end{equation}
Projecting onto the basis $(e_j)_{j \in \mathbb{N}}$, it comes (in contrast with \eref{cj3})
\begin{equation}
   \forall j, \quad \ddot{c}_j^{(3)} + \omega_j^2 c_j^{(3)} - 2i\omega_j(\partial_\tau \alpha_j e^{-i\omega_j t} - \partial_\tau \overline{\alpha}_j e^{i\omega_j t}) = S_j.
\end{equation}
Of course, the introduction of the slow-time did not remove the resonant terms $e^{\pm i \omega_j t}$ on the right-hand side, but
we are now free to enforce them to vanish by imposing
\begin{equation}
   -2i\omega_j \partial_\tau\alpha_j = (\tn{component } e^{-i\omega_j t} \tn{ of } S_j) = \sum_{klm} S_{jklm}\overline{\alpha}_k \alpha_l \alpha_m,
   \label{ttf1}
\end{equation}
where $S_{jklm}$ are real constants representing all the possible resonant channels $\omega_j + \omega_k = \omega_l +
\omega_m$. Other channels like e.g.\ $\omega_j = \omega_k + \omega_l + \omega_m$ can be shown to vanish and a
brute-force calculation of the $S_{jklm}$ coefficients is presented in \cite{Craps14a}. At this point, all functions $c_j^{(3)}$ remain
bounded in time by construction and hence are of little interest, so that we now focus on the $\alpha_j$. Moving to the
exponential complex representation
\begin{equation}
   \alpha_j(\tau) = A_j(\tau)e^{iB_j(\tau)},
   \label{complexalpha}
\end{equation}
with $A_j$ the real amplitude and $B_j$ the real phase, equation \eref{ttf1} becomes
\numparts
\begin{eqnarray}
\label{ttfequationsa}%
\fl 2\omega_j \frac{d A_j}{d \tau} &= \sum_{j+k=l+m}^{\{j,k\} \neq \{l,m\}} S_{jklm}A_k A_l A_m \sin(B_j + B_k - B_l - B_m),\\
\label{ttfequationsb}%
\fl 2\omega_j \frac{d B_j}{d \tau} &= T_j A_j^2 + \sum_{i\neq j} R_{ij}A_j^2 + \frac{1}{A_j}\sum_{j+k=l+m}^{\{j,k\} \neq \{l,m\}} S_{jklm}A_k A_l A_m \cos(B_j + B_k - B_l - B_m),
\end{eqnarray}
\endnumparts
where $\{j,k\} \neq \{l,m\}$ means that neither $j$ nor $k$ is equal to either $l$ or $m$, $T_j = S_{jjjj}$ and $R_{ij} =
S_{ijji} + S_{jiji}$. These equations are called the TTF equations. Alternative denominations are resonant \cite{Bizon15b},
renormalisation flow \cite{Craps14a} or time-averaged equations \cite{Evnin16}.

Defining
\begin{equation}
   N = \sum_{j=0}^\infty \omega_j A_j^2 \quad \tn{and} \quad E = \sum_{j=0}^\infty \omega_j^2 A_j^2,
   \label{NE}
\end{equation}
it was shown in \cite{Craps15a,Buchel15,Yang15} that $N$ and $E$ were conserved quantities, namely
\begin{equation}
   \frac{dN}{d\tau} = 0 \quad \tn{and} \quad \frac{dE}{d\tau} = 0.
\end{equation}
These quantities are then interpreted as $N$ the total particle number (in analogy to quantum mechanics) and $E$ the total energy
of the system. There exists a third conserved quantity, that was uncovered in \cite{Craps15a} and which represents the total
interaction energy between modes. In \cite{Buchel15}, it was argued that if there was a cascade of energy toward high-$j$ modes,
the simultaneous conservation of $E$ and $N$ induced that there was an inverse cascade of particle number toward low-$j$ modes, and
vice versa. This is called the dual cascade phenomenon.

The TTF equations are third order in amplitudes and are valid for durations up to $\tau = O(1)$. It would be possible to
introduce other time-scales like $\varepsilon^4 t$ to go further in time. But usually, evolving the TTF equations up to
$\tau = O(1)$ brings enough information to conclude about the stability.

In general, the TTF equations are evolved numerically. As the number of coupled equations is infinite, a cut-off (or
truncation number) $j_{max}$ has to be enforced. The advantage of these equations is that they are ordinary differential equations
in time, whereas the full EKG system \eref{einsteinscalara}-\eref{einsteinscalarb} is a system of partial differential equations. Within the TTF, the spatial
dependence is entirely encoded on the basis functions $(e_j)_{j \in \mathbb{N}}$, which is somewhat reminiscent of spectral methods in
numerical analysis. The equations are thus less computationally demanding, which contributed to the democratisation of the
AdS non-linear instability study. Incidentally, the number of publications in the field drastically increased after 2014 and
\cite{Balasubramanian14}.

The TTF equations and conservation laws were generalised to non-spherically symmetric scalar field collapse in
\cite{Yang15} with the help of spherical harmonic decomposition. The resonant channels depend on the number of dimensions and the authors
expected the system of equation to display an underlying symmetry that would simplify their systematic determination. An $SU(d)$
symmetry was precisely demonstrated in \cite{Evnin15} and refined in \cite{Evnin16}.

Last but not least, the TTF equations are invariant under the transformation
\begin{equation}
   \alpha_j(\tau) \rightarrow \varepsilon \alpha_j(\tau/\varepsilon^2),
   \label{scalingsym}
\end{equation}
which means that if a solution of amplitude one does something at slow-time $\tau$, then the same solution with amplitude
$\varepsilon$ does the same thing at slow-time $\tau/\varepsilon^2$. This feature was observed in the non-linear case as early as
\cite{Bizon11}. It actually comes out naturally from the TTF equations.

\subsection{The analyticity strip method}
\label{analstrip}

A standard approach is to evolve in time the TTF equations and to determine if the energy spectrum (equation \eref{NE}) becomes
singular, namely if there is an energy flow toward high-$j$ modes so as to get a power-law spectrum instead of an exponential one.
The underlying concept is the one of analyticity strip, described in \cite{Sulem83} and used for the first time in the
AdS instability context in \cite{Bizon13}. It consists in fitting the coefficients $A_j$ (or the energy per mode $E_j =
\omega_j^2 A_j^2$ according to \eref{NE}) as a function of $j$ at each slow-time step by
\begin{equation}
   A_j(\tau) = C(\tau) j^{-\gamma(\tau)}e^{-\rho(\tau)j}.
   \label{analstripeq}
\end{equation}
In practice, the fit is performed on a reduced set of modes that is away from $j = 0$ and $j = j_{max}$
in order to minimise truncation errors. The function $\rho(\tau)$ is the analyticity radius of the solutions. If $\rho$ stays
strictly positive, it means that the solution is regular at all times. If the radius of analyticity hits zero, it is a strong hint that
the solution blows up in a finite time. The existence of a time $\tau_0$ at which $\rho(\tau_0) = 0$ is a necessary but not
sufficient condition for black hole formation (see section \ref{conditions} below).

The TTF equations coupled to the analyticity strip method are thus less effective than fully non-linear evolutions to find
unstable solutions, but they are very good (and computationally cheap) at exploring the AdS sea to find islands of
stability. They also bring a kit of analytical tools to better understand the deep structure of the problem.

\subsection{The two-mode controversy}

One of the first playgrounds of the TTF framework was the two-mode initial data (equation \eref{twomode}). It was first
studied in \cite{Bizon11} as a minimal setting for triggering the instability (recall that a single mode initial data is
non-linearly stable). Subsequently, the authors of \cite{Balasubramanian14} tried to evolve the TTF equations for the
equal-energy two-mode initial data with a cut-off $j_{max} = 47$. Surprisingly they found that their code was showing no sign of
collapse. After a lively debate \cite{Bizon15a,Balasubramanian15,Buchel15,Bizon15b,Green15}, the simulations of
\cite{Bizon15b,Deppe15b} confirmed that the two-mode initial data was really collapsing and that the TTF as
well as the full GR simulations of \cite{Balasubramanian14} suffered from resolution problems and a too small truncation
number.

In particular in \cite{Bizon15b}, the authors carefully scrutinised the two-mode initial data both in full GR and with the TTF
equations with a cut-off $j_{max} = 172$. Not only did they confirm that the two-mode initial data was collapsing in the full theory,
but they also showed the agreement with TTF via the analyticity strip method. Namely, they observed that the analyticity radius
was dropping to zero in a finite slow-time $\tau_\star \simeq 0.509$ and that the exponent $\gamma$ in \eref{analstripeq} was
tending to $2$ at this date. They inferred via the TTF equations \eref{ttfequationsa}-\eref{ttfequationsb} that such behaviours implied a
logarithmic divergence of the phase derivatives $d B_j/d\tau$ ($B_j$ being defined in \eref{complexalpha}) and checked that this
was indeed the case numerically, as pictured in figure \ref{blowup}.

\begin{myfig}
   \includegraphics[width = 0.48\textwidth]{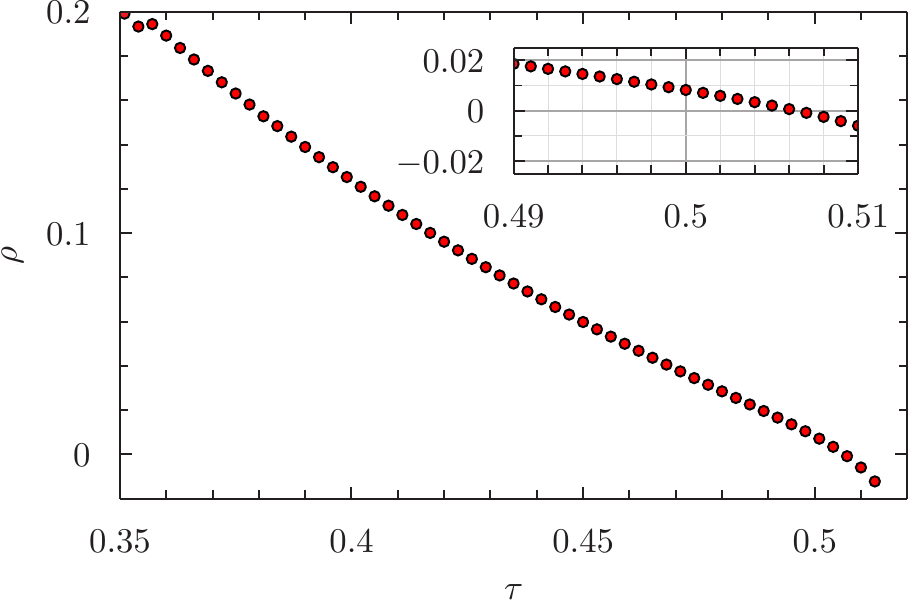}
   \includegraphics[width = 0.49\textwidth]{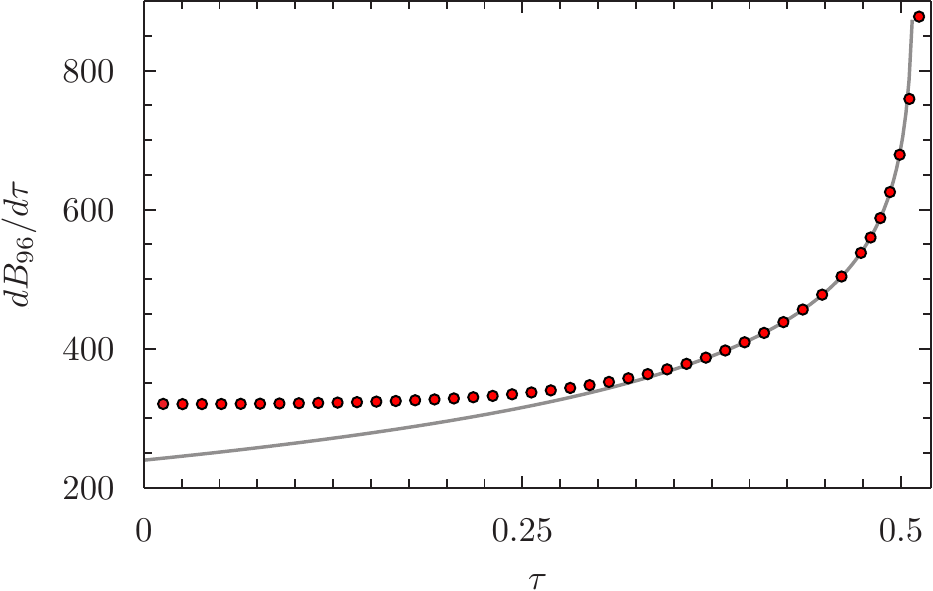}
   \caption{Left: analyticity radius for the TTF evolution of the two-mode initial
   data. It hits zero in a finite slow-time, suggesting a potential consecutive blow-up. Right: slow-time derivative of the $n=96$
   phase $B_n$. A logarithmic blow-up $a \ln(\tau_\star - \tau) + b$ is fitted and is observed for all $n > 20$. Credits:
   \cite{Bizon15b}.}
   \label{blowup}
\end{myfig}

This behaviour was supported by analytical calculations in \cite{Craps15b,Craps15c}, with a discussion about the gauge-dependence
of the result, in particular for different choices of boundary conditions $\delta(t,x = \pi/2)$. Indeed, the two gauges used
in the literature are
\numparts
\begin{eqnarray}
   \delta(t,x=0) &= 0, \quad \tn{Interior Time Gauge (ITG)},\\
   \delta(t,x=\pi/2) &= 0, \quad \tn{Boundary Time Gauge (BTG)},
\end{eqnarray}
\endnumparts
where $\delta$ is the metric coefficient in \eref{ansatzscalar}. The authors of \cite{Craps15b,Craps15c} have shown analytically
that the logarithmic divergence of $dB_j/d\tau$ was suppressed in the BTG. This was confirmed numerically by \cite{Deppe16b}
whose author demonstrated that these features existed only in the ITG gauge. This was given more support in
\cite{Dimitrakopoulos16} whose authors studied precisely the gauge dependence of the TTF equations. They concluded that the gauge
was impacting only the phases $B_j$ but not the amplitudes $A_j$. Furthermore, if one gauge gives singular results and not the
other, it reveals that there is a infinite redshift between the two gauges, so that TTF does become invalid as a
perturbation theory and an instability is triggered. Finally, the results of \cite{Deppe16b} indicated that the logarithmic
blow-up of the phase derivative was completely suppressed in both ITG and BTG gauges in higher dimensions than four.

Based on these results, full GR and TTF finally agreed each other about the two-mode initial data: it is unstable, it
collapses to a black hole, and both frameworks can detect it, either by the vanishing of the blackening factor $A$ of by the
vanishing of the analyticity radius $\rho$. Thanks to the scaling symmetry \eref{scalingsym}, it is established that the two-mode
initial data does collapse for arbitrarily small amplitudes, a regime usually out of reach of numerical simulations.

\begin{myfig}
   \includegraphics[width = 0.47\textwidth]{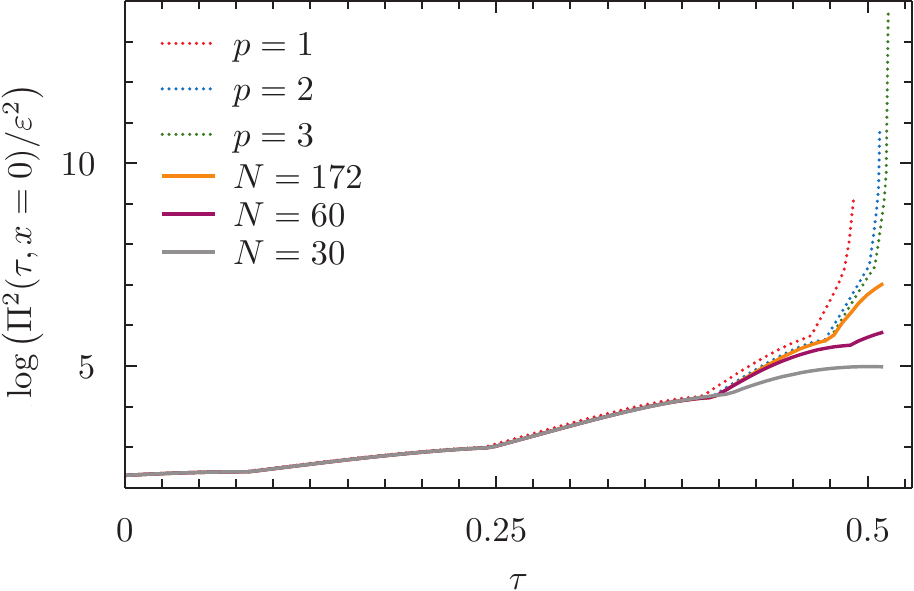}
   \includegraphics[width = 0.51\textwidth]{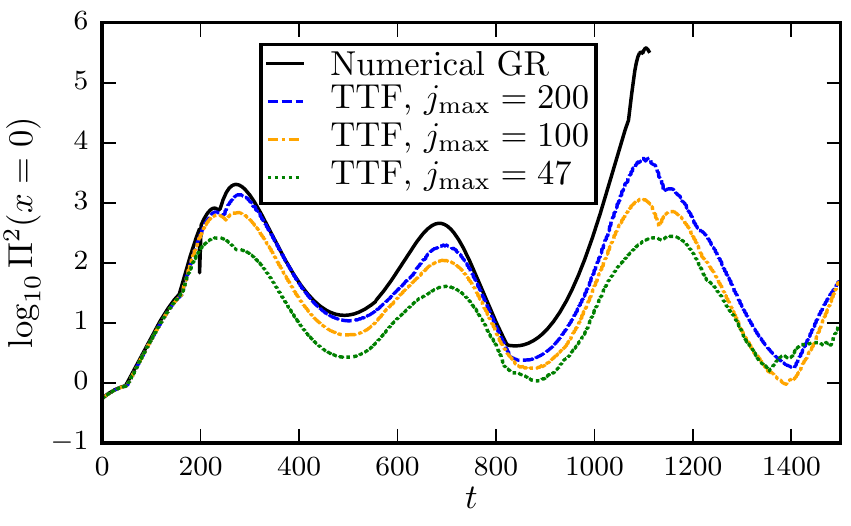}
   \caption{Left: upper envelope of $\Pi^2(t,0)$ in the full GR evolution of
   two-mode initial data with $\varepsilon = (2\pi)^{-3/2}2^{-p}$ for $p = 1,2,3$. The corresponding solutions of the
   truncated TTF system are shown in solid as the truncation number $N$ increases. Right: idem but with a larger
   amplitude $\varepsilon = 0.09$. In this case, three cascades of energy (increasing curve) and two inverse cascade (decreasing
   curve) are visible before collapse. Credits: \cite{Bizon15b,Green15}.}
   \label{agreementttf}
\end{myfig}

The agreement between the two kinds of time evolution can be clearly seen in figure \ref{agreementttf}. For small amplitudes (left
panel), the convergence of the TTF solution to the full GR with increasing truncation number is unambiguous. The case
of amplitude $\varepsilon = 0.09$ (right panel) taken from \cite{Green15} was subject to lively debates. Indeed, if full GR
simulations indicate black hole formation at $t \sim 1080$, TTF disagrees and undergoes an inverse cascade after this date.

The author of \cite{Deppe16b} studied more deeply this point with the help of the analyticity-strip method. He first observed that
the two-mode initial data was unambiguously unstable in 9-dimensional AdS, as can be seen on the left panel of figure
\ref{analtrunc}. This suggested that the instability was favoured in higher dimensions. Back to the 4-dimensional problem, he
observed that the truncation number was slightly impacting the location of the first root of the analyticity radius, as can be
seen on the right panel of figure \ref{analtrunc}. The analyticity radius hits zero at time $t \sim 800$, i.e.\ before the
non-linear collapse at time $t\sim 1080$. The point is: nothing prevents the TTF equations to be evolved in time after the
analyticity radius hits zero, like in the right panel of figure \ref{agreementttf}. However, the physical meaning of the evolution
after this stage should not be considered too seriously. Even if TTF solutions do not blow up at this stage, the only
reliable criterion for instability is the vanishing of the analyticity radius. The fact that it hits zero at time $t\sim 800$ is
already a strong indication of a near subsequent singularity formation.
\begin{myfig}
   \includegraphics[width = 0.49\textwidth]{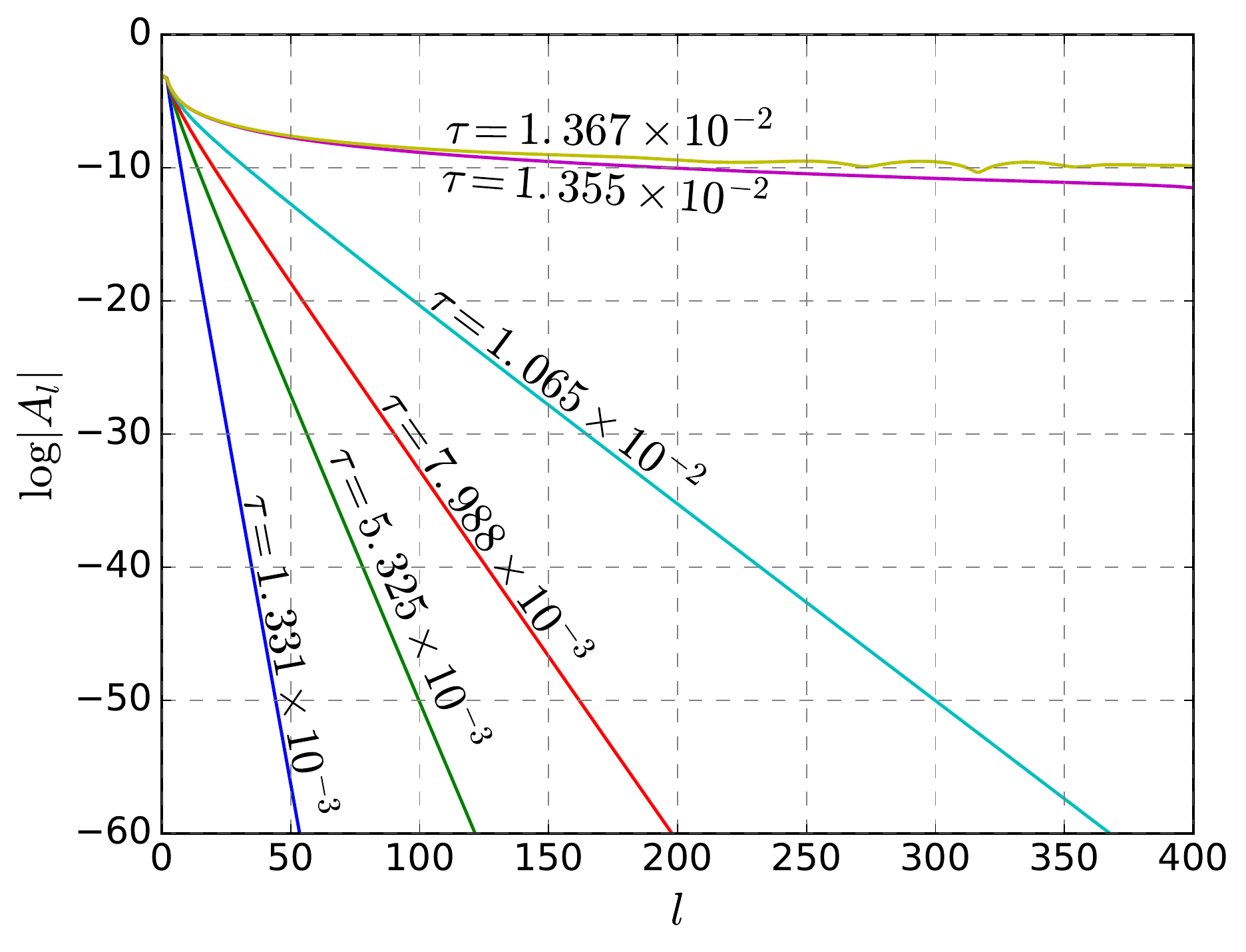}
   \includegraphics[width = 0.49\textwidth]{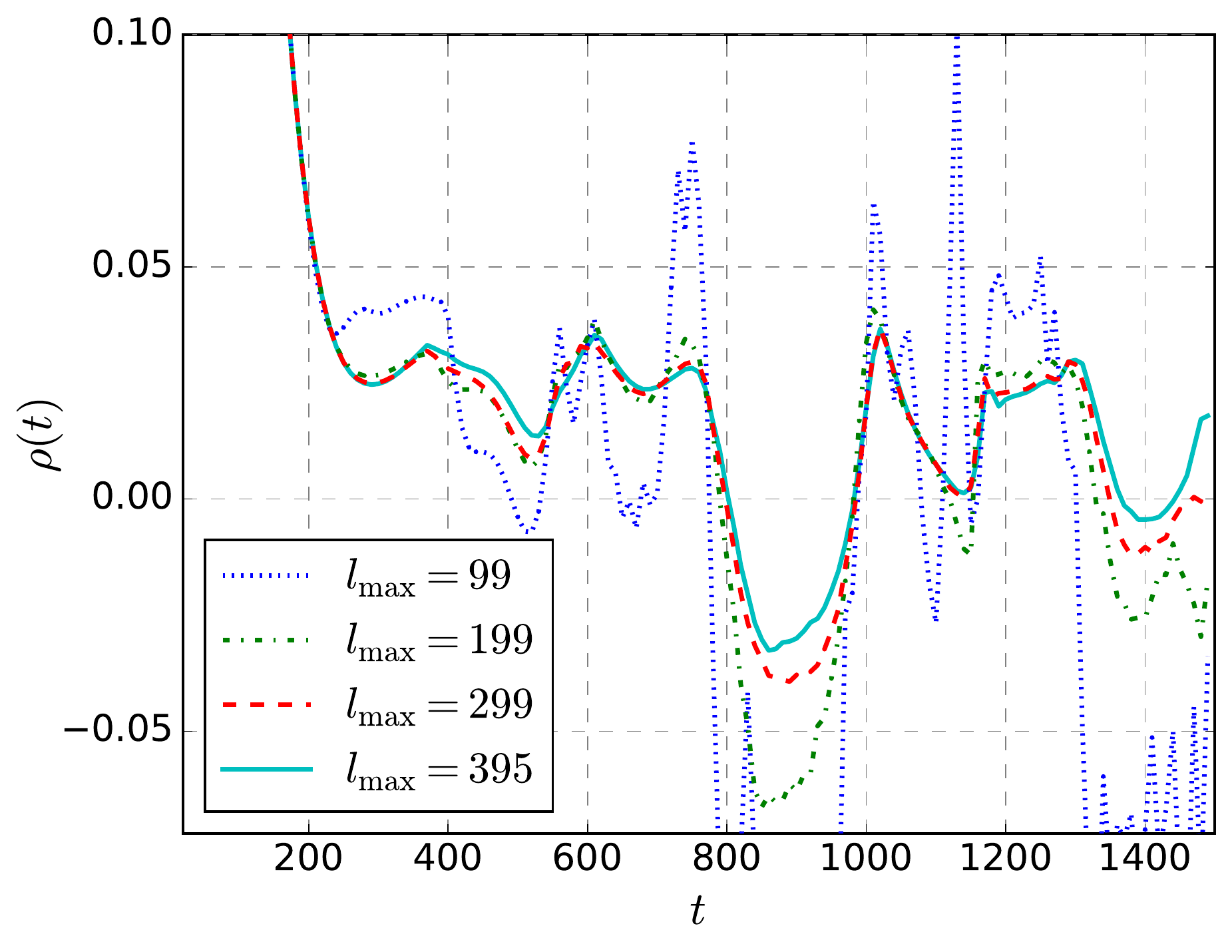}
   \caption{Left: spectra at different slow-times $\tau$ for the
      two-mode initial data in 9-dimensional AdS. The spectrum becomes singular unambiguously at $\tau = 1.356\times 10^{-2}$.
      Right: analyticity radius for the 4-dimensional case. The variability of the roots of $\rho$ with truncation number
      $l_{max}$ indicates that the evolutions suffer from truncation errors to some extent. The time axis is the same as in
      figure \ref{agreementttf} (right panel) where collapse occurs at $t \sim 1080$. Credits: \cite{Deppe16b}.}
   \label{analtrunc}
\end{myfig}

\subsection{Generation of islands of stability with TTF}

Besides being an alternative and cheaper numerical method to study the instability conjecture, the TTF is a valuable tool to
build islands of stability, namely initial data that are non-linearly stable. The first paper employing TTF in the AdS
instability problem was \cite{Balasubramanian14}, where the authors uncovered a new family of periodic and stable solutions by
imposing real amplitudes (equation \eref{complexalpha})
\begin{equation}
   A_j = \varepsilon \frac{e^{-\mu j}}{\omega_j},
   \label{ttfinit}
\end{equation}
with a truncation number $j_{max} = 47$. In figure \ref{qpttf}, it is clearly visible that such initial data remained stable with a
bounded and quasi-periodic departure from the initial conditions. In particular, no turbulent cascade was observed. This was
confirmed in \cite{Basu15} whose authors pushed the truncation number to $j_{max} = 150$ and observed the same stabilisation.

\begin{myfig}
   \includegraphics[width = 0.49\textwidth]{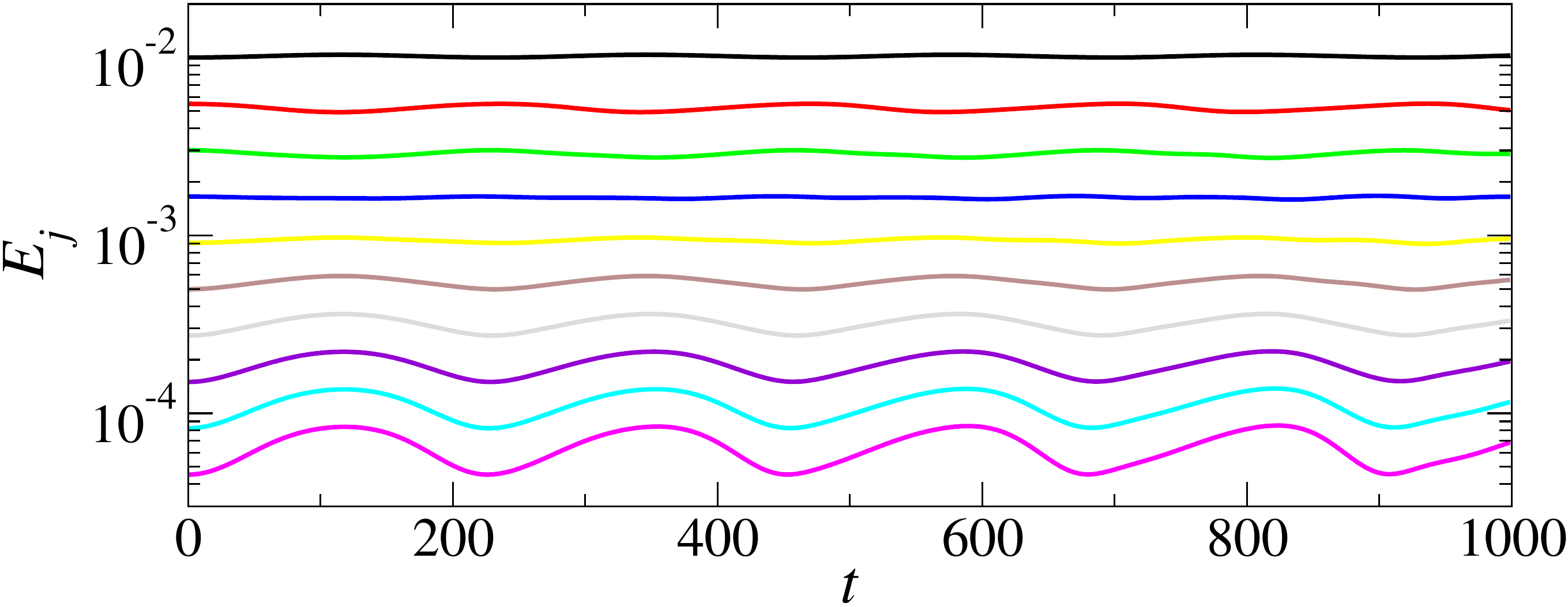}
   \caption{Energy per mode for initial data \eref{ttfinit} with $\mu = 0.3$
   plus an initial random perturbation. The energies in the first modes remain always close to their initial values, signalling a
   stable solution within the duration of the simulation. Credits: \cite{Balasubramanian14}.}
   \label{qpttf}
\end{myfig}

Similarly in \cite{Green15}, a whole two-parameter family of stable quasi-periodic solutions was found, inspired by \cite{Balasubramanian14}
and equation \eref{ttfinit}, namely
\begin{equation}
   \alpha_j(\tau) = a_je^{-ib_j\tau},
\end{equation}
where the parameters $a_j$ and $b_j$ could be finely tuned so that energy flows between modes were perfectly balanced. In figure
\ref{circleqp}, a quasi-periodic solution was perturbed with a small two-mode contribution. For moderate amplitudes of the
perturbation, the phase space of the TTF exhibited a quasi-periodic behaviour, which was reminiscent of the quasi-periodic
solutions of section \ref{tpsol}. The authors were also able to recover the stable/unstable behaviour of Gaussian initial data
(recall that stability is recovered for widths $0.5 \lesssim \sigma \lesssim 8$ ), as depicted in figure \ref{gaussianttf}: stable
initial data featured an exponential spectrum while unstable initial data presented a power-law spectrum.

\begin{myfig}
   \includegraphics[width = 0.49\textwidth]{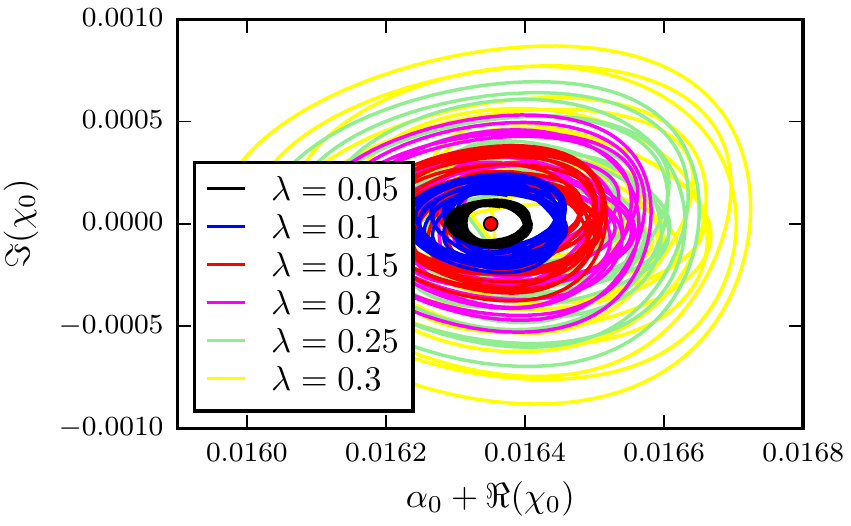}
   \caption{Phase space of initial data interpolating a quasi-periodic solution with the two-mode
      initial data such that $E_j = (1-\lambda)E_j^{QP} + \lambda E_j^{two-mode}$. The quantities $\alpha$ and $\chi$ represents the
      amplitude of the solution. The red dot at the centre is a pure time-periodic solution. When $\lambda \neq 0$, the
      solution merely oscillates in phase space around the periodic solution with no growth of the amplitude.
      Credits: \cite{Green15}.}
   \label{circleqp}
\end{myfig}

\begin{myfig}
   \includegraphics[width = 0.46\textwidth]{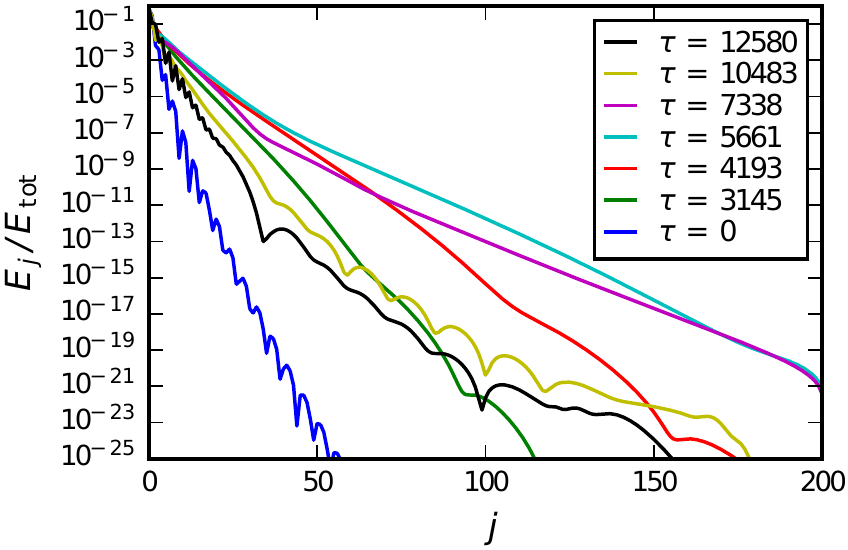}
   \includegraphics[width = 0.49\textwidth]{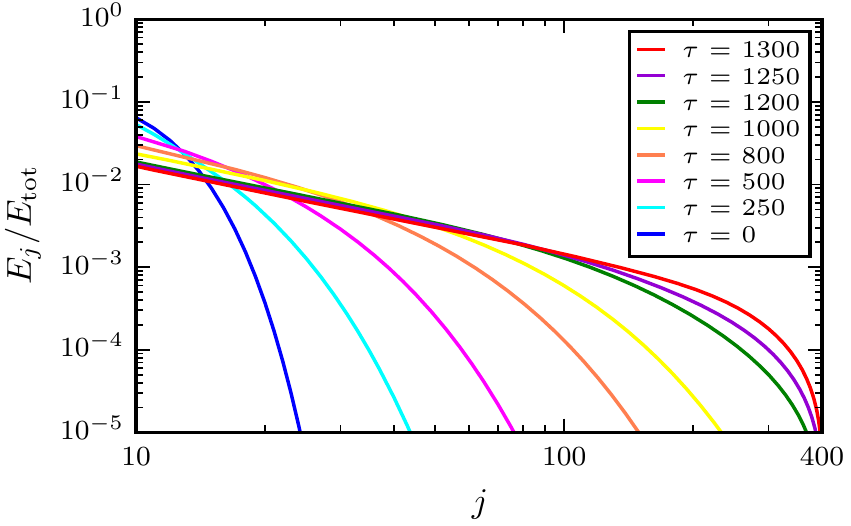}
   \caption{Left: energy spectrum of stable Gaussian initial data $\sigma = 0.4$
      with exponential tail at all slow-times. Right: energy spectrum of unstable Gaussian initial data $\sigma = 1/16$ with a
      power-law tail at large slow-time indicating turbulent cascade and potential black hole formation. Credits: \cite{Green15}.}
   \label{gaussianttf}
\end{myfig}

\section{Structure of the instability}
\label{structure}

So far, we have discussed configurations that were either non-linearly stable or non-linearly unstable. However, the boundary
between these two behaviours is not so clean, and the same family of solution can switch between stability and instability. We
already spotted this phenomenon for the Gaussian initial data, that is generally unstable except for width $0.4 \lesssim \sigma
\lesssim 8$ (section \ref{tpsol}). However, this transition can be much more intricate and chaotic.

\subsection{Chaotic footprints}
\label{chaosfoot}

Apart from Gaussian or two-mode initial data, another family of experiments in AAdS space-times consists in the interaction
of two concentric thin shells of a perfect fluid with a linear equation of state $p = w \rho$ where $p$ is the pressure, $\rho$
the energy density and $w$ a free parameter, that can be constrained by energy conditions (weak, null or strong). This problem is
close to the collapse of a scalar wave packet as several authors \cite{Pretorius00,Abajo14,Silva15,Deppe15a,Deppe15b,Deppe16a}
noticed that broad pulse initial data had the tendency to break off into several sub-pulses interacting with each other.
Furthermore, this setup was the first to exhibit chaotic behaviours in the AdS instability context.

The first study of this problem was initiated in \cite{Mas15}. The authors patched several Schwarzschild metrics together with
Israel junction conditions (see for example \cite{Poisson04}) between each shell and studied the effective potential of
interaction. They concluded that periodic solutions should then exist.

The numerical experiment was performed in \cite{Cardoso16} in a flat space-time enclosed in a cavity and in \cite{Brito16b} in
AdS space-time. This problem could be seen as the simplest two-body problem as it was one-dimensional. Several kinds of
behaviours were observed: prompt collapse, delayed collapse and perpetual oscillatory motion. The confinement seems crucial to
ensure that the shells collide each other repeatedly, thus allowing small effects to build up in time.

Examining the number of crossings of the shells during evolution and before collapse, several striking results emerged. First a
fractal-like structure was clearly visible, as depicted in figure \ref{fractal} (left panel) for initial conditions where the two
shells started at the same positions $R_i$. Like in many chaotic systems, the authors unveiled windows of stability, i.e.\ some
ranges of $R_i$ in which no collapse occurred.

Moreover, for a different set of initial conditions, the mass of the black hole exhibited a critical behaviour in the left
neighbourhood of critical points, namely
\begin{equation}
   M_H - M_{n+1} \propto ( \delta_n - \delta )^\gamma,
\end{equation}
where $\delta$ encodes the mass-energy content of the initial data and $M_n$ was the black hole mass at the critical amplitude
$\delta_n$. The coefficient $\gamma$ was found to be $\sim 0.95$ but actually depended on the equation of state parameter $w$. This
critical behaviour is illustrated in figure \ref{fractal} (right panel).

\begin{myfig}
   \includegraphics[width = 0.44\textwidth]{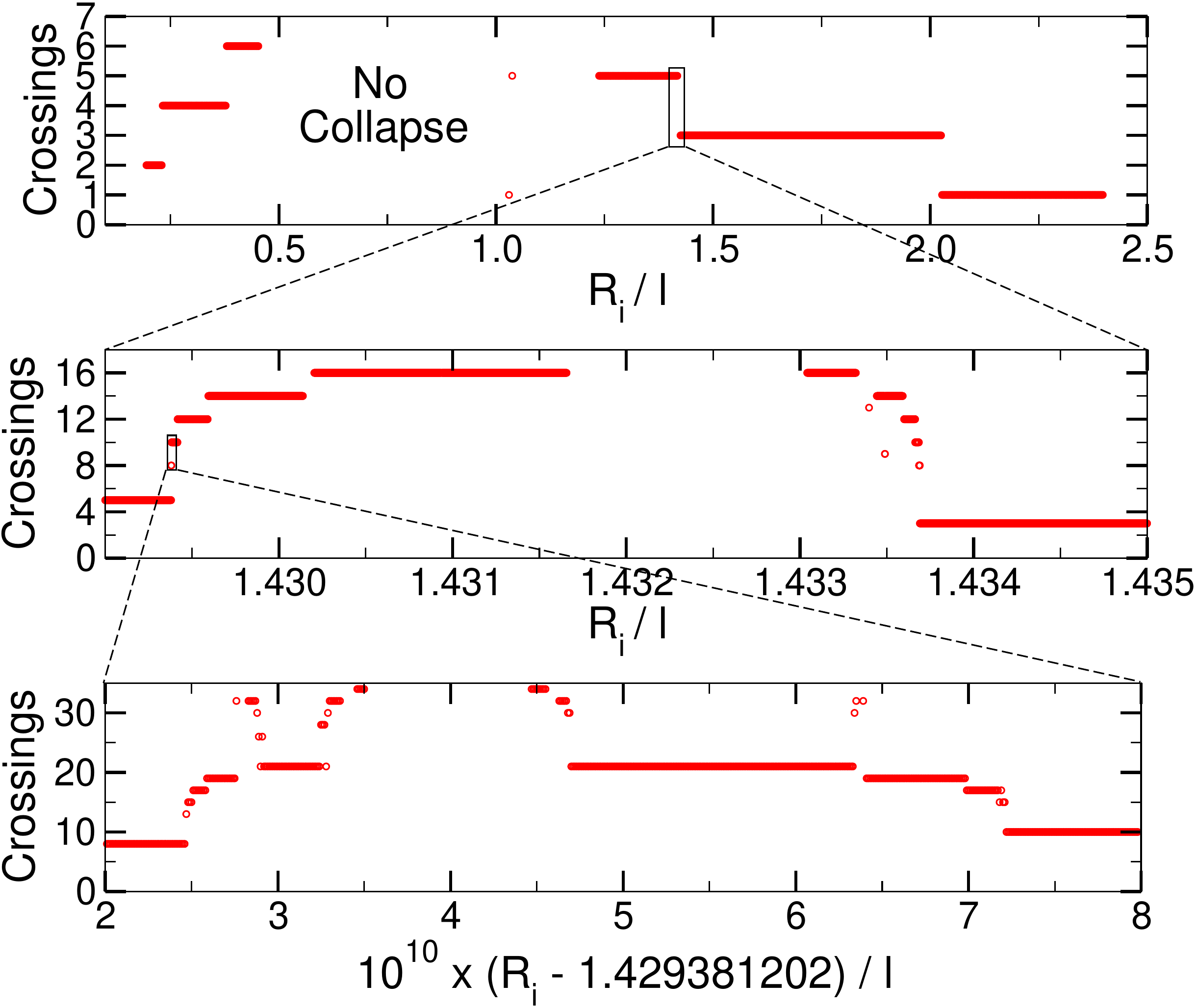}
   \includegraphics[width = 0.54\textwidth]{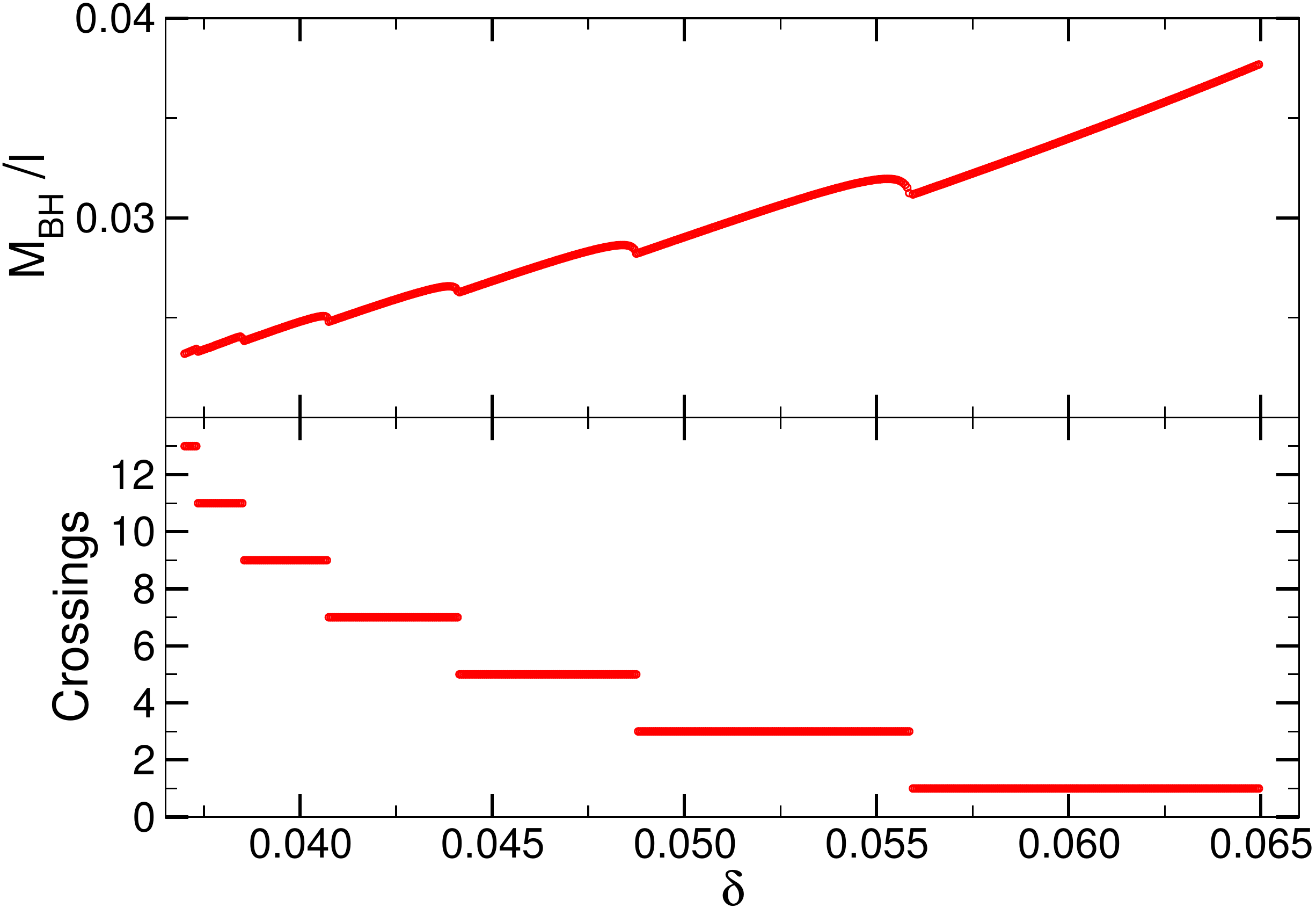}
   \caption{Left: number of crossings between the two
      shells before collapse as a function of $R_i/L$, their common initial position. There are regions where no collapse
      occurs, and a fractal-like structure emerges where an arbitrarily large number of crossings takes place. Right: black hole
      mass and number of crossings as a function of the initial mass-energy content encoded in the $\delta$ parameter for
      a different set of initial conditions. In the left neighbourhood of the $n^{th}$ critical point starting at a mass
      $M_{n+1}$, the mass of the black hole behaves as $M_{H} - M_{n+1} \sim (\delta_n - \delta)^\gamma$. Credits:
      \cite{Brito16b}.}
   \label{fractal}
\end{myfig}

Finally, non-collapsing solutions explored the space of parameters in a chaotic way, as can be seen on figure \ref{chaos}, where
the phase space of the inner and outer shell displayed fractal curves. The thin shell model thus unfolded a very rich structure as
well as simple examples of chaotic islands of stability that were highly sensitive to initial conditions.

\begin{myfig}
   \includegraphics[width = 0.49\textwidth]{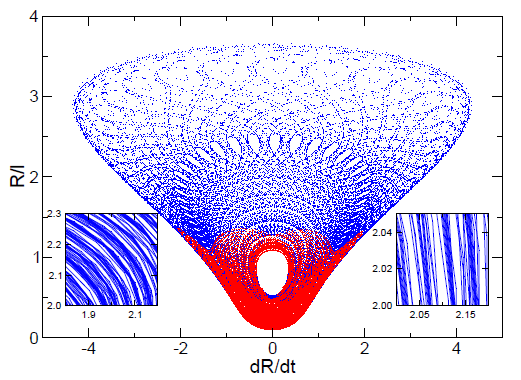}
   \caption{Phase space position-speed of a non-collapsing solution in the thin-shell
      problem. Red depicts the innermost shell's orbit and blue the outermost. Insets show zooms in phase space. Credits: \cite{Brito16b}.}
   \label{chaos}
\end{myfig}

\subsection{Structure of islands of stability}
\label{topoinsta}

As more and more islands of stability were uncovered, a legitimate question was: how large were these islands in the instability sea
? In order to better understand the structure of the instability, let us momentarily consider the simpler case of Minkowski
space-time. This space-time is non-linearly stable when it is slightly perturbed \cite{Christodoulou93}. Thus no black
hole can be formed when the amplitude of the initial data is arbitrarily small. On the other hand, it is well known that many asymptotically
flat self-gravitating objects like neutron stars, white dwarves, boson stars or Proca stars have a maximum mass, i.e.\ 
collapse to a black hole if their mass is too large. Moreover, asymptotically flat electromagnetic or
gravitational geons have no maximum mass \cite{Wheeler55,MartinonPhD}.

All these features can be summed up in an abstract picture, depicted in figure \ref{islandpictureflat}. In a polar
representation, we can insert different families of self-gravitating solutions at different angles and use the radial direction as
a measure of the amplitude (or equivalently the mass) of a particular solution within a given family. On the one hand, at large
amplitudes, many solutions collapse to a black hole, with the notable exception of geons. On the other hand, since Minkowski
space-time is non-linearly stable against small perturbations, no black hole can be formed around the Minkowski background, such
that this space-time lies at the centre of an ``island'' of stability.

\begin{myfig}
   \includegraphics[width = 0.49\textwidth]{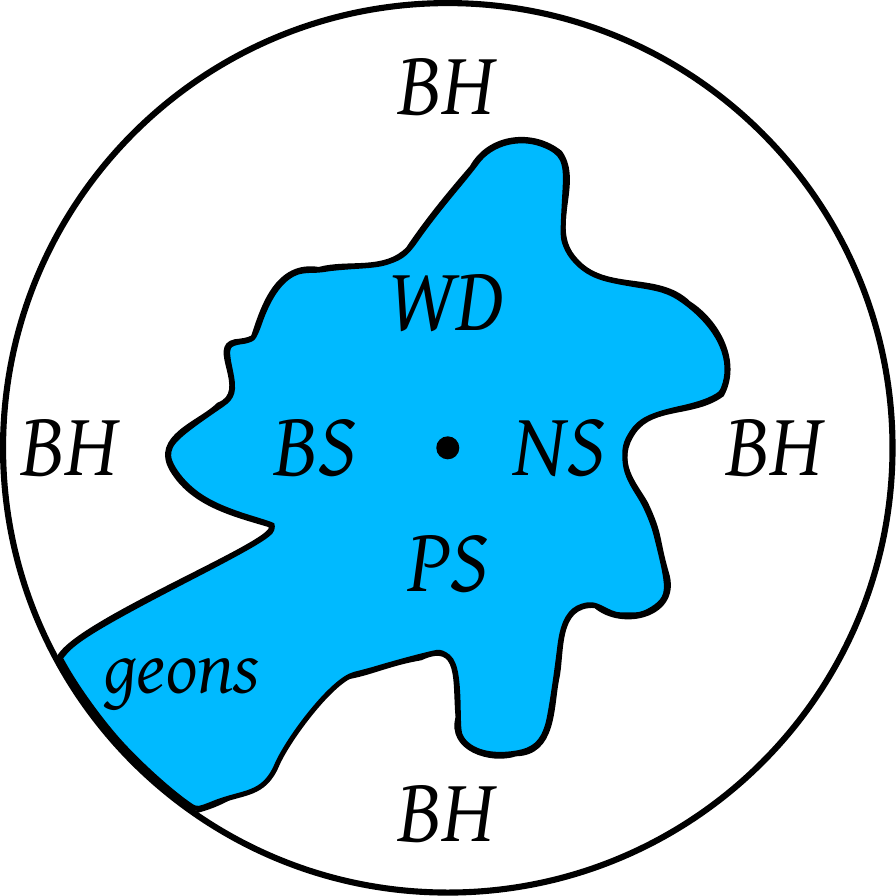}
   \caption{Generic stability diagram of asymptotically flat space-times.
      The radial direction denotes amplitudes of the initial data, so that Minkowski space-time lies at the centre. Angular
      direction is an abstraction that denotes the space of parameters. For example (non-rotating) neutron stars (NS) lie at an
      angle $\theta = 0$, white dwarves (WD) at $\theta = \pi/2$, boson stars (BS) at $\theta = \pi$ and Proca stars (PS) at
      $\theta = 3\pi/2$. We have also inserted electromagnetic geons at $\theta = 5\pi/4$. Since all these objects have a maximum mass,
      they all form a black hole (BH) for sufficiently large amplitudes, which corresponds to the white region. Electromagnetic
      geons on the other hand can have arbitrarily large masses, so they never form a black hole. Since Minkowski space-time is
      non-linearly stable for small perturbations, no black hole can be formed in the neighbourhood of the central Minkowski
      point.}
   \label{islandpictureflat}
\end{myfig}

This kind of diagram is very instructive for studying the structure of the AdS instability. Let us first introduce the following
set of definitions\footnote{The adjective ``generic'' used in this section should not be counfounded with the mathematical
property of genericity, and is used as a synonym of ``almost always''. The distinction with the mathematical property of genericity
is important to perform, though it lacks a rigorous definition of the topology and of a measure in the space of initial data (thanks
to Piotr Bizo\'n for pointing this out).}, taken from \cite{Dimitrakopoulos15a}, and all illustrated in figure \ref{islandpicture}:
\begin{myitem}
   \item ``Generic instability'' means that the set of stable initial conditions (not forming a black hole) shrinks to measure zero
      in the zero-amplitude limit $\varepsilon \rightarrow 0$.
   \item ``Generic stability'' means that the set of unstable initial conditions (forming a black hole) shrinks to measure zero
      in the zero-amplitude limit $\varepsilon \rightarrow 0$.
   \item ``Mixed instability'' means that both sets of initial conditions have non-zero measures in the zero-amplitude limit $\varepsilon \rightarrow 0$.
\end{myitem}
In order to grasp the meaning of these definitions, let us consider a mass isocontour, pictured by a dashed circle in figure
\ref{islandpicture}. If we mentally try to progressively reduce its radius down to zero (zero-amplitude limit), we see
that the circle tends to become completely white on the left diagram (a black hole is always formed), rainbow-like on the central
one (a black hole is never formed), and half-white half-coloured on the right one (black hole may or may not form depending on the
initial data). The three different concepts listed above thus correspond to different colour end states for a limiting isomass circle
whose radius is shrinking to zero.

\begin{myfig}
   \includegraphics[width = 0.98\textwidth]{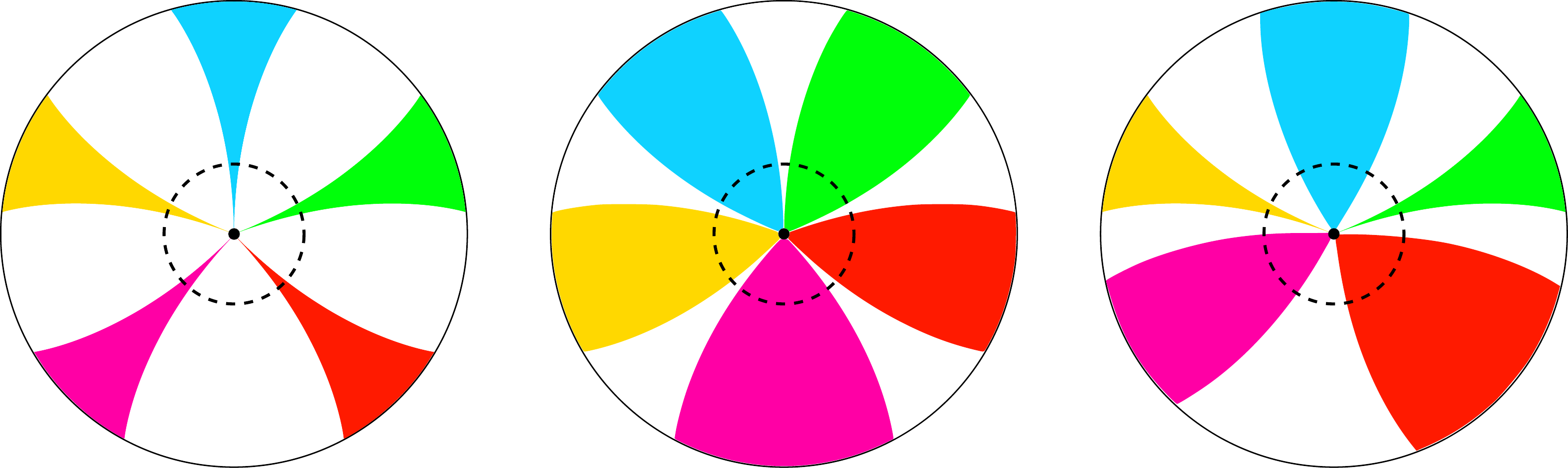}
   \caption{Schematic representation of the generic stability
      definitions. The radial direction denotes amplitudes of the initial data, so that pure vacuum AdS space-time lies at
      the centre. Angular direction is an abstraction that denotes the space of parameters, like the quantum numbers of geons
      (which are non-spherically symmetric periodic solutions, see section \ref{beyondspher}) or the fundamental frequency of
      time-periodic solutions (section \ref{tpsol}), or more generally any functional dependence. Coloured regions correspond to
      initial data that are non-linearly stable and never collapse, different colours corresponding to different families of
      stable solutions. White regions indicate initial data that are non-linearly unstable and collapse to a black hole. The left
      panel describes the ``generic instability'' picture, the middle panel describes the ``generic stability'' picture, and the
      right panel describes a possible ``mixed stability'' picture. On each diagram, we show a mass isocontour with a dashed circle.}
   \label{islandpicture}
\end{myfig}

After the discovery of the weakly turbulent instability \cite{Bizon11} and the perturbative construction of geons \cite{Dias12a}
(non-spherically symmetric time-periodic solutions, see below section \ref{beyondspher}), the
general idea was that AdS space-time was generically unstable \cite{Dias12b}, as can be seen on the left panel of figure
\ref{islands} reproducing that of \cite{Dias12b}. However, the work of \cite{Dimitrakopoulos15b} took full advantage
of the scaling symmetry \eref{scalingsym} of the TTF equations to demonstrate that any non-collapsing solution of amplitude
$\varepsilon$ remained stable in the $\varepsilon \rightarrow 0$ limit in the fully non-linear theory. This forbids the cuspy shape
of the diagram and argues in favour of instability \textit{corners}. In particular, if non-collapsing solutions form a set of
measure non-zero at finite amplitudes, then they persist to be a set of measure non-zero when the amplitude tends to zero.
Figure \ref{islands} illustrates the tension between the original statement of \cite{Dias12b} and the theorems demonstrated in
\cite{Dimitrakopoulos15b}. The former advocates for the ``generic instability'' picture while the latter argues in favour of the
``mixed stability'' hypothesis.

\begin{myfig}
   \includegraphics[width = 0.37\textwidth]{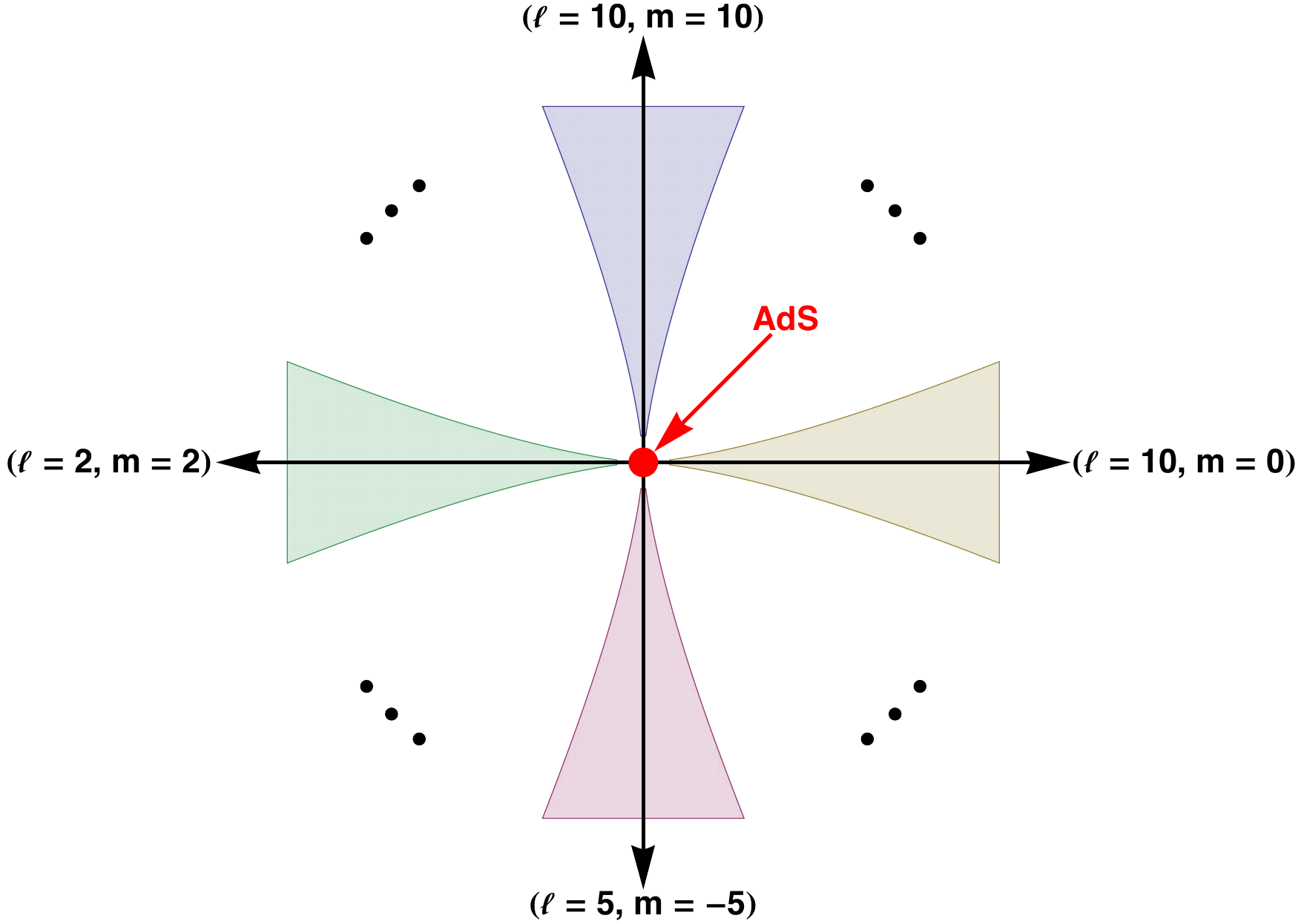}
   \includegraphics[width = 0.28\textwidth]{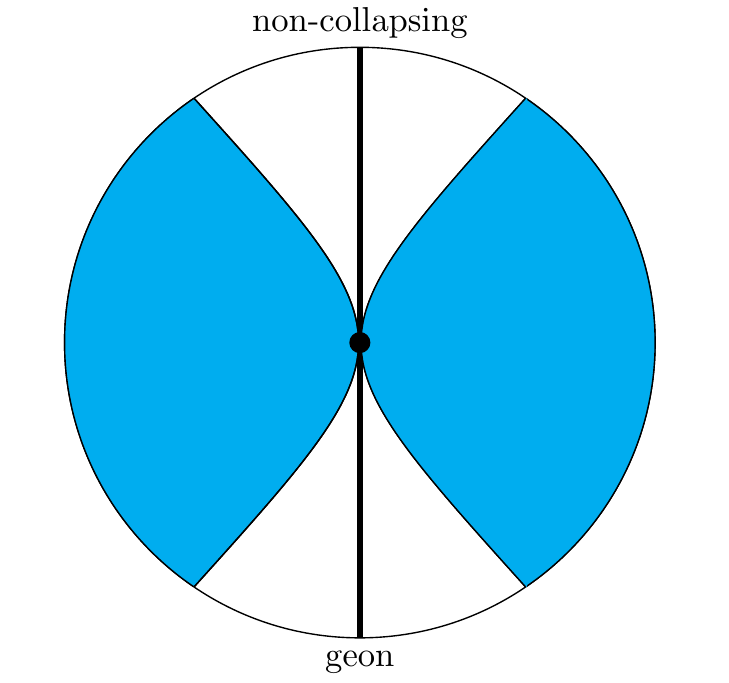}
   \includegraphics[width = 0.28\textwidth]{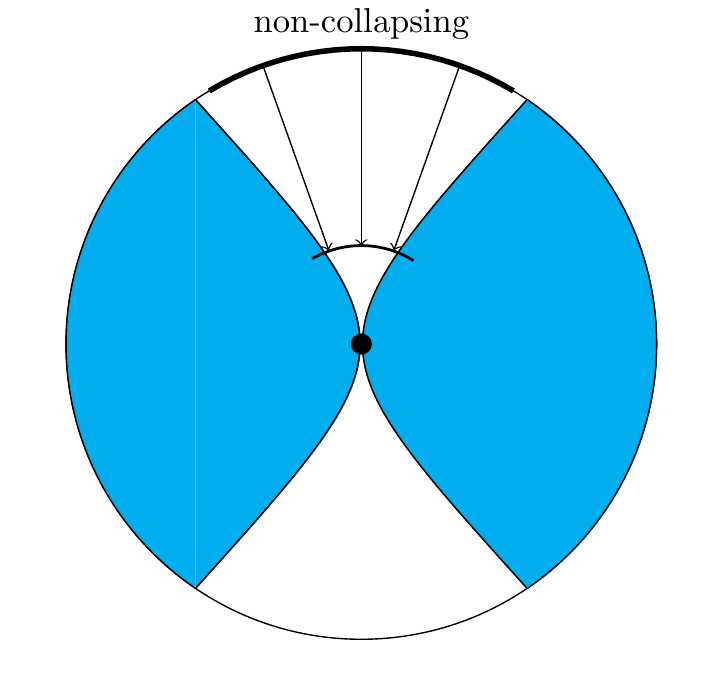}
   \caption{Left panel: diagrammatic picture of islands of stability. Geons solutions lie on the black
      arrows. Shaded regions denote islands of stability around geon solutions. Because perturbation theory about empty AdS
      leads to geons only for a measure zero set of seed solutions in the generic instability picture, each such region has been
      drawn so that empty AdS lies at a cusp. Middle and right panels: phase-space diagrams of the stability island
      conjecture. Initial perturbations in the blue region collapse while the unshaded region represents islands of stability. The
      theorems of \cite{Dimitrakopoulos15b} show that one can transport non-collapsing solutions radially without triggering
      instability. This is in direct contradiction with the cuspy nature of stable regions, that would be better drawn as
      corners. Credits:
      \cite{Dias12b,Dimitrakopoulos15b}.}
   \label{islands}
\end{myfig}

To summarise, the weakly turbulent instability of AdS \cite{Bizon11} ruled out the ``generic stability'' hypothesis. This
lead some people to adopt the ``generic instability'' picture \cite{Dias12b}. But actually, the emergence of numerical islands of
stability in combination with the TTF framework revealed a more appropriate ``mixed stability'' representation, where unlike the
right panel of figure \ref{islands}, all stability regions are stability \textit{corners} \cite{Dimitrakopoulos15b}. If we also
take into account the chaotic behaviour exhibited in section \ref{chaosfoot}, the boundaries between stable and unstable solutions in
the mixed stability picture of figure \ref{islandpicture} might well be fractals.

\section{Conditions for collapse}
\label{conditions}

So far, we have discussed several examples of non-linearly stable or unstable solutions. But given the mixture of possibilities
and the intricate structure of the AdS instability, a legitimate question is: can we give a set of a few necessary or
sufficient conditions for collapse?

\subsection{Competition between focusing and defocusing}
\label{competition}

In hope of giving a mechanism of the instability, intuition argues that self-gravitation of a scalar wave packet tends to
always contract the field every times its typical size is small and self-interaction is large, i.e.\ every times it crosses the
origin. And so on so forth until black hole formation. However, it is a mistake to believe that self-gravitation only contributes to
contracting the scalar field profile. Incidentally, the existence of non-linearly stable solutions suggests that there is at least
one other effect that counterbalances the contraction. Islands of stability should thus result from an endless competition between contraction and
dilatation.

Indeed, in order to understand the islands of stability found in the literature, the authors of \cite{Dimitrakopoulos15a}
developed a perturbative and heuristic argumentation: gravitational self-interaction leads to tidal
deformations which are equally likely to focus or defocus energy. This stresses the potential repulsive nature of gravitation.
A daily illustration of this statement lies in the tidal effects of the Moon onto the Earth that induce a stretching and not a
compacting. The idea is that stable solutions oscillate between focusing and defocusing dynamics. On figure \ref{focusing}, the
simple example of a scalar pulse with one peak region and one extended tail region is carried out (analytical arguments can be
found in \cite{Dimitrakopoulos15a}). On the one hand, if the peak enters first the central region near the origin, the dominant
behaviour is that of contraction. On the other hand, if the tail reaches the origin first, it flattens the peaked region via tidal
interactions. Thus, there is clearly a competition between contraction and dilatation. These two behaviours can act alternatively
in similar proportions like for stable solutions, or fluctuate so much as to finally form a black hole, as depicted in figure
\ref{flowenergy}. It is important to note that these considerations rely on position-space analysis instead of the very popular
energy spectrum analysis.

\begin{myfig}
   \includegraphics[width = 0.49\textwidth]{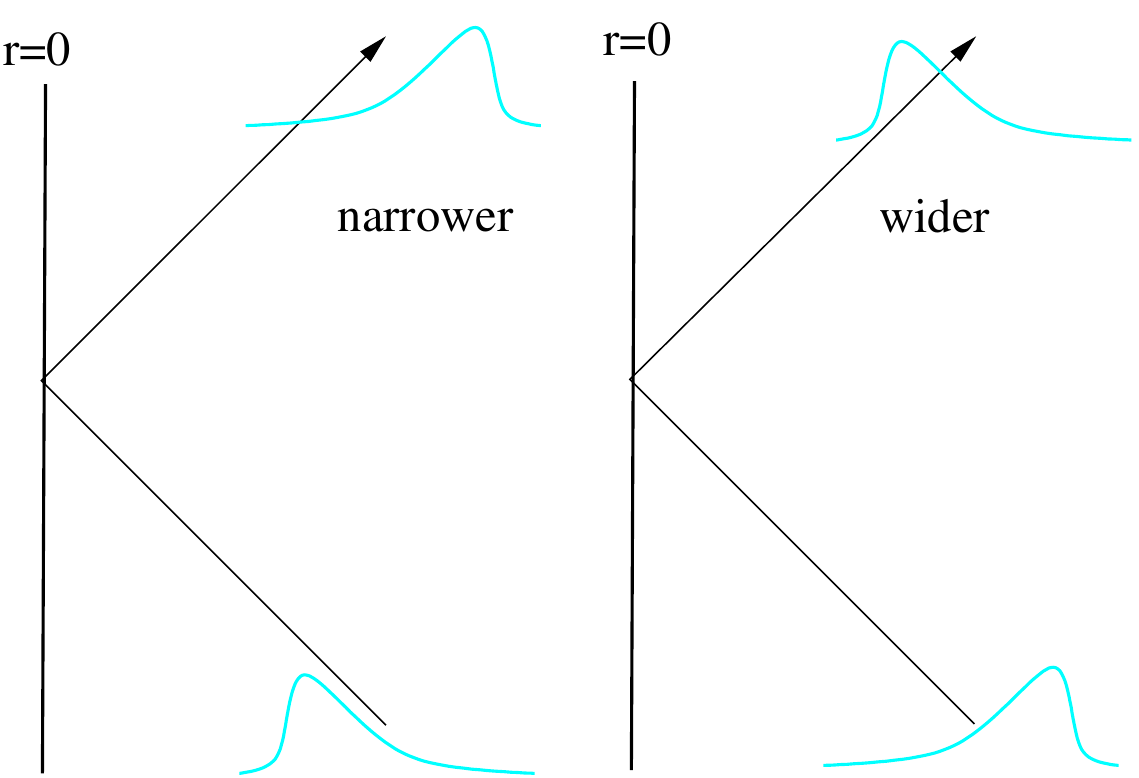}
   \caption{A thin shell with higher energy density in its front comes out
      narrower as it crosses the origin, because of self-gravitating contraction effects. However, the flipped configuration with high
      energy in the tail comes out wider because of tidal effects inducing a dilatation of the profile. Credits: \cite{Dimitrakopoulos15a}.}
   \label{focusing}
\end{myfig}
\begin{myfig}
   \includegraphics[width = 0.49\textwidth]{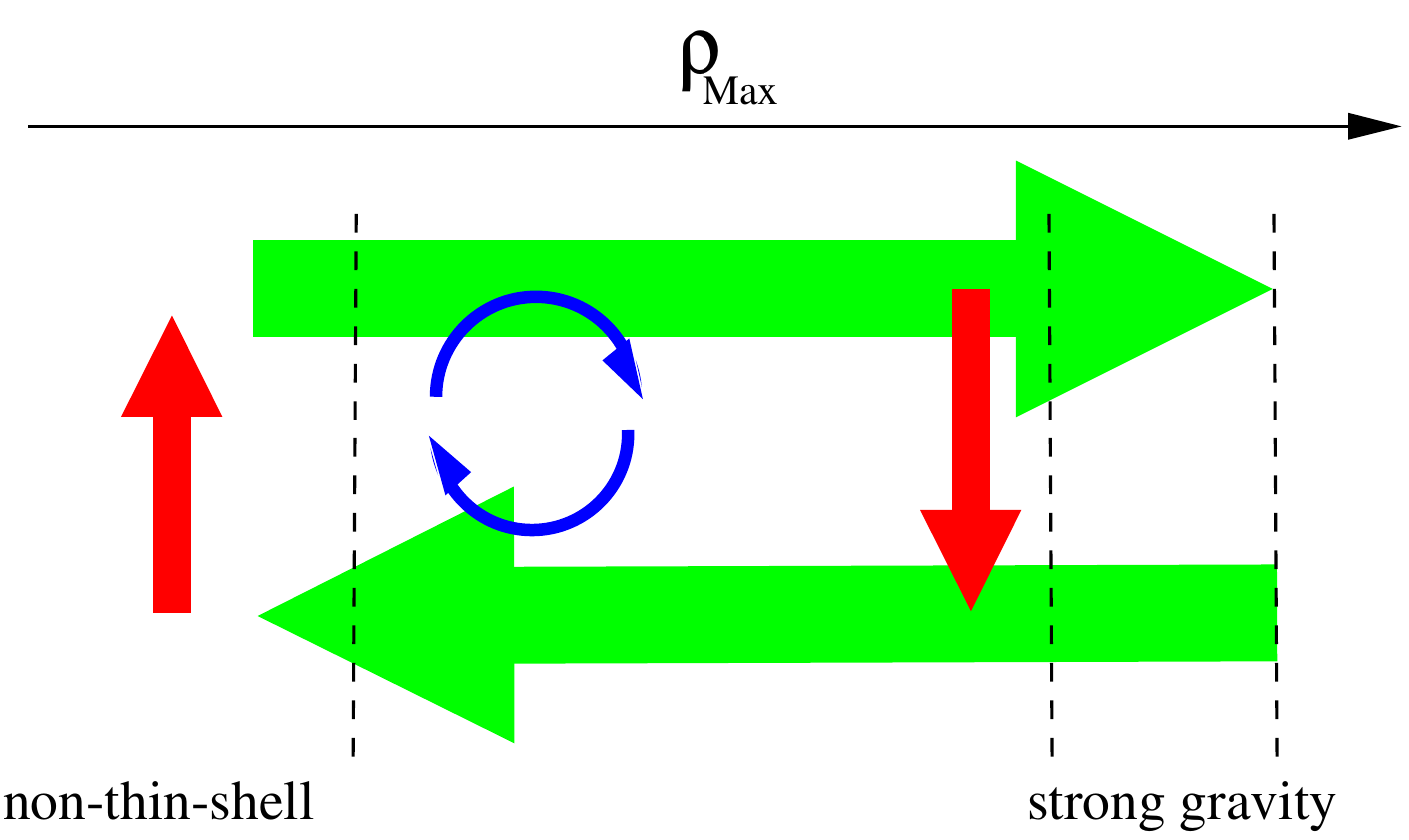}
   \includegraphics[width = 0.49\textwidth]{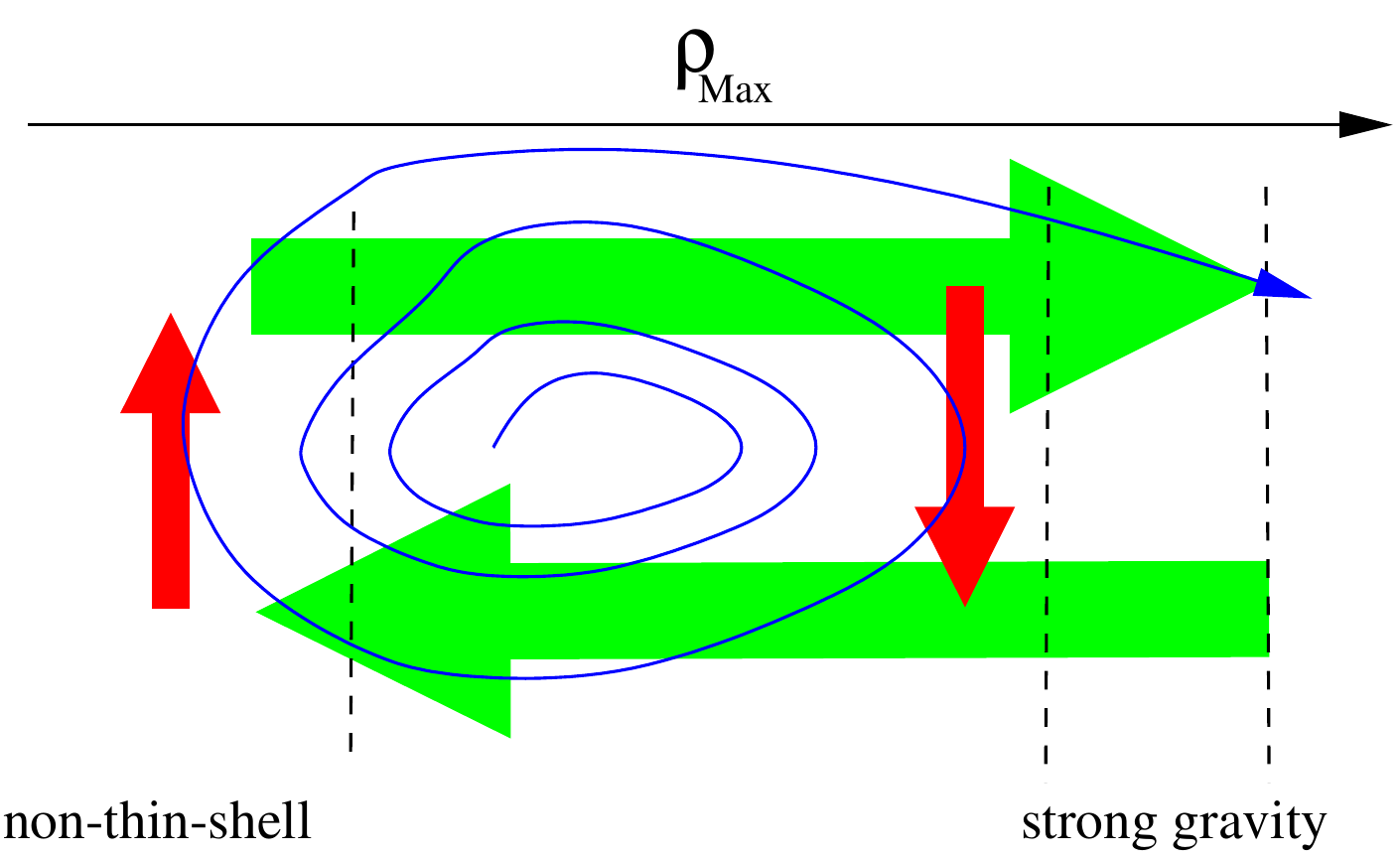}
   \caption{The horizontal axis is the peak energy density of the scalar
      pulse. The green left-oriented arrow represents defocusing effects while the right-oriented one represents focusing effects. There can be
      stable solutions (left panel) that never collapse because of a circular flow pattern (exchange) between these two competitive
      behaviours, while unstable solutions (right panel) fluctuate so much as to form eventually a black hole. Credits:
      \cite{Dimitrakopoulos15a}.}
   \label{flowenergy}
\end{myfig}

\subsection{Phase coherence}

Because of the competition between contraction and tidal dilatation, it can be guessed that collapse
occurs only when the successive contraction stages of the wave packet motion add up coherently during the motion without being
inhibited by the dilatation stages. This is the idea of phase coherence: small effects can build up in time if they are summed
coherently, like a child's swing gathering momentum each time it is pushed forward with the right timing.

The authors of \cite{Freivogel16} took over this argument and showed that a power-law spectrum, usually classified as unstable,
was not a sufficient condition for black hole formation. They uncovered that such a spectrum could indeed belong to a stable
solution if the phases $B_j$ were incoherent. They formulated the condition for phase coherence as (see equations
\eref{ttfequationsa}-\eref{ttfequationsb})
\begin{equation}
   B_j(\tau) = j\gamma(\tau) + \theta(\tau) + \ldots,
   \label{coherent}
\end{equation}
where dots represents anything that goes to zero in the large-$j$ limit. Said differently, the phases between modes are said to be
coherent if they are (asymptotically) equidistant. For example in 4-dimensional AAdS space-times, if the phase coherence
condition \eref{coherent} is satisfied, any spectrum $A_n \sim n^{-\alpha}$ with $\alpha > 3$ remains regular at all times while
black holes form for $\alpha < 2$. Between these two limits, substantial back-reaction on the metric is at work and would need
numerical simulations to conclude about black hole formation. On the other hand, whatever is the value of $\alpha$, any
phase-incoherent initial data never collapses to a black hole in a time $O(\varepsilon^{-2})$. Phase
coherence thus appears as an additional necessary condition for collapse. To prove that this condition was not sufficient,
the authors, by fine-tuning the phases, were able to build power-law spectrum solutions where there was no energy transfer among
the modes and hence non-linear stability. Furthermore they demonstrated that the two-mode initial data were particularly prone to
provide coherent phases and hence to lead to black hole formation. This was also demonstrated for unstable Gaussian initial data
\cite{Dimitrakopoulos17}.

The lesson from \cite{Freivogel16} is thus the following: when employing the TTF equations to probe unstable configurations,
the solution can be reasonably declared unstable if (i) the analyticity radius hits zero in a finite slow-time (section
\ref{analstrip}), (ii) the phase coherence condition \eref{coherent} holds and (iii) the exponent of the power-law
spectrum (recall that an analyticity radius of zero implies a power-law spectrum) should be in some interval (superior to
two in the 4-dimensional case), given in \cite{Freivogel16}.

\subsection{The role of a resonant spectrum}
\label{roleres}

A large part of the literature focused on the massless scalar field in AdS with spherical symmetry. This setting has a linearised
Einstein's operator that is resonant, i.e.\ with equidistant eigenvalues (see section \ref{pertscal}). In order to investigate if
this is a sufficient condition for the instability to be triggered, it is interesting to study different situations where the
resonant character of the spectrum is broken in order to see if the instability is suppressed or maintained. Evolving a massive
scalar field or imposing Neumann boundary conditions in an enclosed cavity are two possible ways of breaking the resonant
spectrum.

For example, massive scalar field initial data in spherical symmetry were evolved in time in the following articles. In \cite{Okawa14a},
initial data were evolved in an asymptotically flat space-time, mimicking a confining mechanism with a $\phi^4$
potential\footnote{For another confining mechanism mimicking that of AdS space-time, see also \cite{Biasi17} that studies
the Gross-Pitaevskii equation with attractive non-linearity in a harmonic potential. In this non-gravitational study, the authors
found that turning off the resonant spectrum always gave rise to a minimum threshold amplitude below which the wave function never
becomes singular.}. Even if delayed collapse (i.e.\ after several ``bounces'' off the potential barrier) was observed, there was always a
finite threshold below which no collapse occurred. In \cite{Okawa14b}, the confining mechanism was enforced within a flat enclosed
cavity with Dirichlet or Neumann boundary conditions. In the case of Dirichlet boundary conditions, and denoting by $\mu$ the mass
of the scalar field, the spectrum of the linear operator was
\begin{equation}
   \omega_j = \sqrt{\mu^2 + \frac{j^2\pi^2}{R^2}},
\end{equation}
and thus only asymptotically resonant, i.e.\ resonant only for infinite wave-numbers. Finally in \cite{Okawa15}, the standard AdS setup was studied with a massive scalar
field. One important result of these works was that time evolutions with non-resonant spectrum collapsed earlier than in the fully
resonant case, as illustrated in figure \ref{nonresonant}. Furthermore, the authors recovered that in AdS space-time, be the
spectrum resonant or not, all the original features of the massless scalar case were present (instability for very small
amplitudes and finely-tuned islands of stability). These results thus suggested that a resonant spectrum was not a necessary
condition for collapse.

\begin{myfig}
   \includegraphics[width = 0.49\textwidth]{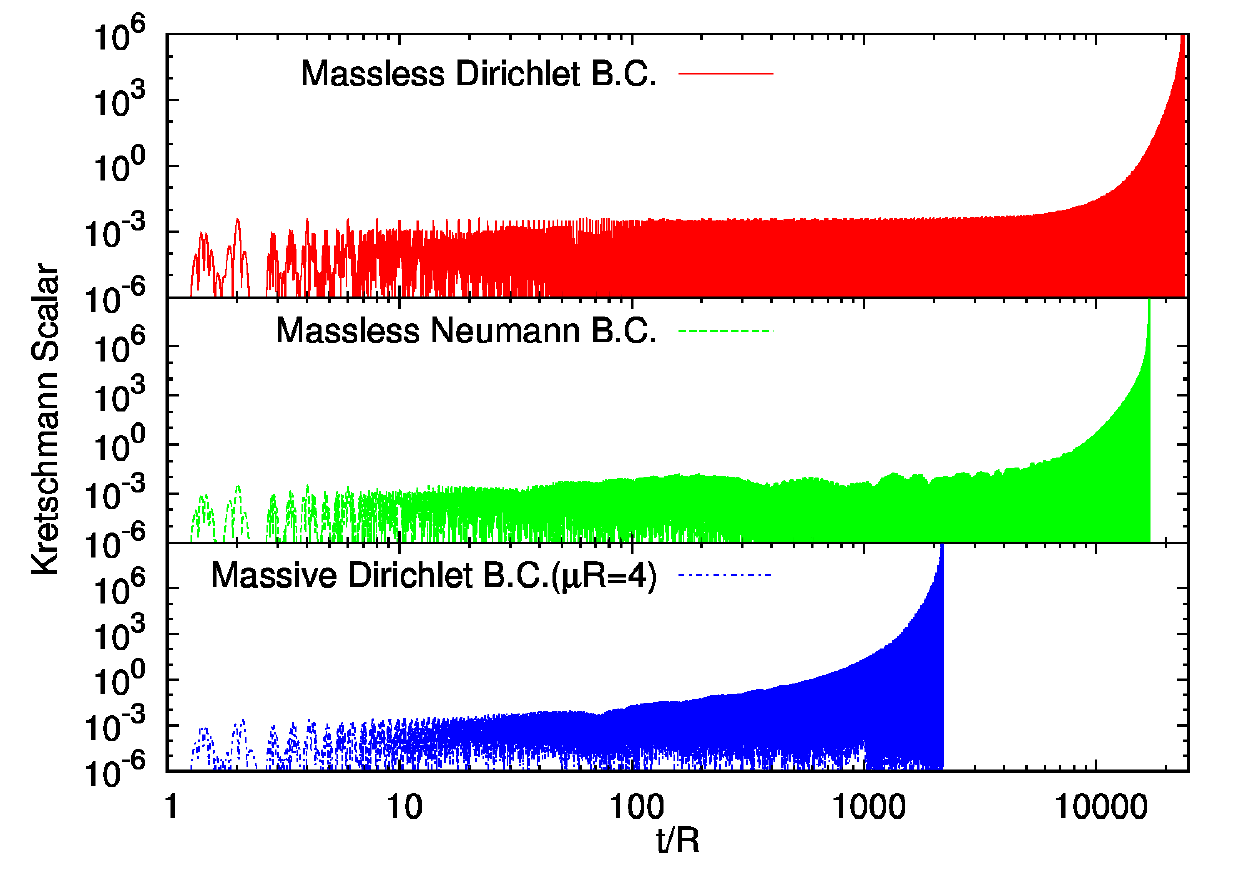}
   \caption{Evolution in time of the Kretschmann scalar for three identical Gaussian
      initial data in a flat enclosed cavity. Top: massless scalar field with Dirichlet boundary conditions, the spectrum is fully
      resonant. Middle: massless scalar field with Neumann boundary conditions, the spectrum is asymptotically resonant. Bottom:
      massive scalar field with Dirichlet boundary conditions, the spectrum is non-resonant. Credits: \cite{Okawa14b}.}
   \label{nonresonant}
\end{myfig}

In order to further investigate this point, it is also possible to investigate the dynamics of a massless scalar field in a flat
enclosed cavity, as initially suggested in \cite{Maliborski12}. At the time, the authors shared the opinion that the resonant
spectrum was not a necessary condition for the instability. However, the careful analysis beyond spherical symmetry of
\cite{Dias12b} (see below section \ref{beyondspher}) came out soon after and established an opposite conclusion. In order to solve
this contradiction between analytical and numerical studies, the flat enclosed cavity was scrutinised once more in
\cite{Maliborski14}, whose authors pursued their initial work \cite{Maliborski12}. Decreasing the amplitude of the initial data,
they did find this time that, with Neumann boundary conditions, there existed a threshold $\varepsilon_0$ below which no collapse
occurred. The case of Neumann boundary conditions was particular precisely because the spectrum of the linear operator was only
asymptotically resonant. Indeed, the eigen frequencies obey $\tan(\omega_j R) = \omega_j R$, where $R$ is the radius of the
cavity. In the large-$j$ limit, it comes
\begin{equation}
   \omega_j = \frac{\pi}{R}\left( j + \frac{1}{2} \right) + O\left( \frac{1}{j} \right),
\end{equation}
so that these frequencies are only asymptotically equidistant. This setup was generalised to the charged scalar field in
\cite{Ponglertsakul16} with qualitatively identical results: no collapse below a certain threshold of amplitude. This was also analytically supported by arguments within the
Kolmogorov-Arnold-Moser (KAM) theory in \cite{Menon16}. Indeed, non-linear dynamics theorems argue that when the spectrum is not
perfectly resonant, there always exists a finite (but possibly arbitrarily small and difficult to spot numerically) threshold
$\varepsilon_0$ below which the instability is suppressed.

Thus the three numerical studies \cite{Okawa14a,Okawa14b,Okawa15} dealing with massive scalar fields had opposite conclusions to the ones of
\cite{Maliborski14,Ponglertsakul16} whose authors used a massless scalar field. However, thanks to the analytical studies
\cite{Dias12b,Menon16}, the role of a resonant spectrum is now well understood to be that of a necessary (but not sufficient)
condition for the instability. 

At this point, it is wise to point out two numerical limitations: (i) the amplitude of numerical
initial data cannot be arbitrarily small and (ii) numerical simulations run for a finite amount of time. Thus, if an unstable
solution is spotted in a numerical simulation, it is rigorously impossible to state that it remains unstable in the zero-amplitude
limit (in contrast with stable solutions whose stability is preserved by decreasing the amplitude, as explained in section
\ref{topoinsta} via the TTF equations). In the same vein, if a stable solution is observed to be stable for a finite (even if very
long) time in numerical simulations, it is rigorously impossible to state that it remains stable in the infinite-time limit. The
different numerical setups and precisions used in the literature may explain the apparent disputes between several authors. This
is why analytical frameworks like KAM theory, perturbations, analyticity strip and TTF are invaluable tools to get more insight in
the structure of the instability.

For the sake of completeness, let us mention that a lot more new classes of islands of stability were uncovered in the massive
scalar case. For example in \cite{Fodor14,Fodor15}, time-periodic solutions (or breathers) were constructed perturbatively and
numerically. Moreover, by fine-tuning the mass of the scalar field, whole families of non-linearly stable Gaussian initial data
were found in \cite{Deppe15a}. Finally in \cite{Okawa15}, initial data made of three Gaussian wave packets were observed to remain
non-linearly stable by continuously exchanging energy between its constitutive parts. This is in full agreement with the
focusing/defocusing mechanism discussed in section \ref{competition}.

\section{Instability when no black hole is allowed}
\label{nobh}

So far, we always confounded non-linear instability and black hole (or apparent horizon) formation. How would the instability be
expressed if, by one way or another, black hole formation was simply forbidden? There are at least two simple cases when black
hole formation is not allowed: 3-dimensional AAdS space-times and Einstein-Gauss-Bonnet (EGB) gravity.

\subsection{3-dimensional AAdS space-times}

The particular case of 3-dimensional AAdS space-times was studied in \cite{Bizon13}. As mentioned
in section \ref{critical}, an AAdS 3-dimensional (or BTZ) black hole has a minimum mass $M_0$, so that initial data with total mass $M < M_0$ can
never form this kind of singularity. Thus, the natural turbulent cascade cut-off of black hole formation does not
prevail any more, and what is observed is an exponential growth of the Sobolev norm $H_2(t) = \| \phi''(t,x) \|_2$ as well
as an exponential decrease of the analyticity radius, shown in figure \ref{sobolev}. Unlike the higher dimensional cases, the
analyticity radius never reaches zero.

\begin{myfig}
   \includegraphics[width = 0.49\textwidth]{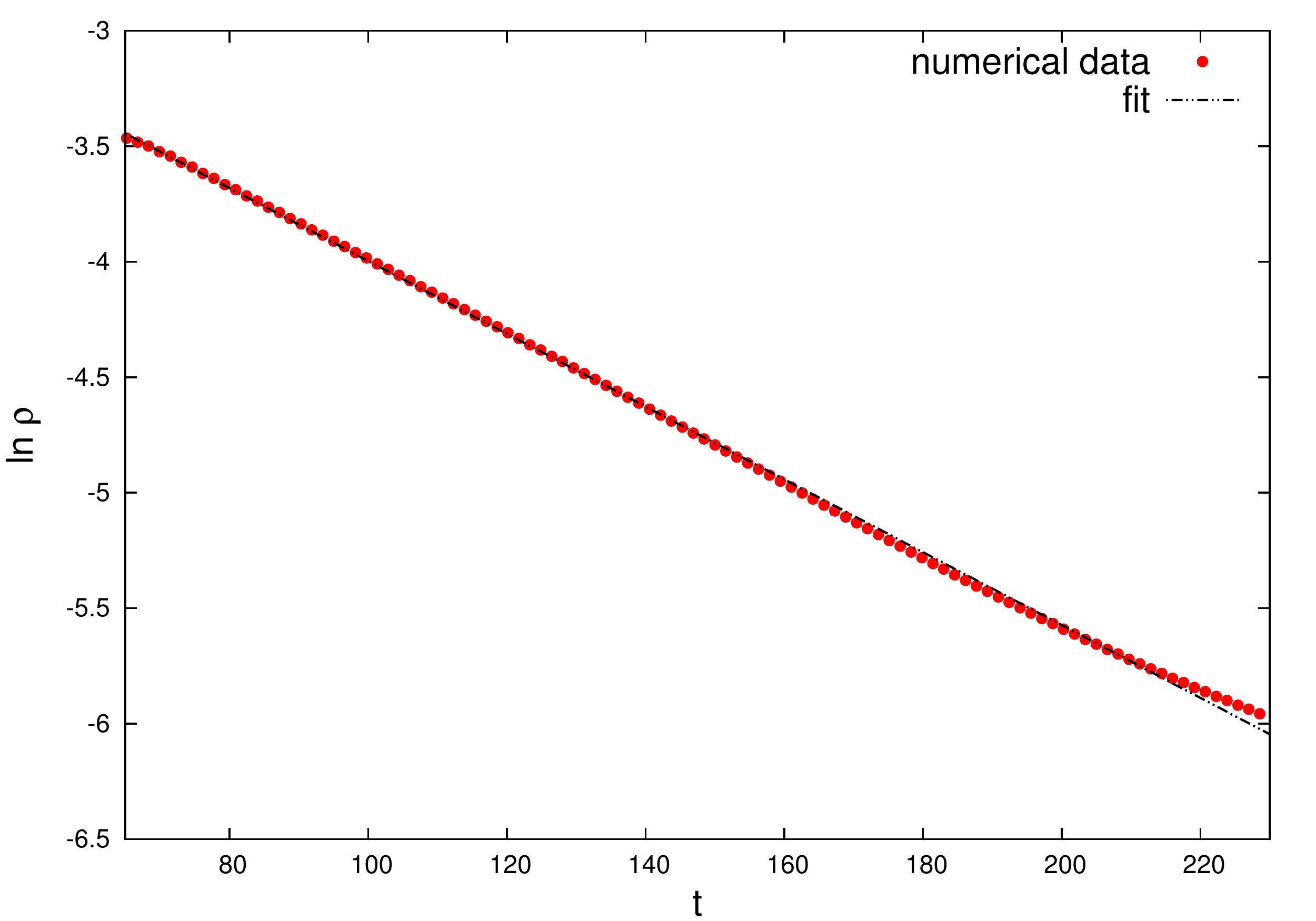}
   \includegraphics[width = 0.49\textwidth]{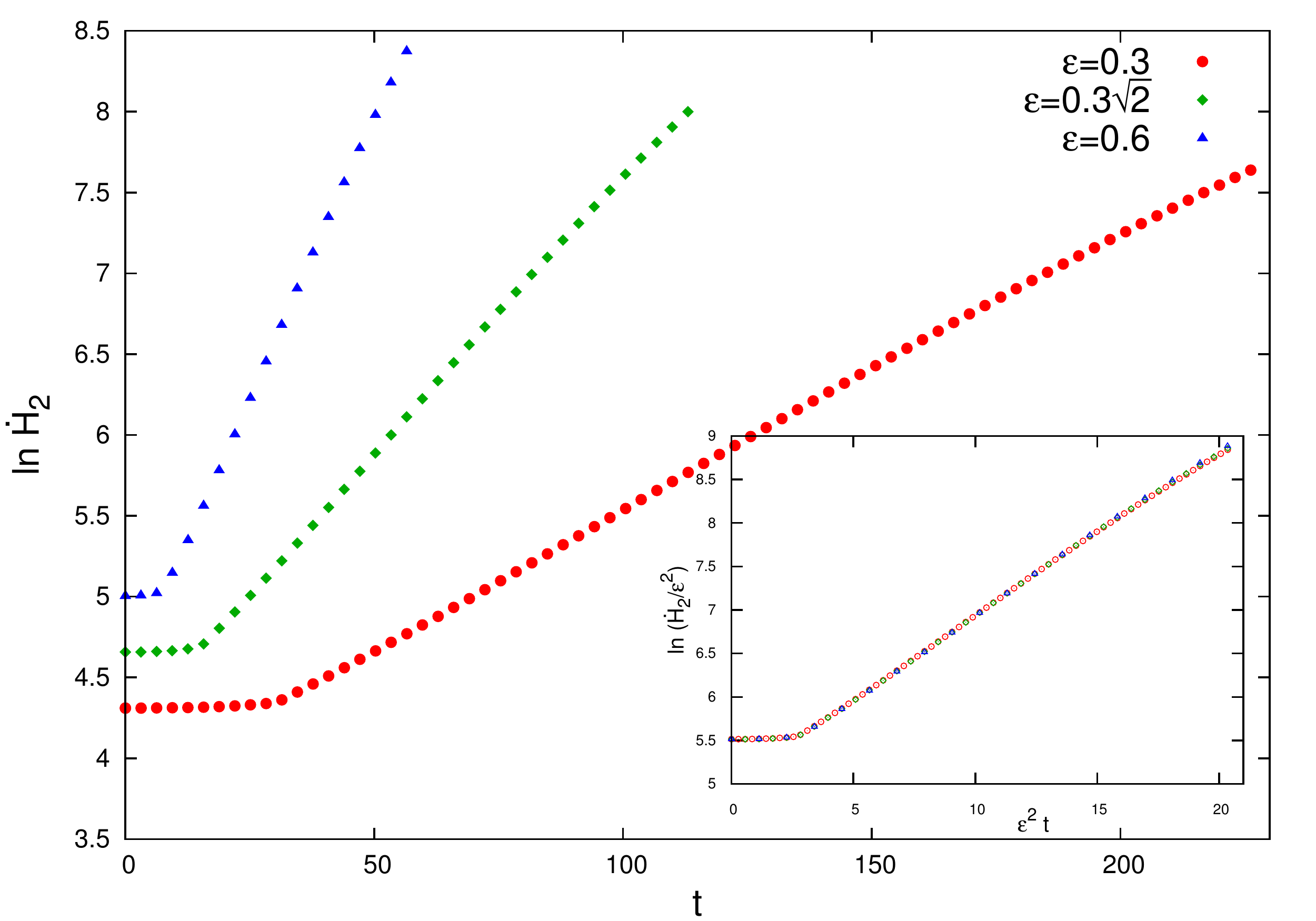}
   \caption{Left: analyticity radius $\rho$ of a Gaussian initial
      data in 3-dimensional AdS space-time. The decrease is exponential, but $\rho$ never hits zero. Right: Sobolev norm $H_2(t) =
      \| \phi''(t,x) \|_2$ of the scalar field for three different amplitudes of the Gaussian initial data. The inset
      shows the same curves in terms of the slow-time $\varepsilon^2 t$, so that it is clear that the blow up is $\propto
      \exp(O(\varepsilon^2 t))$. Credits: \cite{Bizon13}.}
   \label{sobolev}
\end{myfig}

The instability conjecture has thus a different flavour in three dimensions \cite{Bizon14}:

\begin{conjecture}[Anti-de Sitter instability in 3 dimensions]
Small smooth perturbations of 3-dimensional AdS remain smooth at all times but their radius of analyticity shrinks to
zero exponentially fast.
\end{conjecture}

Said differently, even if no black hole is formed for sufficiently small amplitude initial data, the turbulent cascade still
occurs as the perturbations do not remain small in any reasonable norm that captures the turbulent behaviour.

\subsection{Einstein-Gauss-Bonnet gravity}

The extension of the EKG setup to 5-dimensional EGB gravity was performed in \cite{Deppe15a,Deppe16a}, adding the
following term to the Lagrangian of the theory
\begin{equation}
   \lambda(R^2 - 4R_{\mu\nu}R^{\mu\nu} + R_{\mu\nu\rho\sigma}R^{\mu\nu\rho\sigma}),
\end{equation}
where $\lambda$ is the Gauss-Bonnet (GB) parameter. In this theory, a limiting black hole of radius $r_H\rightarrow 0$ has a
minimum mass $M_0 = \lambda/2$. This implies that, in analogy with the 3-dimensional AdS case, no black hole can form if the
initial data has mass lower than $M_0$.

The authors chose to study Gaussian initial data, for which the amplitude corresponding to $M_0$ is $\varepsilon = 21.86$. However, no black
hole formation was observed for amplitudes below $\varepsilon = 36$, whose last collapsing solution bounced 24 times. This means
that there was a whole range of amplitudes where black hole formation was theoretically possible but yet did not occur. Running a simulation with
an amplitude in this range but turning off the GB term did indeed reignite collapse. The intuition was that the GB
term contributes largely to defocusing mechanisms that resists black hole formation. The instability was thus suppressed more
severely than expected.  The comparison of collapses with and without the GB term is clearly visible on
figure \ref{gaussbonnet}, where the structure of the instability seems much more chaotic in the GB gravity.

\begin{myfig}
   \includegraphics[width = 0.49\textwidth]{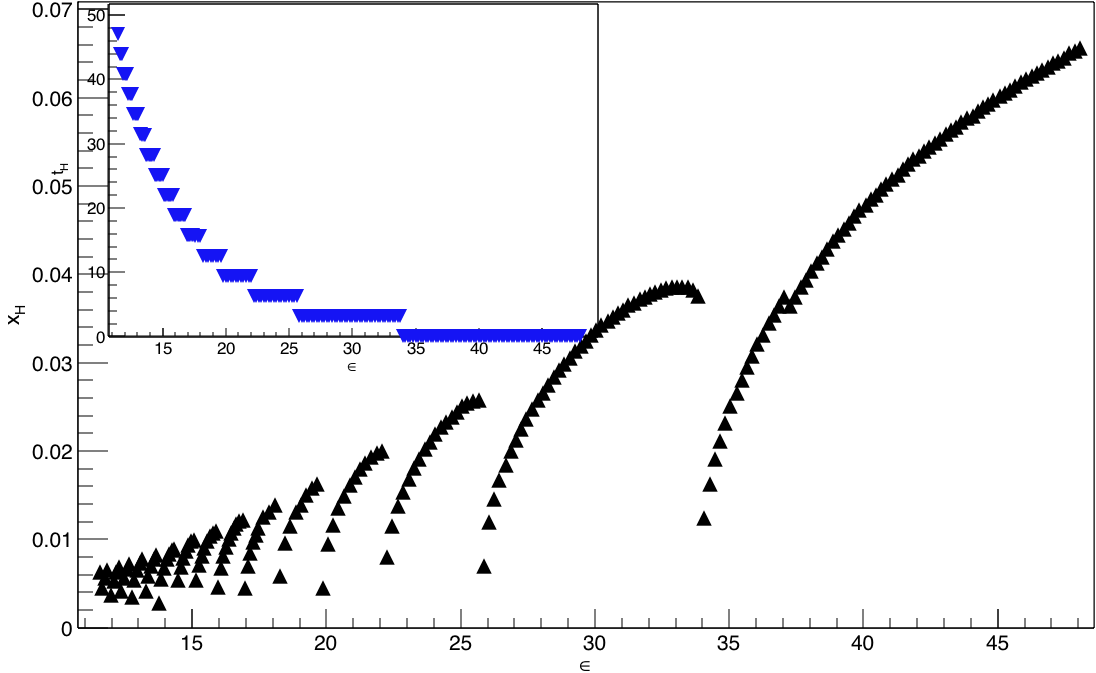}
   \includegraphics[width = 0.49\textwidth]{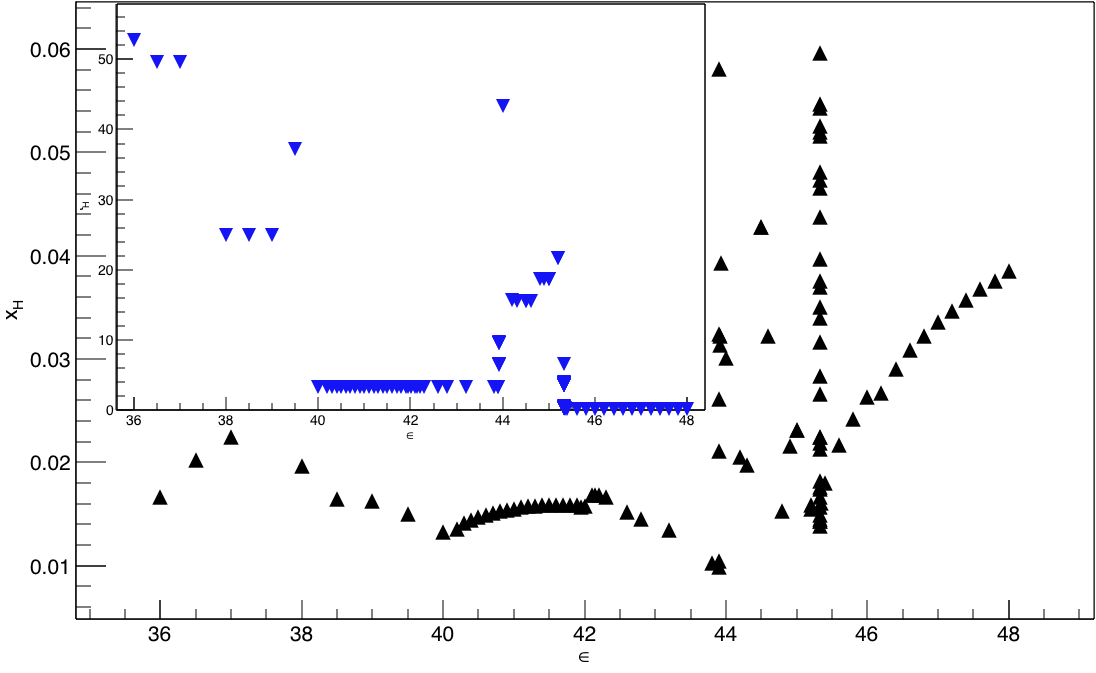}
   \caption{Horizon radius $x_H$ as a function of the initial amplitude of the
      Gaussian scalar wave packet in Einstein's (left) and EGB gravity (right). Insets show the black hole
      formation times $t_H$. Credits: \cite{Deppe15a}.}
   \label{gaussbonnet}
\end{myfig}

\section{Beyond spherical symmetry}
\label{beyondspher}

In the literature, the large majority of studies in the field of the instability conjecture assumed
spherical symmetry. Birkhoff's theorem states that at the exterior of a spherically symmetric non-rotating body the metric
should match the one of Schwarzschild-AdS space-time. Thus no gravitational waves can be emitted in such space-times and gravitational dynamics were thus
induced with a scalar field. This was the simplest setting we can imagine, and also the cheapest computationally speaking. The
first investigations beyond spherical symmetry were dedicated to find purely gravitational time-periodic configurations called
geons. They can be seen as generalisations of the time-periodic solutions of section \ref{tpsol}. The cost of numerical evolutions of
initial data increasing significantly with the number of dimensions, there are very few probes of the instability beyond spherical
symmetry to date. However, this seems a promising area of research in a near future.

\subsection{Geons}
\label{geonsec}

The first study beyond spherical symmetry was dedicated to AAdS geons in \cite{Dias12a}, which was also the first attempt to build
solutions that resisted black hole formation. The authors solved the vacuum Einstein's equation with a negative cosmological
constant perturbatively with the help of the Kodama-Ishibashi-Seto formalism (see \cite{Kodama00,Kodama03,Kodama04,Ishibashi04}). The
solutions, called geons, could be described by four parameters which were the amplitude and three quantum numbers $(n,l,m)$ that
corresponds to the number of radial nodes $n$ and the spherical harmonic seed $Y_m^l$.

Recalling the scalar field knowledge, and in the same spirit of time-periodic solutions (section \ref{tpsol}), it is known that
single-mode excitations do not suffer from secular resonances. In the gravitational sector, the Poincar\'e-Lindstedt method was
also successful in removing all secular resonances appearing at third and fifth order in the case of an $(l,m,n) = (2,2,0)$ seed.
This suggested that fully non-linear $(2,2,0)$ geons could be constructed at arbitrary orders and could thus provide the first
island of stability that was not spherically symmetric.

In these configurations, the angular momentum provides a natural barrier against black hole formation, and the rotation is
described by a helical Killing vector. Geons thus play the role of the non-linear time-periodic functions of
\cite{Maliborski13b,Fodor15,Kim15} in the gravitational sector, i.e.\ fundamental non-linearly stable modes of vibrations of
AdS space-time. The fully non-linear numerical construction of gravitational $(l,m,n) = (2,2,0)$ geons was initiated in \cite{Horowitz15} in the
harmonic gauge with the help of the de-Turck method and spectral discretisation\footnote{See \cite{Dias16b} for an excellent
review on these two popular numerical methods applied to stationary solutions. See also chapter 6 of \cite{MartinonPhD} and
appendix A of \cite{Martinon17} for a connection between de-Turck and 3+1 gauges.}. These geons were also obtained in \cite{Martinon17} in a maximal slicing-spatial harmonic gauge. The two papers
\cite{Horowitz15,Martinon17} disagree to some extent, as illustrated in figure \ref{compare}. In particular, their numerical
solutions display significant differences in the determination of the rotation frequency $\omega$ (which is determined by the
quantum numbers, see equation \eref{omegas} below) and the angular momentum $J$.  However, the perturbative approach at $5^{th}$
order shows that the curve should be convex (see \cite{Martinon17,Fodor17} for detailed computations), like on the right panel of figure \ref{compare}.

\begin{myfig}
   \includegraphics[width = 0.98\textwidth]{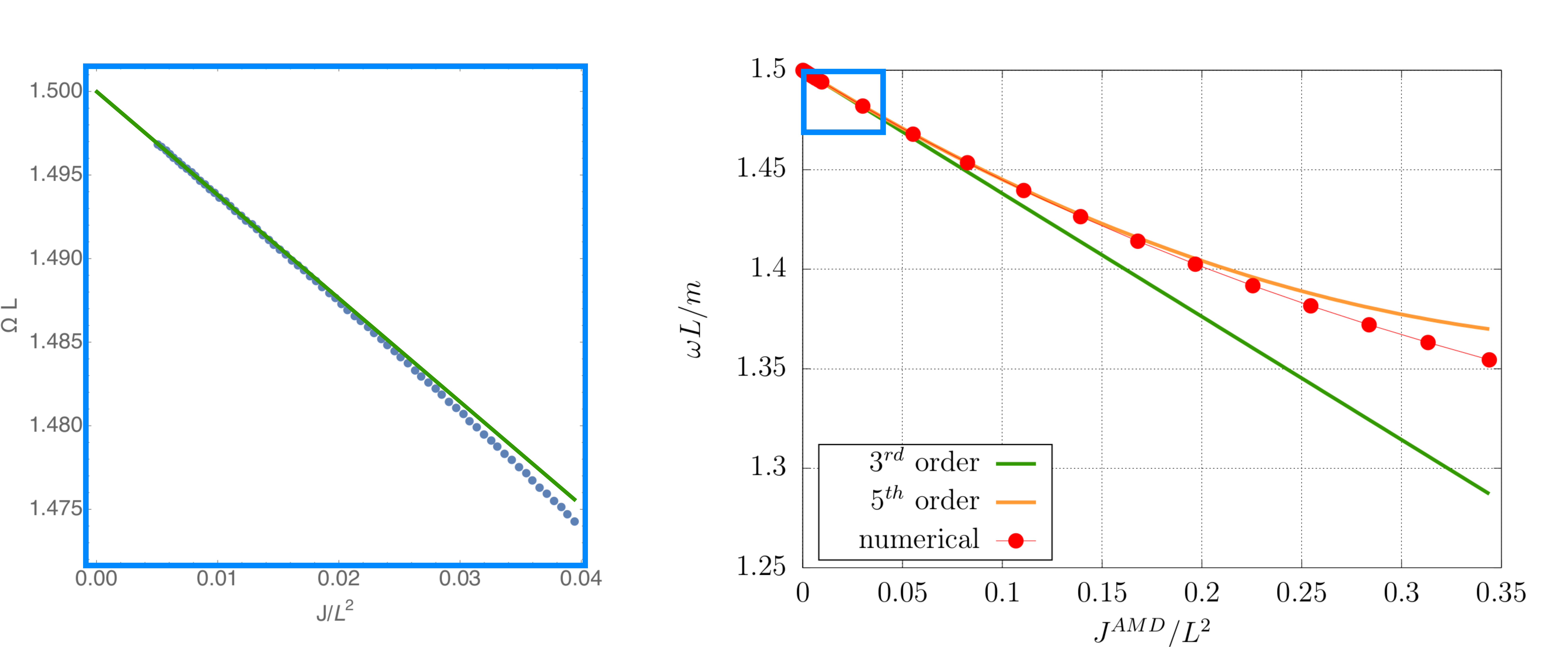}
   \caption{Left panel: $\omega-J$ plane taken from the numerical simulations of
      \cite{Horowitz15}. Blue points are numerical results while the solid green curve denotes $3^{rd}$ order results. Right panel:
      $\omega-J$ plane computed in \cite{Martinon17}. The range of amplitudes probed on the left panel is reported on the right
      panel with the help of a light blue frame at scale. Credits: \cite{Horowitz15,Martinon17}.}
   \label{compare}
\end{myfig}

If the existence of geons stands on firm arguments, their non-linear stability is still a partially unanswered question. One
expects that, akin to the time-periodic solutions of the scalar sector (section \ref{tpsol}), geons constitute non-linearly stable
attractors. In \cite{Dias12b}, some perturbative arguments in favour of their non-linear stability were given. The conclusion of
this work was that geons and boson stars (the latter being constructed for the first time in AdS in \cite{Astefanesei03}) were non-linearly stable
unless they were embedded in very high dimensional space-times or if perturbations had low differentiability, i.e.\ were far from
being $\mathcal{C}^\infty$. However, the proof was missing some theorems that were proven only in the analogous case of the
non-linear Schr\"odinger equation. So this result was half-way between a demonstration and a conjecture. The equivalent of two-mode
initial data for geons was also considered perturbatively, but this time irremovable secular resonances did emerge, like in the
scalar case.

Not much work has been done beyond the $(l,m,n) = (2,2,0)$ geon before \cite{Dias16a,Dias17a} (and more recently \cite{Fodor17}),
whose authors greatly extended the perturbative results of \cite{Dias12a}. They worked at fixed quantum numbers $(l,m,n)$ and
obtained the allowed linear frequencies
\numparts
\begin{eqnarray}
\label{omegas}%
   \omega_S &= l + 1 + 2n,\\
\label{omegav}%
   \omega_V &= l + 2 + 2n,
\end{eqnarray}
\endnumparts
$n$ being a positive integer, and the labels $S$ and $V$ corresponding to scalar-type or vector-type perturbations in the
Kodama-Ishibashi-Seto formalism. For these single-mode geon excitations, the authors found many configurations with irremovable
secular resonances, indicating that non-spherically symmetric systems could be even more unstable than spherically symmetric ones.
This claim was tempered though in \cite{Rostworowski16} whose author argued that these resonances were indeed removable if,
instead of a single mode, a linear combination of modes sharing the same frequency $\omega$ and azimuthal number $m$ were
considered. This amounted to work at fixed frequency and azimuthal number $(\omega,m)$ instead of fixed $(l,m,n)$, $\omega$ being
a degenerate function of $l$ and $n$ according to \eref{omegas}-\eref{omegav}. This claim was made stronger in
\cite{Rostworowski17a,Rostworowski17b} where the author found that, at the perturbative level, the number of possible geon
excitations with a given frequency $\omega$ was precisely equal to the multiplicity of the frequency, suggesting that all secular
resonances could be removed properly. An illustration of these arguments is given in figure \ref{geoncontroversy}. The fully
non-linear numerical study \cite{Martinon17} put an end to the debate by explicitly constructing non-linear \textit{excited} geon
solutions, thus confirming the arguments of both \cite{Dias16a,Dias17a} (non-existence of excited $(l,m,n) = (2,2,1)$ geons) and
\cite{Rostworowski16,Rostworowski17a,Rostworowski17b} (existence of three excited families $(\omega,m) = (5,2)$).

\begin{myfig}
   \includegraphics[width = 0.49\textwidth]{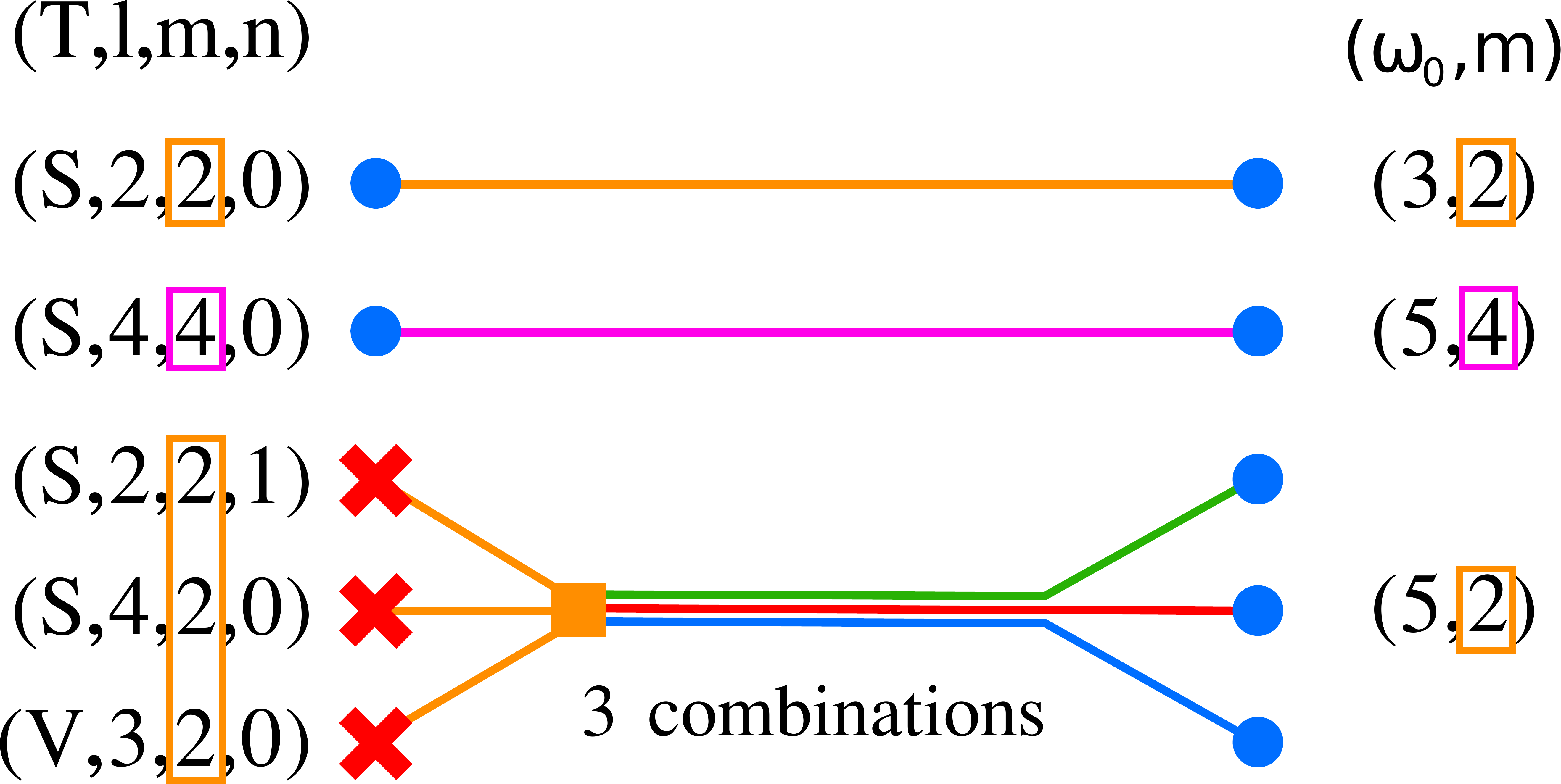}
   \caption{Two approaches for building geons perturbatively. $S$ and $V$ stand for scalar-type and vector-type perturbations respectively.
      The first point of view consists in looking for non-linear extensions of fixed $(l,m,n)$ single-mode geon seeds. This is
      represented by the left part of the picture. In this case, many single-modes appear to suffer from irremovable secular
      resonances (red crosses). The second point of view, on the right, considers non-linear extensions of fixed $(\omega,m)$ geon seeds,
      $\omega$ being a degenerated function of $l$ and $n$ (equations \eref{omegas}-\eref{omegav}). In this case, it turns out that all secular
      resonances can be removed (blue spheres). The non-linear geon extensions thus obtained are merely linear combinations of
      single-modes geons. The number of allowed combinations matches precisely the degeneracy of the frequency $\omega$ at fixed
      $m$.}
   \label{geoncontroversy}
\end{myfig}

These numerical solutions of geons are pictured in figure \ref{killinggallery}. Since they all feature a Killing vector
$\partial_{t'}$ (i.e.\ they are stationary in a co-rotating frame), the figure shows its norm, given by $g_{t't'}$. After
regularisation and subtraction of the AdS background, what is left is the geon contribution $\widetilde{h}_{t't'}$. The shape of
the isocontours increases in complexity for excited geons.

Without a doubt, many other families will be constructed in the future, with other symmetries than these helically
symmetric ones. A straightforward extension of this work would be to construct $m=0$ axisymmetric geons, that have no $\varphi$
dependence, but are periodic in time. Other families of geons could be constructed too, e.g.\ scalar-type modes with odd quantum
numbers $l$ and $m$. These geons, however, have less symmetries than the even-even ones, and are thus potentially more greedy in
terms of numerical resolution.

\begin{myfig}
   \includegraphics[width = 0.29\textwidth]{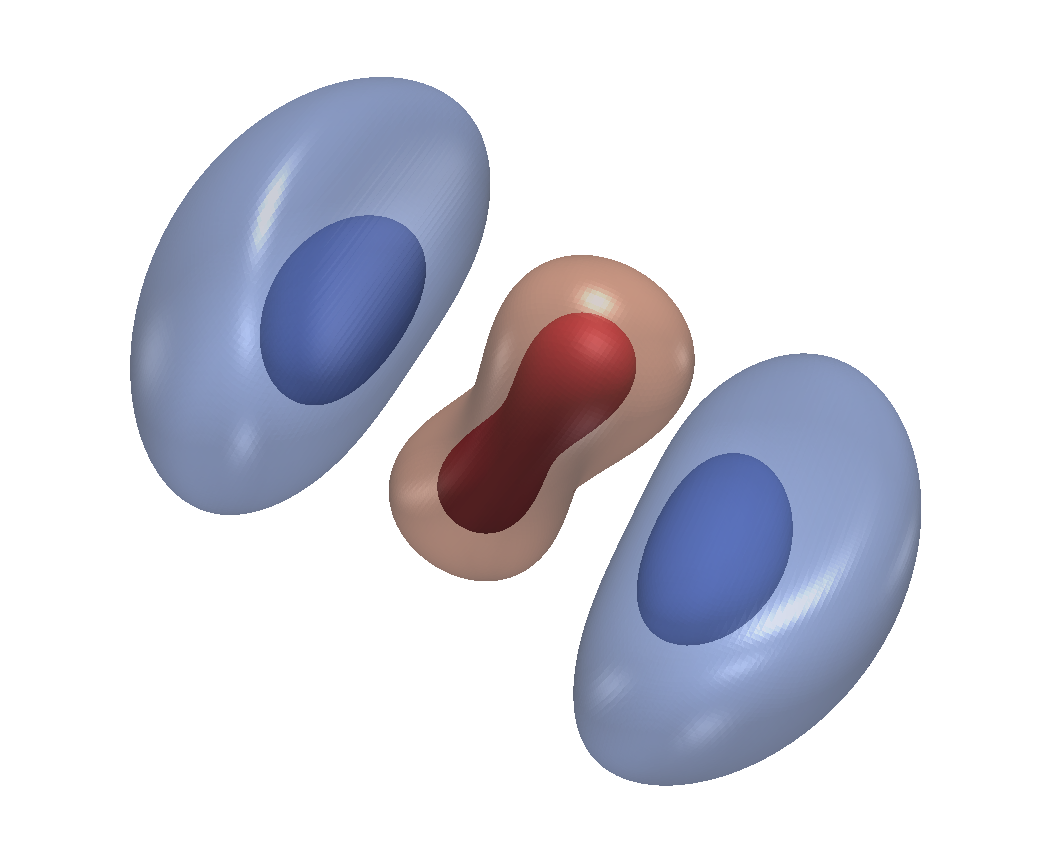}
   \includegraphics[width = 0.29\textwidth]{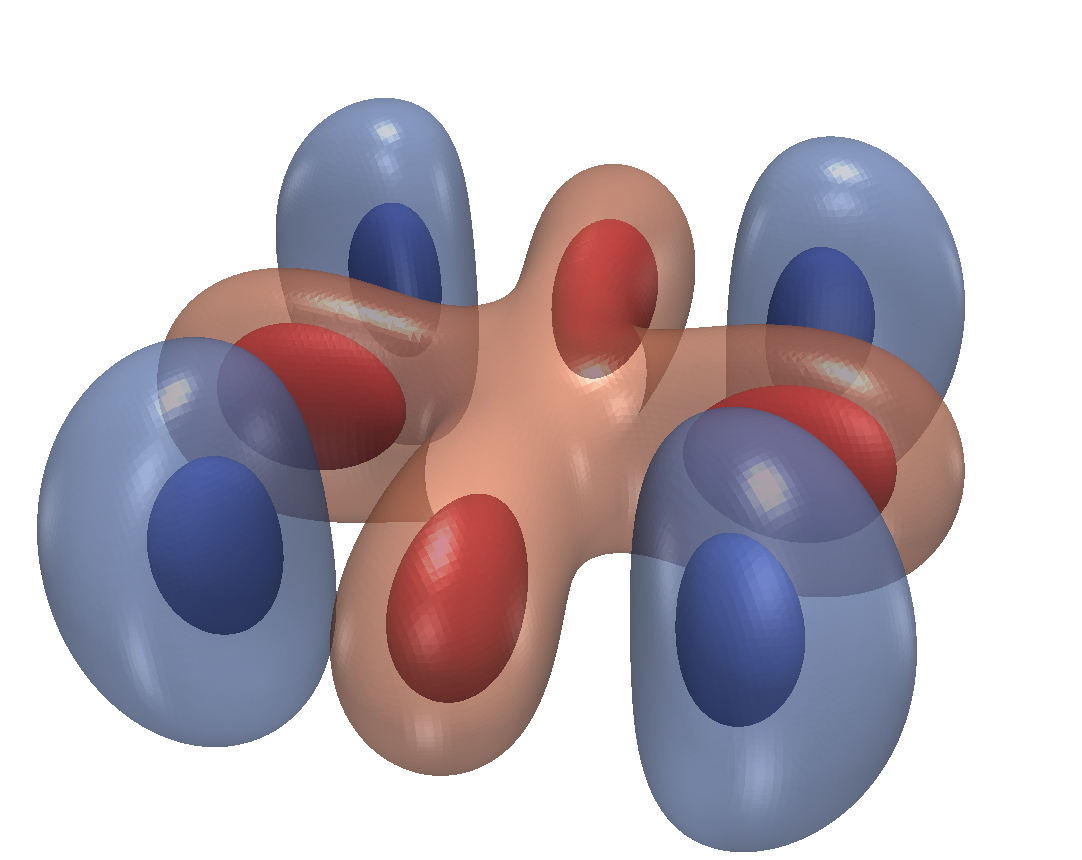}\\
   \includegraphics[width = 0.29\textwidth]{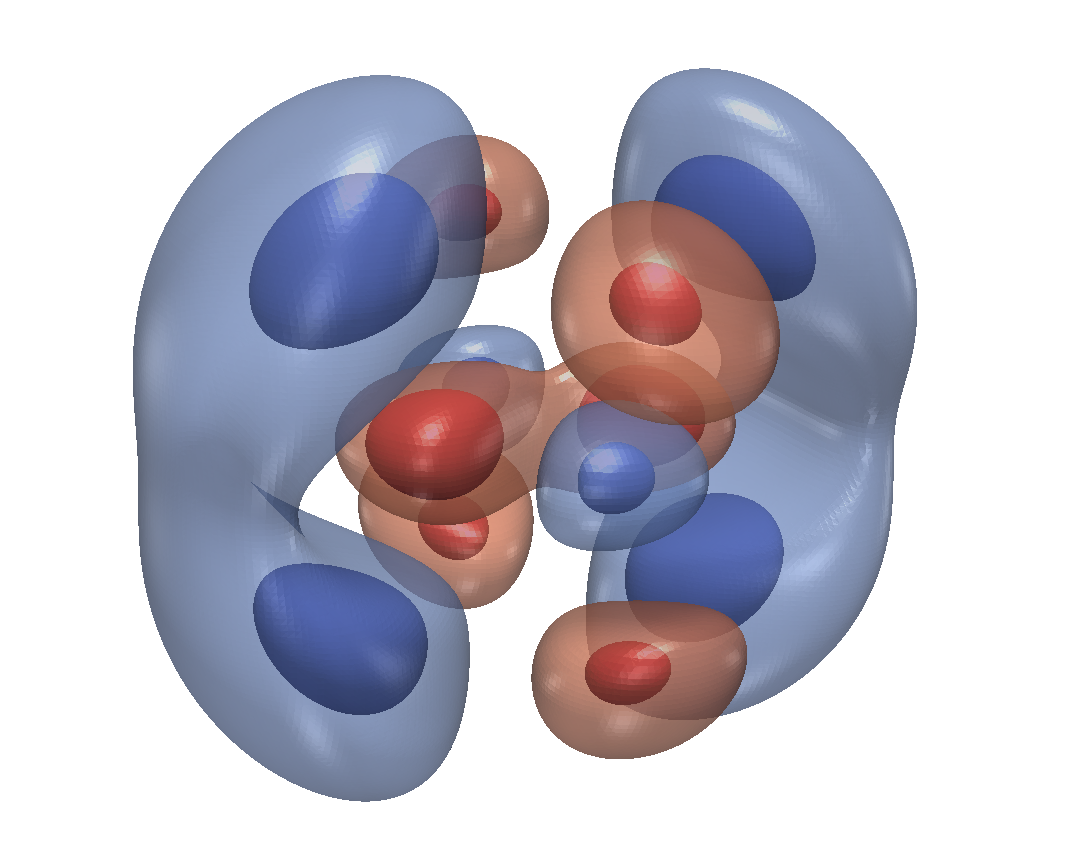}\\
   \includegraphics[width = 0.29\textwidth]{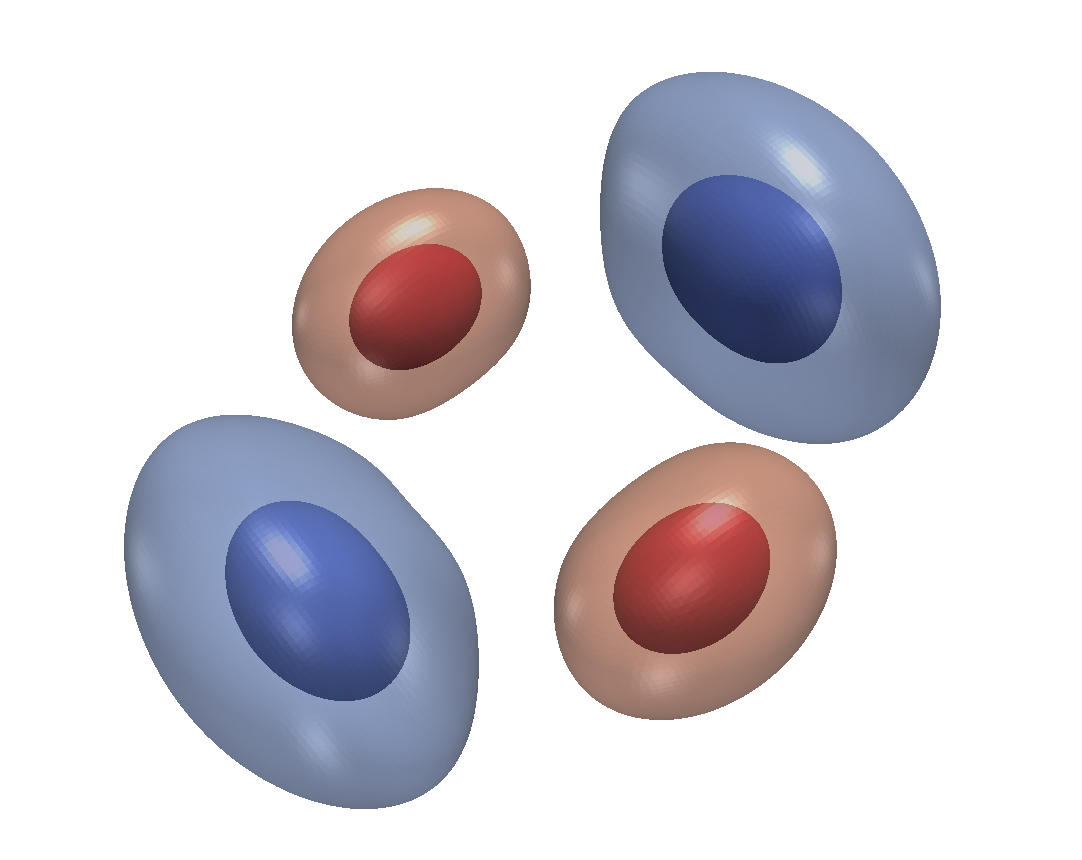}
   \includegraphics[width = 0.29\textwidth]{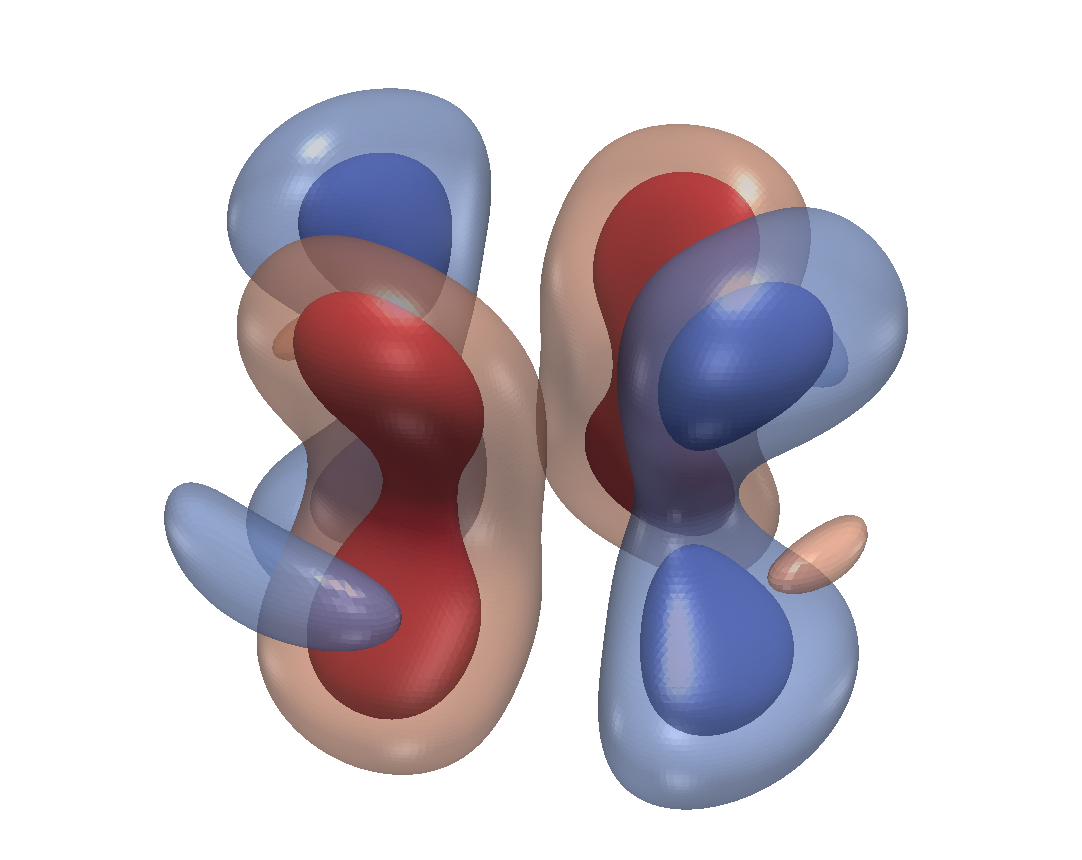}
   \caption{Isocontours of $\widetilde{h}_{t't'}$. They correspond to
   the regularised norm of the helical Killing vector $\partial_{t'}$ in the co-rotating frame of geons, after background
   subtraction. All five families of geons are presented, namely the scalar $(l,m,n)=(2,2,0),(4,4,0)$ modes and the three excited families
   I-III, in this order. Red isocontours denote positive contributions to the norm.}
   \label{killinggallery}
\end{myfig}

\subsection{Black holes surrounded by geons}
\label{bhgeon}

A question of fundamental interest is how geons are linked to black holes in AAdS space-times. Precisely, the black resonators of
\cite{Dias15} were configurations where a rotating black hole lied at the centre of an $l=m=2$ geon. However, they were shown to
be unstable against $m>2$ superradiant modes (see also\footnote{The main result of \cite{Green16} was that black holes with
ergoregions in AdS were linearly unstable for perturbations of rotation speeds above the Hawking-Reall bound of superradiance
\cite{Hawking00} (see \cite{Brito15a} for a review on superradiance in astrophysics). To make a long story short, superradiant
instability, or black hole bomb, happens when a rotating black hole absorbs a wave and radiates away another one with much higher
amplitude via a Penrose process. If a mirror is placed around the black hole so as to send this energetic wave back into the black
hole, the process can repeat indefinitely and back-react substantially on the metric even if the initial perturbation was small
\cite{Press72}. This mechanism is of course at play in AdS space-time because of the reflective boundary conditions (see e.g.\
\cite{Dias13,Cardoso14} for dedicated reviews).} \cite{Green16}). The reason is that black resonators have a geon as their
zero-horizon radius limit and then merge with Kerr (precisely at the onset of superradiance) when the gravitational hair
vanishes. Black resonators were candidates for a putative stationary endpoint of the superradiant instability, but the
authors \cite{Niehoff16} demonstrated that it was impossible, and that maybe the stationary endpoint of superradiance may not
exist at all. These peculiar black holes exhibit only one single Killing vector, and thus constitute the generalisation the scalar
hairy black holes obtained previously in \cite{Dias11}. These single Killing field black holes with scalar hair displayed a boson
star as a zero-horizon radius limit. They then merged with Myers-Perry black holes (see \cite{Emparan08} for a review) at the
onset of superradiance, i.e.\ when the scalar hair vanishes. 

\begin{myfig}
   \includegraphics[width = 0.49\textwidth]{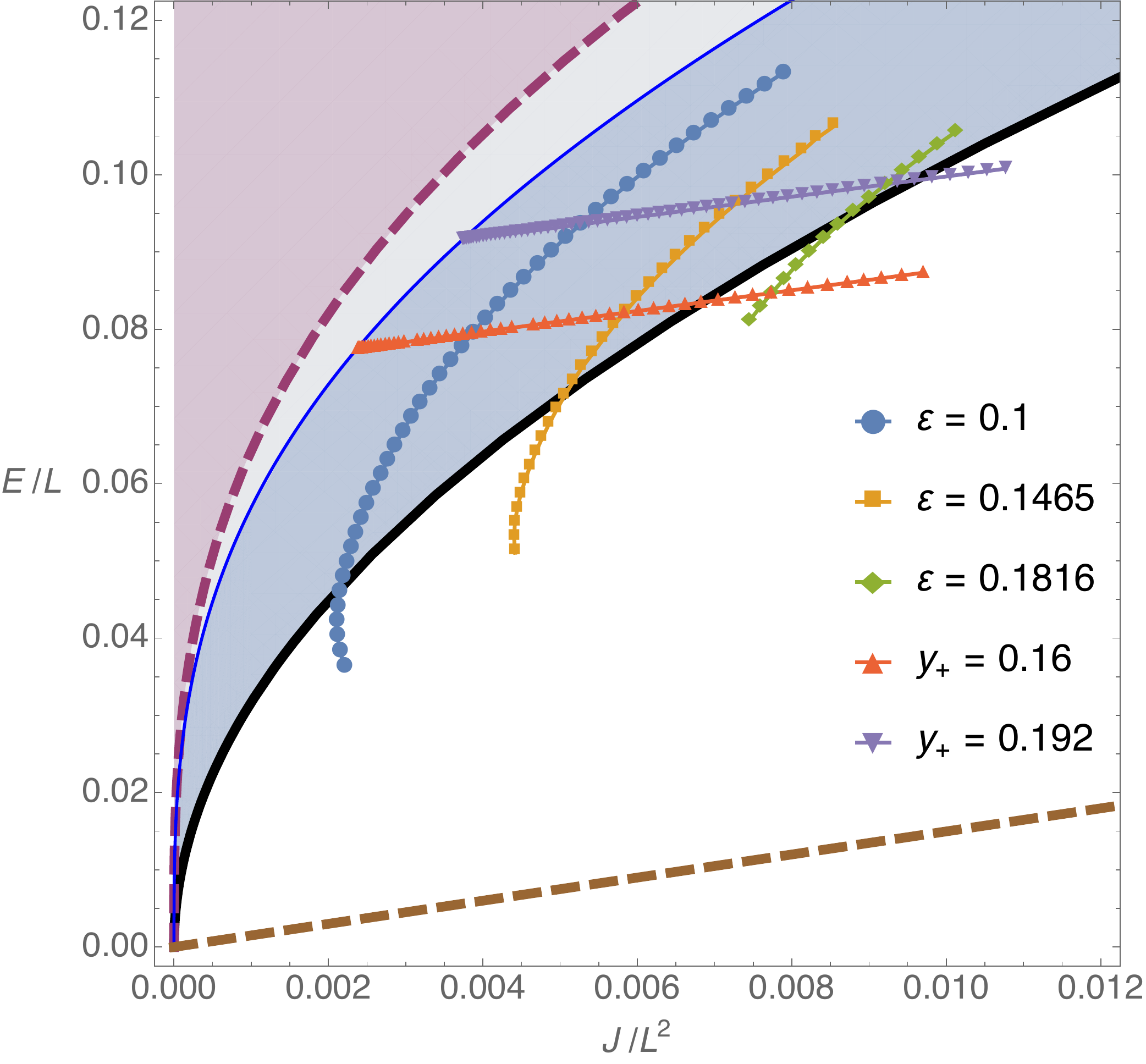}
   \caption{Mass-angular momentum diagram of black resonators. The thick black line depicts
      extremal Kerr-AdS such that non-extremal ones lie above. The dashed purple region delimits black holes with rotation
      frequency $\Omega_H L \leq 1$. The onset of superradiant instability for the $l = m = 2$ mode is denoted by the thin blue
      line. The bottom dashed line represent $l = m = 2$ geons. Data points represent black resonators obtained numerically. Credits: \cite{Dias15}.}
   \label{blackres}
\end{myfig}

As close counterparts of AAdS gravitational geons, let us mention the Einstein-Maxwell-AdS spinning
solitons obtained in \cite{Herdeiro16a,Herdeiro16c} and the AAdS Proca stars made of a massive spin-1 Abelian field of
\cite{Duarte16}. Proca stars were spherically symmetric and shown to be stable against radial perturbations. As for the Einstein-Maxwell
solitons, they could also be dressed with a black hole at their centre, in much the same way as the scalar hairy black holes. Since no symmetry was assumed, the obtained black holes could have a
multipolar structure, depicted in figure \ref{bhmultipole}, unlike in Minkowski space-time. The stability of these solutions was not
investigated though.

\begin{myfig}
   \includegraphics[width = 0.22\textwidth]{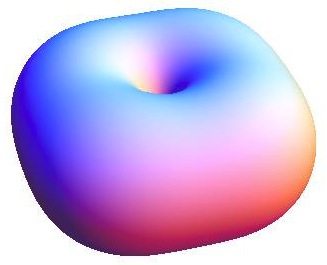}
   \includegraphics[width = 0.22\textwidth]{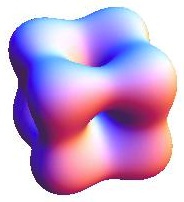}
   \includegraphics[width = 0.22\textwidth]{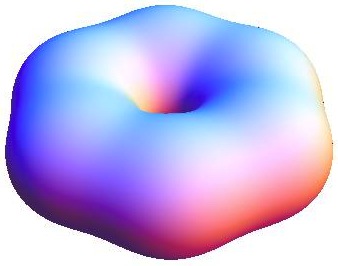}\\
   \includegraphics[width = 0.22\textwidth]{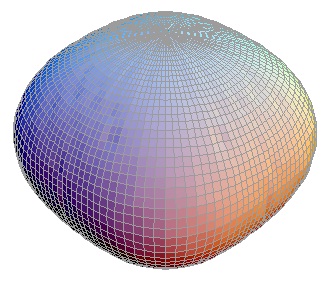}
   \includegraphics[width = 0.22\textwidth]{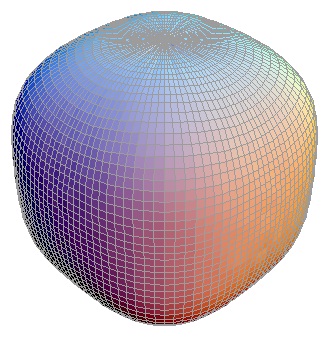}
   \includegraphics[width = 0.22\textwidth]{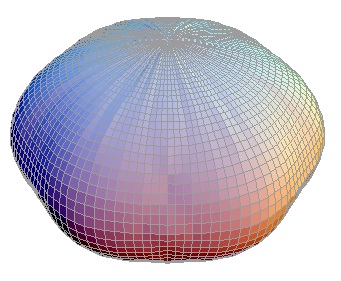}
   \caption{Top: surface of constant energy density for the
      Einstein-Maxwell-AdS solitons with regular electric multipoles. Bottom: isometrics embedding for the horizon of
      AdS-electro-vacuum black holes. The same multipole (from left to right) are used: $(l,m) = (2,2),(3,2),(3,3)$. The
      topology of horizon matches that of the energy isocontours of the surrounding soliton. Credits: \cite{Herdeiro16c}.}
   \label{bhmultipole}
\end{myfig}

As a final remark on black holes, it is important to mention that the authors of \cite{Holzegel13b} have studied analytically the
non-linear stability of Schwarzschild-AdS black holes with no spherical symmetry assumption on the perturbations and conjectured
that these black holes were non-linearly unstable. Thus, what was granted in spherical symmetry seems to break down when no
symmetry is involved.

\subsection{AdS instability beyond spherical symmetry}

Very recently, the first ever numerical evolution of a massless scalar field beyond spherical symmetry was investigated in
\cite{Bantilan17}. The authors chose to work in 5-dimensional AdS space-time with an $SO(3)$ symmetry, such that dynamics
occurred only in the $(t,x,y)$ directions (compared to $(t,r)$ for spherical symmetry). The numerical evolution was also restricted
to zero angular momentum initial data, which was parametrised by two amplitudes $A$ and $B$, $A$ being a spherically symmetric
component and $B$ a first harmonic excitation. The scalar field $\phi$ initial data then read
\begin{equation}
   \phi(\rho,\chi) = A f(\rho) + B g(\gamma) \cosh \chi,
\end{equation}
where $\rho$ was a radial coordinate, $\chi$ an angular coordinate, and functions $f$ and $g$ were piecewise $C^2$ functions.
The spherically symmetric case is recovered whenever $B$ is zero. The main results of this study are summarised in figure
\ref{nonspherical}. At fixed mass, non-spherically symmetric initial data undergo fewer bounces before black hole formation. At
fixed number of bounces, the non-spherically symmetric initial data need less mass to collapse. This lead the authors to
conjecture that the instability was even more prominent beyond spherical symmetry, as already suggested by \cite{Dias16a,Dias17a}.
These results are somewhat difficult to compare to the spherically symmetric knowledge since no low amplitude limit is probed
numerically. 

\begin{myfig}
   \includegraphics[width = 0.49\textwidth]{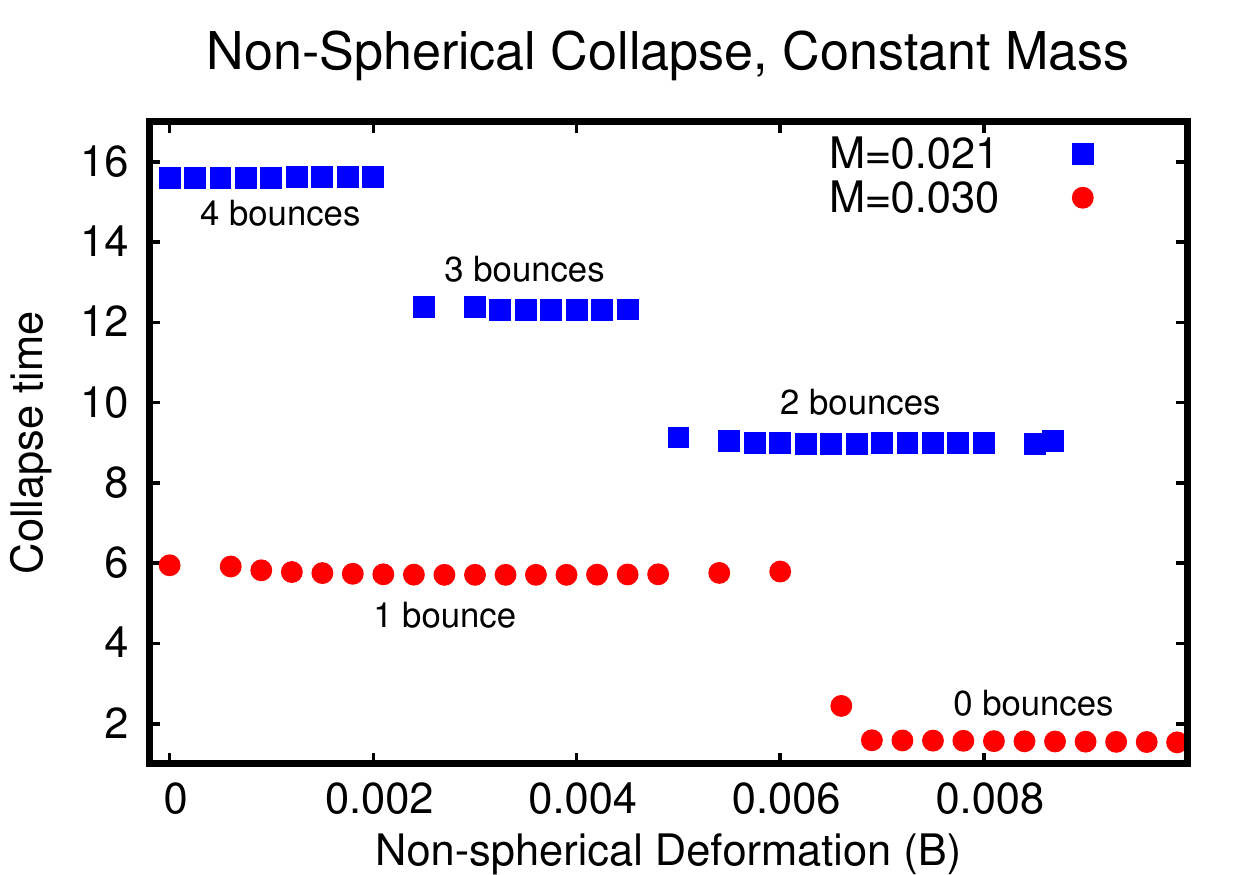}
   \includegraphics[width = 0.49\textwidth]{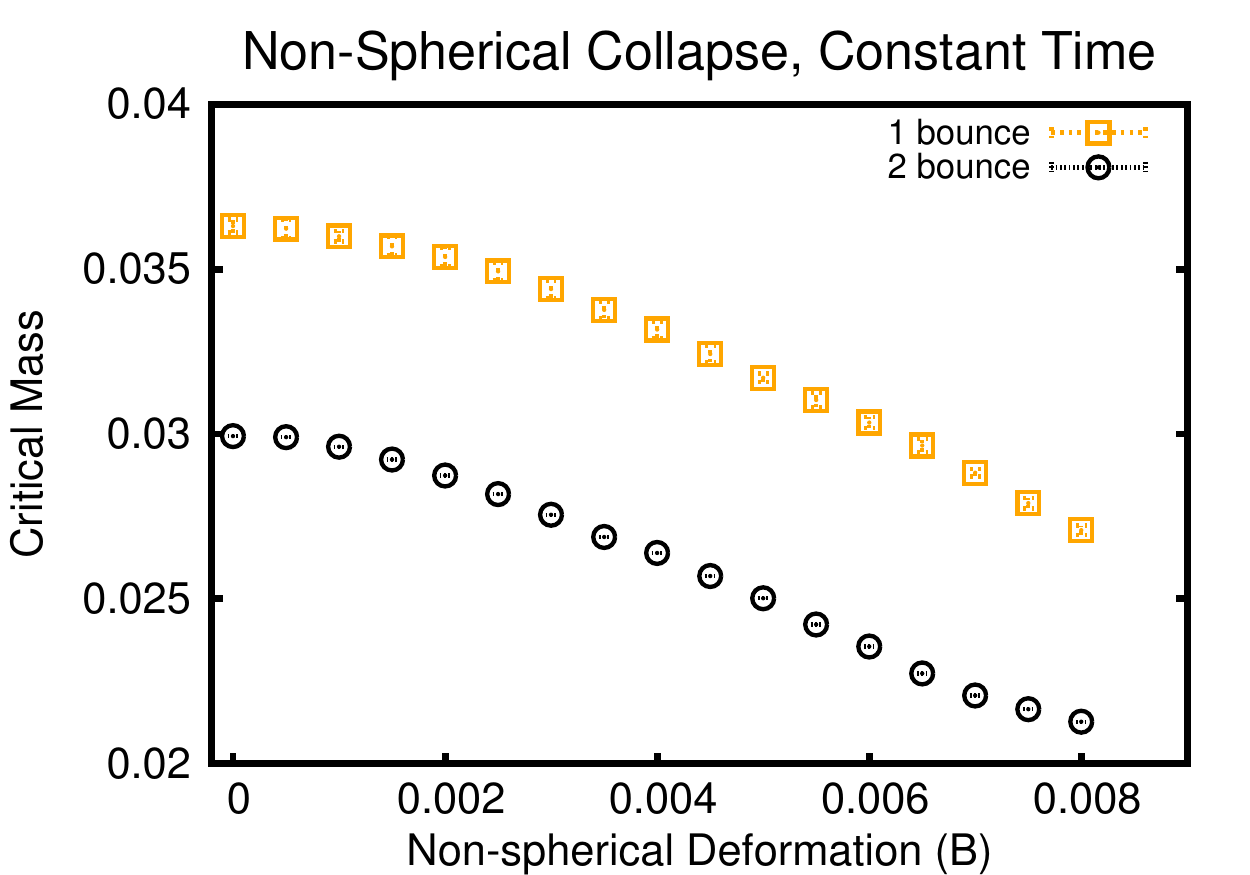}
   \caption{Left: Collapse time as a function of the non-spherically symmetric
      deformation of the initial scalar profile, for fixed total mass $M = 0.021$ (blue squares) and $M = 0.030$ (red circles).
      Right: Maximum mass for which a black hole is formed after $N$ bounces, $N$ being 1 (yellow squares) or 2 (black circles),
      as a function of the non-spherical deformation. Credits: \cite{Bantilan17}.}
   \label{nonspherical}
\end{myfig}

Almost simultaneously, the authors of \cite{Choptuik17} performed simulations of a massless scalar field with an azimuthal number $m=1$
ansatz. They compared two families of initial data, one with zero angular momentum $J$ and one with $J \simeq 0.155 E$, $E$ being
the total mass. They observed numerically that non-zero angular momentum simulations were still unstable at low amplitudes.
However, it did take more time to form a black hole compared to the zero angular momentum case. This is illustrated in figure
\ref{twoinit}. The end state of the stability could be a Myers-Perry black hole or the rotating hairy black holes of \cite{Dias11} discussed in section
\ref{bhgeon}.

\begin{myfig}
   \includegraphics[width = 0.49\textwidth]{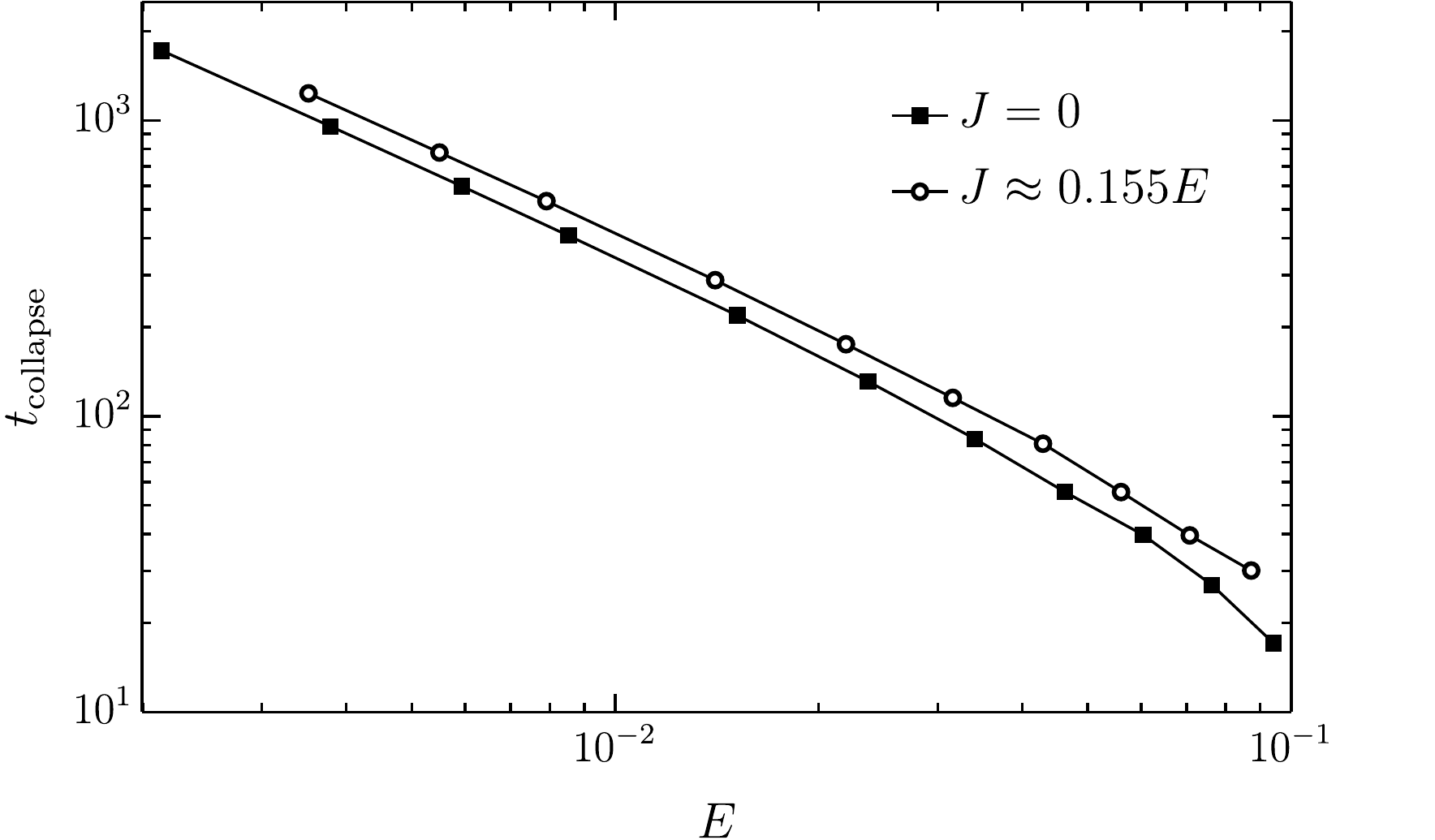}
   \caption{Time of collapse as a function of the initial energy content $E$ for two
      kinds of initial data: one with zero angular momentum $J=0$ and one with $J \simeq 0.155 E$. The latter simulations do not
      suppress the instability but delay it. Credits: \cite{Choptuik17}.}
   \label{twoinit}
\end{myfig}

Thus, \cite{Bantilan17} argues that no spherical symmetry speed up the instability, while \cite{Choptuik17} advocates that angular
momentum delays black hole formation. This suggests that the structure of the instability beyond spherical symmetry is quite
involved, and not very well-known for the moment. These results are very promising for the future. For the time being, the qualitative understanding of the AdS instability is not
challenged by angular momentum or asymmetric considerations, but it is much too soon to consider it a definitive statement. The
freedom of initial data being much larger in higher dynamical dimensions, and much more computationally expansive, we expect that
progress in this area of research, however exciting, will be much slower than in the spherically symmetric case (which literally boomed in
about a few years).

\section{The CFT interpretation}
\label{cftinterp}

Of course, one of the main motivations to study the AdS instability is the AdS-CFT correspondence. What does the
instability means on the CFT side? And what are the quantum observables that may be impacted by the instability?

\subsection{Dual thermodynamics}

From a thermodynamical point of view, black hole formation is usually understood as thermalisation of the dual system in the
CFT side, and the weakly turbulent cascade came as no surprise for the AdS-CFT community who expected any
thermodynamical system to thermalise \cite{Hubeny15}. However, several authors cautioned that this was not always true
\cite{Abajo14,Silva15,Dimitrakopoulos15a}. Indeed, black holes obtained in numerical simulations are always classical. But if one
considers the Hawking radiation \cite{Bekenstein73,Hawking75}, these black holes formed after several bounces could be themselves
thermodynamically unstable and may evaporate, partially or totally, according the Hawking-Page phase transition \cite{Hawking82}.
The thermalised state could thus be described by a smaller black hole in equilibrium with its Hawking radiation or by a thermal
gas in AdS space-time. This means that the final classical black holes observed in the simulations, if too small, are not
thermalised states, but just pre-thermalisation steps in the dual theory, that only achieve equilibrium on a longer timescale.
This is strikingly reminiscent of revivals and equilibrations that were observed experimentally in some isolated Bose-Einstein
condensates \cite{Gring12,Trotzky12,Kinoshita06}. Others discussions of CFT duals in the context of the AdS instability can be
found in \cite{Garfinkle11,Dias12a,Garfinkle12,Balasubramanian14}.

As first noticed in \cite{Dias12a,Gentle12,Buchel13}, the existence of non-linearly stable solutions is even more surprising, since they are dual to systems that never
thermalise in the CFT side, and do not exhibit black hole formation even if their mass is above the
Hawking-Page limit. A potential heuristic explanation is that such systems may form a meta-stable state that can survive by
continuously exchanging energy between its constitutive parts, as observed in
\cite{Pretorius00,Abajo14,Silva15,Deppe15a,Deppe15b,Deppe16a,Okawa15,Brito16b}, which is translated in a gravitational balance
between focusing and defocusing effects in the gravitational AdS dual (discussed in section \ref{competition}).

\subsection{Dual systems}

Several CFT interpretations are available in the literature. For example, the authors of \cite{Abajo14,Silva15} focused on
the CFT duals of direct and delayed (i.e.\ with bounces) collapses in AdS space-time. They computed the entanglement entropy of the
dual system and observed that this quantity was oscillating in the latter case. The authors also suggested that the
maximum of matter energy distribution in AdS was dual to the density of strongly correlated excitations in the field
theory.

In the hard wall model of \cite{Craps14c,Silva16}, the authors observed that non-collapsing solutions induced a modulated
oscillation of the boundary operator $\langle \mathcal{O} \rangle$, and the time scale of the modulation was precisely
$O(\varepsilon^{-2})$. The oscillations were interpreted in terms of conversions of a collection of glueballs into another
collection (and back).

Finally, from an AdS-CFT point of view and since a linearised graviton can be interpreted as a spin-2 excitation, gravitational
geons are equivalent to a Bose-Einstein condensate of spin-2 field excitations \cite{Dias12a}, namely glueballs excitations. This
is illustrated in figure \ref{dual}.

\begin{myfig}
   \includegraphics[width = 0.49\textwidth]{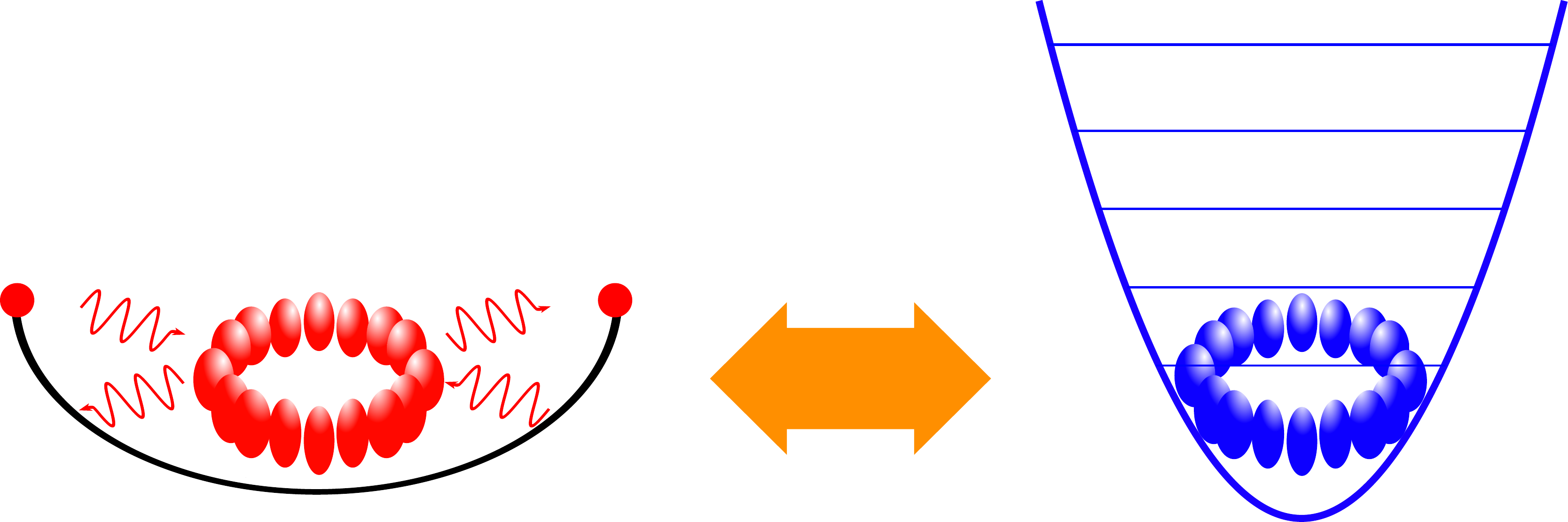}
   \caption{Left: diagrammatic representation of a gravitational geon in an AAdS space-time. Right:
   dual system living on the boundary. It is a Bose-Einstein condensate of a spin-2 field, namely glueball excitations.}
   \label{dual}
\end{myfig}

Because of the holographic principle, the dual system lives always in a space-time that is one dimension less than the bulk
space-time. For example, a 4-dimensional geon is dual to a 3-dimensional Bose-Einstein condensate of glueball excitations. The
latter, however, can well be interpreted as a 4-dimensional system with one symmetry (or invariance), so much as to reduce the
\textit{effective} dimensionality to three.

\section{Conclusion}

Since the original work of Bizo\'n and Rostworowski in 2011 \cite{Bizon11}, much progress has been made to understand the
AdS instability. In table \ref{adsrecap}, we have tried to recap all the different aspects of the problem. The statements
are mainly a short synopsis of the previous sections of this review.

\begin{mystab}
   \vspace*{17cm}
\begin{small}
\begin{tabular}{llllcc}
\hline
field               & initial data            & statement                                                                             & year & original reference           \\
\hline
gravitons           & -                       & linear stability of AdS                                                               & 1981 & \cite{Abbott82}           \\
gravitons           & -                       & AdS ``rigidity'' theorem                                                              & 2006 & \cite{Anderson06}         \\
massless scalar     & Gaussian                & non-linear instability of a Gaussian wave packet                                      & 2011 & \cite{Bizon11}            \\
massless scalar     & single-mode             & non-linear stability of a single-mode excitation                                      & 2011 & \cite{Bizon11}            \\
massless scalar     & two-mode                & non-linear instability of a two-mode excitation with turbulent cascade                & 2011 & \cite{Bizon11}            \\
massless scalar     & -                       & secular resonances on timescale $O(\varepsilon^{-2})$                                 & 2011 & \cite{Bizon11}            \\
massless scalar     & Gaussian                & instability due to the resonant spectrum in all dimensions                            & 2011 & \cite{Jalmuzna11}         \\
massless scalar     & Gaussian                & instability recovered in flat space-time enclosed in a cavity                         & 2011 & \cite{Maliborski12}       \\
massless scalar     & Gaussian                & critical phenomena in 3-dimensional AdS space-time                                    & 2000 & \cite{Pretorius00}        \\
massless scalar     & Gaussian                & critical phenomena in 4-dimensional AdS space-time                                    & 2003 & \cite{Husain03}           \\
massless scalar     & Gaussian                & critical phenomena for an arbitrary number of bounces                                 & 2016 & \cite{Olivan16a}          \\
perfect fluid       & -                       & singularity theorems in AdS space-time                                                & 2012 & \cite{Ishibashi12}        \\
null dust           & -                       & demonstration of the instability conjecture for radial Einstein-Vlasov                & 2017 & \cite{Moschidis17a}       \\
massless scalar     & -                       & first time-periodic solution                                                          & 2010 & \cite{Basu10}             \\
massless scalar     & time-periodic           & first numerical evolution of an island of stability                                   & 2013 & \cite{Maliborski13b}      \\
massless scalar     & Gaussian                & non-linear stability of a Gaussian profile of width $0.4 \lesssim \sigma \lesssim 8$  & 2013 & \cite{Buchel13}           \\
massless scalar     & exponential spectrum    & TTF equations can probe the instability conjecture                                    & 2014 & \cite{Balasubramanian14}  \\
massless scalar     & oscillatory spectrum    & TTF can help generating non-linearly stable solutions                                 & 2015 & \cite{Green15}            \\
perfect fluid       & thin shells             & fractal signature of collapse and chaotic islands of stability                        & 2016 & \cite{Brito16b}           \\
massless scalar     & Gaussian injection      & dichotomy of (in)stability in the hard wall model                                     & 2014 & \cite{Craps14c}           \\
massive scalar      & -                       & Boson stars in AdS space-time                                                         & 2003 & \cite{Astefanesei03}      \\
gravitons           & spherical harmonic seed & perturbative construction of gravitational geons in AdS space-time                    & 2012 & \cite{Dias12a}            \\
gravitons           & spherical harmonic seed & numerical construction of gravitational geons in AdS space-time                       & 2015 & \cite{Horowitz15}         \\
photons             & spherical harmonic seed & Einstein-Maxwell spinning solitons in AdS space-time                                  & 2012 & \cite{Herdeiro16a}        \\
massive Proca       & -                       & Proca stars in AdS space-time                                                         & 2016 & \cite{Duarte16}           \\
-                   & -                       & AdS instability obeys the ``mixed stability'' picture                                 & 2015 & \cite{Dimitrakopoulos15b} \\
massless scalar     & -                       & competition between focusing and defocusing effects                                   & 2015 & \cite{Dimitrakopoulos15a} \\
massless scalar     & phase coherent          & phase coherence with power-law spectrum as a sufficient condition for collapse        & 2016 & \cite{Freivogel16}        \\
massive scalar      & Gaussian                & a resonant spectrum is a necessary condition for the instability                      & 2012 & \cite{Dias12b}            \\
massless scalar     & Gaussian                & instability in 3-dimensional AdS means blow-up of the Sobolev norm                    & 2013 & \cite{Bizon13}            \\
massless scalar     & Gaussian                & instability less systematic in GB gravity                                             & 2015 & \cite{Deppe15a}           \\
massless scalar     & Gaussian shell          & AdS instability as a pre-thermalisation process with Hawking-Page phase transition    & 2014 & \cite{Abajo14}            \\
massless scalar     & non-spherical           & first numerical evolution of the instability beyond spherical symmetry                & 2017 & \cite{Bantilan17,Choptuik17} \\
\hline
\end{tabular}
\end{small}
\caption{Summary of the recent progress in the field of AdS instability.}
\label{adsrecap}
\end{mystab}

Why and how black holes form in AdS space-time appears as troublesome questions, much more difficult to answer than in Minkowski
or de Sitter space-times. This make AdS a very interesting tool to probe the very deep nature of gravitational dynamics, and
motivates the use of ever improved analytical and numerical tools. The non-linear stability of AdS, that was almost ignored for a
decade after the advent of the AdS-CFT correspondence, has now become a very active field of research. 

Given the intricate and chaotic distribution of islands of stability, it seems very challenging to provide a mathematical
demonstration of the instability conjecture, even if some progress has been made recently \cite{Moschidis17a,Moschidis17b}.
Moreover, from a pure numerical point of view, it is wise to recall that numerical simulations are limited in two ways: they
cannot run for arbitrarily long times, and they cannot probe arbitrarily small amplitudes. Even if these difficulties can somewhat
be circumvented by some theorems in some particular cases (section \ref{topoinsta}), these caveats should always be kept in mind
in order to temper some conclusions.

The vast majority of the literature focused on spherical symmetry, very often with a massless scalar field and Gaussian initial
data. The few breakthroughs beyond spherical symmetry were carried out in the case of gravitational geons (section
\ref{beyondspher}) and more recently in time evolutions \cite{Bantilan17,Choptuik17}. The qualitative understanding of the
instability is, for the time being, not really challenged by these additional degrees of freedom, but this is only the beginning.

One can get dizzy when thinking of all the possibilities brought by additional dynamical dimensions, the parameter space being
huge compared to spherically symmetric cases. Our intuition is that a good starting point is given by geons, since they form the
simplest and fundamental eigenmodes of AdS beyond spherical symmetry, and are stripped from any matter field. In the same spirit
of \cite{Green15}, it could be interesting to evolve in time geon initial data and to measure how much can they be perturbed
before jumping to unstable configurations.

Beyond black hole formation, the concept of islands of stability can take on various aspects. One of interest, discussed in
\cite{Dias12b}, is called non-coalescing black hole binaries. In these configurations, two black holes orbit each other in AdS
space-time. Because of the reflective boundary conditions, there is no decay of mass or angular momentum like in astrophysical
binaries. Instead, we could imagine that the binary is in equilibrium with its own radiation, giving an exactly periodic solution.
Even if two singularities are at play, non-coalescing binaries could well deserve the denomination island
of stability, within an ocean of merging configurations. In this regard, the dichotomy would not lie in the presence/absence of a
singularity, but in the presence/absence of merger. This example illustrates that the concept of stability in AdS space-time could
well be enlarged, giving rise, without a doubt, to a profusion of new physical properties yet to be discovered.

\ack

We would like to thank Vitor Cardoso, Andrzej Rostworowski, Christos Charmousis, Lo\"ic Villain, Nathalie Deruelle, Gyula Fodor,
Peter Forg\'acs, Philippe Grandcl\'ement, Oleg Evnin, Piotr Bizo\'n, Oscar J. C. Dias and Jorge E. Santos for useful and interesting discussions.

\newpage

\section*{References}

\bibliographystyle{unsrt}
\bibliography{biblio}

\begin{thebibliography}{100}

\bibitem{Christodoulou93}
D.~{Christodoulou} and S.~{Klainerman}.
\newblock {\em {The global nonlinear stability of the Minkowski space}}.
\newblock 1993.

\bibitem{Friedrich86}
H.~{Friedrich}.
\newblock {On the existence of n-geodesically complete or future complete
  solutions of Einstein's field equations with smooth asymptotic structure}.
\newblock {\em Communications in Mathematical Physics}, 107:587--609, December
  1986.

\bibitem{Abbott82}
L.~F. {Abbott} and S.~{Deser}.
\newblock {Stability of gravity with a cosmological constant}.
\newblock {\em Nuclear Physics B}, 195:76--96, February 1982.

\bibitem{Henneaux85}
M.~{Henneaux} and C.~{Teitelboim}.
\newblock {Asymptotically anti-de Sitter spaces}.
\newblock {\em Communications in Mathematical Physics}, 98:391--424, September
  1985.

\bibitem{Ashtekar84}
A.~{Ashtekar} and A.~{Magnon}.
\newblock {Asymptotically anti-de Sitter space-times}.
\newblock {\em Classical and Quantum Gravity}, 1:L39--L44, July 1984.

\bibitem{Ashtekar00}
A.~{Ashtekar} and S.~{Das}.
\newblock {Asymptotically anti-de Sitter spacetimes: conserved quantities}.
\newblock {\em Classical and Quantum Gravity}, 17:L17--L30, January 2000.

\bibitem{Wald00}
R.~M. {Wald} and A.~{Zoupas}.
\newblock {General definition of ``conserved quantities'' in general relativity
  and other theories of gravity}.
\newblock {\em \prd}, 61(8):084027, April 2000.

\bibitem{Papadimitriou05}
I.~{Papadimitriou} and K.~{Skenderis}.
\newblock {Thermodynamics of asymptotically locally AdS spacetimes}.
\newblock {\em Journal of High Energy Physics}, 8:004, August 2005.

\bibitem{Hollands05a}
S.~{Hollands}, A.~{Ishibashi}, and D.~{Marolf}.
\newblock {Comparison between various notions of conserved charges in
  asymptotically AdS spacetimes}.
\newblock {\em Classical and Quantum Gravity}, 22:2881--2920, July 2005.

\bibitem{Ashtekar14}
A.~{Ashtekar} and V.~{Petkov}.
\newblock {\em {Springer Handbook of Spacetime}}.
\newblock 2014.

\bibitem{Avis78}
S.~J. {Avis}, C.~J. {Isham}, and D.~{Storey}.
\newblock {Quantum field theory in anti-de Sitter space-time}.
\newblock {\em \prd}, 18:3565--3576, November 1978.

\bibitem{Breitenlohner82}
P.~{Breitenlohner} and D.~Z. {Freedman}.
\newblock {Stability in gauged extended supergravity.}
\newblock {\em Annals of Physics}, 144:249, 1982.

\bibitem{Maldacena98}
J.~M. {Maldacena}.
\newblock {The Large N Limit of Superconformal Field Theories and
  Supergravity}.
\newblock {\em Advances in Theoretical and Mathematical Physics}, 2:231, 1998.

\bibitem{Maldacena99}
Juan Maldacena.
\newblock The large-n limit of superconformal field theories and supergravity.
\newblock {\em International Journal of Theoretical Physics}, 38(4):1113--1133,
  1999.

\bibitem{Witten98}
E.~{Witten}.
\newblock {Anti-de Sitter space and holography}.
\newblock {\em Advances in Theoretical and Mathematical Physics}, 2:253--291,
  1998.

\bibitem{Aharony00}
O.~{Aharony}, S.~S. {Gubser}, J.~{Maldacena}, H.~{Ooguri}, and Y.~{Oz}.
\newblock {Large N field theories, string theory and gravity}.
\newblock {\em \physrep}, 323:183--386, January 2000.

\bibitem{Hubeny15}
V.~E. {Hubeny}.
\newblock {The AdS/CFT correspondence}.
\newblock {\em Classical and Quantum Gravity}, 32(12):124010, June 2015.

\bibitem{Dafermos06}
M.~Dafermos and G.~{Holzegel}.
\newblock {Dynamic instability of solitons in 4+1-dimensional gravity with
  negative cosmological constant}.
\newblock \url{https://dpmms.cam.ac.uk/~md384/ADSinstability.pdf}, 2006.
\newblock Accessed: 2017-03-01.

\bibitem{Anderson06}
M.~T. {Anderson}.
\newblock {On the uniqueness and global dynamics of AdS spacetimes}.
\newblock {\em Classical and Quantum Gravity}, 23:6935--6953, December 2006.

\bibitem{Bizon11}
P.~{Bizo{\'n}} and A.~{Rostworowski}.
\newblock {Weakly Turbulent Instability of Anti-de Sitter Spacetime}.
\newblock {\em Physical Review Letters}, 107(3):031102, July 2011.

\bibitem{Holzegel12}
G.~{Holzegel} and J.~{Smulevici}.
\newblock {Self-Gravitating Klein-Gordon Fields in Asymptotically
  Anti-de-Sitter Spacetimes}.
\newblock {\em Annales Henri Poincar{\'e}}, 13:991--1038, May 2012.

\bibitem{Holzegel13a}
G.~{Holzegel} and J.~{Smulevici}.
\newblock {Stability of Schwarzschild-AdS for the Spherically Symmetric
  Einstein-Klein-Gordon System}.
\newblock {\em Communications in Mathematical Physics}, 317:205--251, January
  2013.

\bibitem{Maliborski13c}
M.~{Maliborski} and A.~{Rostworowski}.
\newblock {Turbulent Instability of Anti-De Sitter Space-Time}.
\newblock {\em International Journal of Modern Physics A}, 28:1340020,
  September 2013.

\bibitem{Abraham93}
A.~M. {Abrahams} and C.~R. {Evans}.
\newblock {Critical behavior and scaling in vacuum axisymmetric gravitational
  collapse}.
\newblock {\em Physical Review Letters}, 70:2980--2983, May 1993.

\bibitem{Choptuik93}
M.~W. {Choptuik}.
\newblock {Universality and scaling in gravitational collapse of a massless
  scalar field}.
\newblock {\em Physical Review Letters}, 70:9--12, January 1993.

\bibitem{Garfinkle12}
D.~{Garfinkle}, L.~A. {Pando Zayas}, and D.~{Reichmann}.
\newblock {On field theory thermalization from gravitational collapse}.
\newblock {\em Journal of High Energy Physics}, 2:119, February 2012.

\bibitem{Jalmuzna11}
J.~{Ja{\l}mu{\.z}na}, A.~{Rostworowski}, and P.~{Bizo{\'n}}.
\newblock {AdS collapse of a scalar field in higher dimensions}.
\newblock {\em \prd}, 84(8):085021, October 2011.

\bibitem{Bizon14}
P.~{Bizo{\'n}}.
\newblock {Is AdS stable?}
\newblock {\em General Relativity and Gravitation}, 46:1724, May 2014.

\bibitem{Friedrich14}
H.~{Friedrich}.
\newblock {On the AdS stability problem}.
\newblock {\em Classical and Quantum Gravity}, 31(10):105001, May 2014.

\bibitem{Greenwood10}
E.~{Greenwood}, D.~C. {Dai}, and D.~{Stojkovic}.
\newblock {Time-dependent fluctuations and particle production in cosmological
  de Sitter and anti-de Sitter spaces}.
\newblock {\em Physics Letters B}, 692:226--231, September 2010.

\bibitem{Garfinkle11}
D.~{Garfinkle} and L.~A. {Pando Zayas}.
\newblock {Rapid thermalization in field theory from gravitational collapse}.
\newblock {\em \prd}, 84(6):066006, September 2011.

\bibitem{Buchel12}
A.~{Buchel}, L.~{Lehner}, and S.~L. {Liebling}.
\newblock {Scalar collapse in AdS spacetimes}.
\newblock {\em \prd}, 86(12):123011, December 2012.

\bibitem{Liebling13}
S.~L. {Liebling}.
\newblock {Nonlinear collapse in the semilinear wave equation in AdS space}.
\newblock {\em \prd}, 87(8):081501, April 2013.

\bibitem{Abajo14}
J.~{Abajo-Arrastia}, E.~{da Silva}, E.~{Lopez}, J.~{Mas}, and A.~{Serantes}.
\newblock {Holographic relaxation of finite size isolated quantum systems}.
\newblock {\em Journal of High Energy Physics}, 5:126, May 2014.

\bibitem{Silva15}
E.~{da Silva}, E.~{Lopez}, J.~{Mas}, and A.~{Serantes}.
\newblock {Collapse and revival in holographic quenches}.
\newblock {\em Journal of High Energy Physics}, 4:38, April 2015.

\bibitem{Maliborski12}
M.~{Maliborski}.
\newblock {Instability of Flat Space Enclosed in a Cavity}.
\newblock {\em Physical Review Letters}, 109(22):221101, November 2012.

\bibitem{Oliveira13}
H.~P. {de Oliveira}, L.~A. {Pando Zayas}, and E.~L. {Rodrigues}.
\newblock {Kolmogorov-Zakharov Spectrum in AdS Gravitational Collapse}.
\newblock {\em Physical Review Letters}, 111(5):051101, August 2013.

\bibitem{Pretorius00}
F.~{Pretorius} and M.~W. {Choptuik}.
\newblock {Gravitational collapse in 2+1 dimensional AdS spacetime}.
\newblock {\em \prd}, 62(12):124012, December 2000.

\bibitem{Husain03}
V.~{Husain}, G.~{Kunstatter}, B.~{Preston}, and M.~{Birukou}.
\newblock {Anti-de Sitter gravitational collapse}.
\newblock {\em Classical and Quantum Gravity}, 20:L23--L29, February 2003.

\bibitem{Banados92}
M.~{Banados}, C.~{Teitelboim}, and J.~{Zanelli}.
\newblock {Black hole in three-dimensional spacetime}.
\newblock {\em Physical Review Letters}, 69:1849--1851, September 1992.

\bibitem{Olivan16a}
D.~{Santos-Oliv{\'a}n} and C.~F. {Sopuerta}.
\newblock {New Features of Gravitational Collapse in Anti-de Sitter
  Spacetimes}.
\newblock {\em Physical Review Letters}, 116(4):041101, January 2016.

\bibitem{Olivan16b}
D.~{Santos-Oliv{\'a}n} and C.~F. {Sopuerta}.
\newblock {Moving closer to the collapse of a massless scalar field in
  spherically symmetric anti-de Sitter spacetimes}.
\newblock {\em \prd}, 93(10):104002, May 2016.

\bibitem{Hawking73}
S.~W. {Hawking} and G.~F.~R. {Ellis}.
\newblock {\em {The large-scale structure of space-time.}}
\newblock 1973.

\bibitem{Ishibashi12}
A.~{Ishibashi} and K.~{Maeda}.
\newblock {Singularities in asymptotically anti-de Sitter spacetimes}.
\newblock {\em \prd}, 86(10):104012, November 2012.

\bibitem{Moschidis17a}
G.~{Moschidis}.
\newblock {The Einstein--null dust system in spherical symmetry with an inner
  mirror: structure of the maximal development and Cauchy stability}.
\newblock {\em ArXiv e-prints}, April 2017.

\bibitem{Moschidis17b}
G.~{Moschidis}.
\newblock {A proof of the instability of AdS for the Einstein--null dust system
  with an inner mirror}.
\newblock {\em ArXiv e-prints}, April 2017.

\bibitem{Dias12b}
{\'O}.~J.~C. {Dias}, G.~T. {Horowitz}, D.~{Marolf}, and J.~E. {Santos}.
\newblock {On the nonlinear stability of asymptotically anti-de Sitter
  solutions}.
\newblock {\em Classical and Quantum Gravity}, 29(23):235019, December 2012.

\bibitem{Craps14c}
B.~{Craps}, E.~J. {Lindgren}, A.~{Taliotis}, J.~{Vanhoof}, and H.~{Zhang}.
\newblock {Holographic gravitational infall in the hard wall model}.
\newblock {\em \prd}, 90(8):086004, October 2014.

\bibitem{Silva16}
E.~{da Silva}, E.~{Lopez}, J.~{Mas}, and A.~{Serantes}.
\newblock {Holographic quenches with a gap}.
\newblock {\em Journal of High Energy Physics}, 6:172, June 2016.

\bibitem{Krishnan16}
C.~{Krishnan}, A.~{Raju}, and P.~N.~B. {Subramanian}.
\newblock {Dynamical boundary for anti-de Sitter space}.
\newblock {\em \prd}, 94(12):126011, December 2016.

\bibitem{Astefanesei03}
D.~{Astefanesei} and E.~{Radu}.
\newblock {Boson stars with negative cosmological constant}.
\newblock {\em Nuclear Physics B}, 665:594--622, August 2003.

\bibitem{Basu10}
P.~{Basu}, J.~{Bhattacharya}, S.~{Bhattacharyya}, R.~{Loganayagam},
  S.~{Minwalla}, and V.~{Umesh}.
\newblock {Small hairy black holes in global AdS spacetime}.
\newblock {\em Journal of High Energy Physics}, 10:45, October 2010.

\bibitem{Gentle12}
S.~A. {Gentle}, M.~{Rangamani}, and B.~{Withers}.
\newblock {A soliton menagerie in AdS}.
\newblock {\em Journal of High Energy Physics}, 5:106, May 2012.

\bibitem{Dias12c}
{\'O}.~J.~C. {Dias}, P.~{Figueras}, S.~{Minwalla}, P.~{Mitra}, R.~{Monteiro},
  and J.~E. {Santos}.
\newblock {Hairy black holes and solitons in global AdS$_{5}$}.
\newblock {\em Journal of High Energy Physics}, 8:117, August 2012.

\bibitem{Dias17b}
{\'O}.~J.~C. {Dias} and R.~{Masachs}.
\newblock {Hairy black holes and the endpoint of AdS$_{4}$ charged
  superradiance}.
\newblock {\em Journal of High Energy Physics}, 2:128, February 2017.

\bibitem{Maliborski13b}
M.~{Maliborski} and A.~{Rostworowski}.
\newblock {Time-Periodic Solutions in an Einstein AdS-Massless-Scalar-Field
  System}.
\newblock {\em Physical Review Letters}, 111(5):051102, August 2013.

\bibitem{Buchel13}
A.~{Buchel}, S.~L. {Liebling}, and L.~{Lehner}.
\newblock {Boson stars in AdS spacetime}.
\newblock {\em \prd}, 87(12):123006, June 2013.

\bibitem{Maliborski13a}
M.~{Maliborski} and A.~{Rostworowski}.
\newblock {A comment on ``Boson stars in AdS'' arXiv:1307.2875}.
\newblock {\em ArXiv e-prints}, July 2013.

\bibitem{Arias16}
R.~{Arias}, J.~{Mas}, and A.~{Serantes}.
\newblock {Stability of charged global AdS$_{4}$ spacetimes}.
\newblock {\em Journal of High Energy Physics}, 9:24, September 2016.

\bibitem{Kim15}
N.~{Kim}.
\newblock {Time-periodic solutions of massive scalar fields in dynamical AdS
  background: Perturbative constructions}.
\newblock {\em Physics Letters B}, 742:274--278, March 2015.

\bibitem{Fodor15}
G.~{Fodor}, P.~{Forg{\'a}cs}, and P.~{Grandcl{\'e}ment}.
\newblock {Self-gravitating scalar breathers with a negative cosmological
  constant}.
\newblock {\em \prd}, 92(2):025036, July 2015.

\bibitem{Maliborski14}
M.~{Maliborski} and A.~{Rostworowski}.
\newblock {What drives AdS spacetime unstable?}
\newblock {\em \prd}, 89(12):124006, June 2014.

\bibitem{MartinonPhD}
G.~{Martinon}.
\newblock {Gravitational systems in asymptotically anti-de Sitter space-times,
  arXiv:1707.00952}.
\newblock July 2017.

\bibitem{Fermi55}
Enrico Fermi, J~Pasta, and S~Ulam.
\newblock Studies of nonlinear problems.
\newblock {\em Los Alamos Report LA-1940}, 978, 1955.

\bibitem{Benettin08}
G.~{Benettin}, A.~{Carati}, L.~{Galgani}, and A.~{Giorgilli}.
\newblock {The Fermi-Pasta-Ulam Problem and the Metastability Perspective}.
\newblock In G.~{Gallavotti}, editor, {\em The Fermi-Pasta-Ulam Problem},
  volume 728 of {\em Lecture Notes in Physics, Berlin Springer Verlag}, page
  151, 2008.

\bibitem{Balasubramanian14}
V.~{Balasubramanian}, A.~{Buchel}, S.~R. {Green}, L.~{Lehner}, and S.~L.
  {Liebling}.
\newblock {Holographic Thermalization, Stability of Anti-de Sitter Space, and
  the Fermi-Pasta-Ulam Paradox}.
\newblock {\em Physical Review Letters}, 113(7):071601, August 2014.

\bibitem{Craps14a}
B.~{Craps}, O.~{Evnin}, and J.~{Vanhoof}.
\newblock {Renormalization group, secular term resummation and AdS
  (in)stability}.
\newblock {\em Journal of High Energy Physics}, 10:48, October 2014.

\bibitem{Craps15a}
B.~{Craps}, O.~{Evnin}, and J.~{Vanhoof}.
\newblock {Renormalization, averaging, conservation laws and AdS
  (in)stability}.
\newblock {\em Journal of High Energy Physics}, 1:108, January 2015.

\bibitem{Buchel15}
A.~{Buchel}, S.~R. {Green}, L.~{Lehner}, and S.~L. {Liebling}.
\newblock {Conserved quantities and dual turbulent cascades in anti-de Sitter
  spacetime}.
\newblock {\em \prd}, 91(6):064026, March 2015.

\bibitem{Bizon15b}
P.~{Bizo{\'n}}, M.~{Maliborski}, and A.~{Rostworowski}.
\newblock {Resonant Dynamics and the Instability of Anti-de Sitter Spacetime}.
\newblock {\em Physical Review Letters}, 115(8):081103, August 2015.

\bibitem{Evnin16}
O.~{Evnin} and R.~{Nivesvivat}.
\newblock {AdS perturbations, isometries, selection rules and the Higgs
  oscillator}.
\newblock {\em Journal of High Energy Physics}, 1:151, January 2016.

\bibitem{Yang15}
I.-S. {Yang}.
\newblock {Missing top of the AdS resonance structure}.
\newblock {\em \prd}, 91(6):065011, March 2015.

\bibitem{Evnin15}
O.~{Evnin} and C.~{Krishnan}.
\newblock {A hidden symmetry of AdS resonances}.
\newblock {\em \prd}, 91(12):126010, June 2015.

\bibitem{Sulem83}
C.~{Sulem}, P.-L. {Sulem}, and H.~{Frisch}.
\newblock {Tracing complex singularities with spectral methods}.
\newblock {\em Journal of Computational Physics}, 50:138--161, April 1983.

\bibitem{Bizon13}
P.~{Bizo{\'n}} and J.~{Ja{\l}mu{\.z}na}.
\newblock {Globally Regular Instability of 3-Dimensional Anti-De Sitter
  Spacetime}.
\newblock {\em Physical Review Letters}, 111(4):041102, July 2013.

\bibitem{Bizon15a}
P.~{Bizo{\'n}} and A.~{Rostworowski}.
\newblock {Comment on ``Holographic Thermalization, Stability of Anti-de Sitter
  Space, and the Fermi-Pasta-Ulam Paradox''}.
\newblock {\em Physical Review Letters}, 115(4):049101, July 2015.

\bibitem{Balasubramanian15}
V.~{Balasubramanian}, A.~{Buchel}, S.~R. {Green}, L.~{Lehner}, and S.~L.
  {Liebling}.
\newblock {Balasubramanian et al. Reply:}.
\newblock {\em Physical Review Letters}, 115(4):049102, July 2015.

\bibitem{Green15}
S.~R. {Green}, A.~{Maillard}, L.~{Lehner}, and S.~L. {Liebling}.
\newblock {Islands of stability and recurrence times in AdS}.
\newblock {\em \prd}, 92(8):084001, October 2015.

\bibitem{Deppe15b}
N.~{Deppe} and A.~R. {Frey}.
\newblock {Classes of stable initial data for massless and massive scalars in
  Anti-de Sitter spacetime}.
\newblock {\em Journal of High Energy Physics}, 12:4, December 2015.

\bibitem{Craps15b}
B.~{Craps}, O.~{Evnin}, and J.~{Vanhoof}.
\newblock {Ultraviolet asymptotics and singular dynamics of AdS perturbations}.
\newblock {\em Journal of High Energy Physics}, 10:79, October 2015.

\bibitem{Craps15c}
B.~{Craps}, O.~{Evnin}, P.~{Jai-akson}, and J.~{Vanhoof}.
\newblock {Ultraviolet asymptotics for quasiperiodic AdS$_{4}$ perturbations}.
\newblock {\em Journal of High Energy Physics}, 10:80, October 2015.

\bibitem{Deppe16b}
N.~{Deppe}.
\newblock {On the stability of anti-de Sitter spacetime, arXiv:1606.02712}.
\newblock {\em ArXiv e-prints}, June 2016.

\bibitem{Dimitrakopoulos16}
F.~V. {Dimitrakopoulos}, B.~{Freivogel}, J.~F. {Pedraza}, and I.-S. {Yang}.
\newblock {Gauge dependence of the AdS instability problem}.
\newblock {\em \prd}, 94(12):124008, December 2016.

\bibitem{Basu15}
P.~{Basu}, C.~{Krishnan}, and P.~N. {Bala Subramanian}.
\newblock {AdS (in)stability: Lessons from the scalar field}.
\newblock {\em Physics Letters B}, 746:261--265, June 2015.

\bibitem{Deppe15a}
N.~{Deppe}, A.~{Kolly}, A.~{Frey}, and G.~{Kunstatter}.
\newblock {Stability of Anti-de Sitter Space in Einstein-Gauss-Bonnet Gravity}.
\newblock {\em Physical Review Letters}, 114(7):071102, February 2015.

\bibitem{Deppe16a}
N.~{Deppe}, A.~{Kolly}, A.~R. {Frey}, and G.~{Kunstatter}.
\newblock {Black hole formation in AdS Einstein-Gauss-Bonnet gravity}.
\newblock {\em Journal of High Energy Physics}, 10:87, October 2016.

\bibitem{Mas15}
J.~{Mas} and A.~{Serantes}.
\newblock {Oscillating shells in Anti-de Sitter space}.
\newblock {\em International Journal of Modern Physics D}, 24:1542003, April
  2015.

\bibitem{Poisson04}
Eric Poisson.
\newblock {\em A relativist's toolkit: the mathematics of black-hole
  mechanics}.
\newblock Cambridge university press, 2004.

\bibitem{Cardoso16}
V.~{Cardoso} and J.~V. {Rocha}.
\newblock {Collapsing shells, critical phenomena, and black hole formation}.
\newblock {\em \prd}, 93(8):084034, April 2016.

\bibitem{Brito16b}
R.~{Brito}, V.~{Cardoso}, and J.~V. {Rocha}.
\newblock {Interacting shells in AdS spacetime and chaos}.
\newblock {\em \prd}, 94(2):024003, July 2016.

\bibitem{Wheeler55}
J.~A. {Wheeler}.
\newblock {Geons}.
\newblock {\em Physical Review}, 97:511--536, January 1955.

\bibitem{Dimitrakopoulos15a}
F.~V. {Dimitrakopoulos}, B.~{Freivogel}, M.~{Lippert}, and I.-S. {Yang}.
\newblock {Position space analysis of the AdS (in)stability problem}.
\newblock {\em Journal of High Energy Physics}, 8:77, August 2015.

\bibitem{Dias12a}
{\'O}.~J.~C. {Dias}, G.~T. {Horowitz}, and J.~E. {Santos}.
\newblock {Gravitational turbulent instability of anti-de Sitter space}.
\newblock {\em Classical and Quantum Gravity}, 29(19):194002, October 2012.

\bibitem{Dimitrakopoulos15b}
F.~{Dimitrakopoulos} and I.-S. {Yang}.
\newblock {Conditionally extended validity of perturbation theory: Persistence
  of AdS stability islands}.
\newblock {\em \prd}, 92(8):083013, October 2015.

\bibitem{Freivogel16}
B.~{Freivogel} and I.-S. {Yang}.
\newblock {Coherent cascade conjecture for collapsing solutions in global AdS}.
\newblock {\em \prd}, 93(10):103007, May 2016.

\bibitem{Dimitrakopoulos17}
F.~V. {Dimitrakopoulos}, B.~{Freivogel}, and J.~F. {Pedraza}.
\newblock {Fast and Slow Coherent Cascades in Anti-de Sitter Spacetime,
  arXiv:1612.04758}.
\newblock {\em ArXiv e-prints}, December 2016.

\bibitem{Okawa14a}
H.~{Okawa}, V.~{Cardoso}, and P.~{Pani}.
\newblock {Collapse of self-interacting fields in asymptotically flat
  spacetimes: Do self-interactions render Minkowski spacetime unstable?}
\newblock {\em \prd}, 89(4):041502, February 2014.

\bibitem{Biasi17}
Anxo~F. Biasi, Javier Mas, and Angel Paredes.
\newblock {Delayed collapses of Bose-Einstein condensates in relation to
  anti-de Sitter gravity}.
\newblock {\em Phys. Rev.}, E95(3):032216, 2017.

\bibitem{Okawa14b}
H.~{Okawa}, V.~{Cardoso}, and P.~{Pani}.
\newblock {Study of the nonlinear instability of confined geometries}.
\newblock {\em \prd}, 90(10):104032, November 2014.

\bibitem{Okawa15}
H.~{Okawa}, J.~C. {Lopes}, and V.~{Cardoso}.
\newblock {Collapse of massive fields in anti-de Sitter spacetime,
  arXiv:1504.05203}.
\newblock {\em ArXiv e-prints}, April 2015.

\bibitem{Ponglertsakul16}
S.~{Ponglertsakul}, S.~R. {Dolan}, and E.~{Winstanley}.
\newblock {Stability of gravitating charged-scalar solitons in a cavity}.
\newblock {\em \prd}, 94(2):024031, July 2016.

\bibitem{Menon16}
D.~S. {Menon} and V.~{Suneeta}.
\newblock {Necessary conditions for an AdS-type instability}.
\newblock {\em \prd}, 93(2):024044, January 2016.

\bibitem{Fodor14}
G.~{Fodor}, P.~{Forg{\'a}cs}, and P.~{Grandcl{\'e}ment}.
\newblock {Scalar field breathers on anti-de Sitter background}.
\newblock {\em \prd}, 89(6):065027, March 2014.

\bibitem{Kodama00}
H.~{Kodama}, A.~{Ishibashi}, and O.~{Seto}.
\newblock {Brane world cosmology: Gauge-invariant formalism for perturbation}.
\newblock {\em \prd}, 62(6):064022, September 2000.

\bibitem{Kodama03}
H.~{Kodama} and A.~{Ishibashi}.
\newblock {A Master Equation for Gravitational Perturbations of Maximally
  Symmetric Black Holes in Higher Dimensions}.
\newblock {\em Progress of Theoretical Physics}, 110:701--722, October 2003.

\bibitem{Kodama04}
H.~{Kodama} and A.~{Ishibashi}.
\newblock {Master Equations for Perturbations of Generalized Static Black Holes
  with Charge in Higher Dimensions}.
\newblock {\em Progress of Theoretical Physics}, 111:29--73, January 2004.

\bibitem{Ishibashi04}
A.~{Ishibashi} and R.~M. {Wald}.
\newblock {Dynamics in non-globally-hyperbolic static spacetimes: III. Anti-de
  Sitter spacetime}.
\newblock {\em Classical and Quantum Gravity}, 21:2981--3013, June 2004.

\bibitem{Horowitz15}
Gary~T. Horowitz and Jorge~E. Santos.
\newblock {Geons and the Instability of Anti-de Sitter Spacetime}.
\newblock {\em Surveys Diff. Geom.}, 20:321--335, 2015.

\bibitem{Dias16b}
{\'O}.~J.~C. {Dias}, J.~E. {Santos}, and B.~{Way}.
\newblock {Numerical methods for finding stationary gravitational solutions}.
\newblock {\em Classical and Quantum Gravity}, 33(13):133001, July 2016.

\bibitem{Martinon17}
G.~Martinon, G.~Fodor, P.~Grandcl{\'e}ment, and P.~Forg{\'a}cs.
\newblock {Gravitational geons in asymptotically anti-de Sitter spacetimes}.
\newblock {\em Classical and Quantum Gravity}, 34(12):125012, June 2017.

\bibitem{Fodor17}
G.~{Fodor} and P.~{Forg{\'a}cs}.
\newblock {Anti-de Sitter geon families, arXiv:1708.09228}.
\newblock {\em ArXiv e-prints}, August 2017.

\bibitem{Dias16a}
{\'O}.~J.~C. {Dias} and J.~E. {Santos}.
\newblock {AdS nonlinear instability: moving beyond spherical symmetry}.
\newblock {\em Classical and Quantum Gravity}, 33(23):23LT01, December 2016.

\bibitem{Dias17a}
O.~J.~C. {Dias} and J.~E. {Santos}.
\newblock {AdS nonlinear instability: breaking spherical and axial symmetries,
  arXiv:1705.03065}.
\newblock {\em ArXiv e-prints}, May 2017.

\bibitem{Rostworowski16}
Andrzej Rostworowski.
\newblock Comment on 'ads nonlinear instability: moving beyond spherical
  symmetry' (2016 class. quantum grav . 33
  [http://https://doi.org/10.1088/1361-6382/aa71cc] 23lt01 ).
\newblock {\em Classical and Quantum Gravity}, 34(12):128001, 2017.

\bibitem{Rostworowski17a}
A.~{Rostworowski}.
\newblock {Higher order perturbations of anti-de Sitter space and time-periodic
  solutions of vacuum Einstein equations}.
\newblock {\em \prd}, 95(12):124043, June 2017.

\bibitem{Rostworowski17b}
A.~{Rostworowski}.
\newblock {Towards a theory of nonlinear gravitational waves: a systematic
  approach to nonlinear gravitational perturbations in vacuum}.
\newblock {\em ArXiv e-prints}, May 2017.

\bibitem{Dias15}
{\'O}.~J.~C. {Dias}, J.~E. {Santos}, and B.~{Way}.
\newblock {Black holes with a single Killing vector field: black resonators}.
\newblock {\em Journal of High Energy Physics}, 12:171, December 2015.

\bibitem{Green16}
S.~R. {Green}, S.~{Hollands}, A.~{Ishibashi}, and R.~M. {Wald}.
\newblock {Superradiant instabilities of asymptotically anti-de Sitter black
  holes}.
\newblock {\em Classical and Quantum Gravity}, 33(12):125022, June 2016.

\bibitem{Hawking00}
S.~W. {Hawking} and H.~S. {Reall}.
\newblock {Charged and rotating AdS black holes and their CFT duals}.
\newblock {\em \prd}, 61(2):024014, January 2000.

\bibitem{Brito15a}
Richard Brito, Vitor Cardoso, and Paolo Pani.
\newblock {\em Superradiance}, volume 906.
\newblock Springer, 2015.

\bibitem{Press72}
W.~H. {Press} and S.~A. {Teukolsky}.
\newblock {Floating Orbits, Superradiant Scattering and the Black-hole Bomb}.
\newblock {\em \nat}, 238:211--212, July 1972.

\bibitem{Dias13}
{\'O}.~J.~C. {Dias} and J.~E. {Santos}.
\newblock {Boundary conditions for Kerr-AdS perturbations}.
\newblock {\em Journal of High Energy Physics}, 10:156, October 2013.

\bibitem{Cardoso14}
V.~{Cardoso}, {\'O}.~J.~C. {Dias}, G.~S. {Hartnett}, L.~{Lehner}, and J.~E.
  {Santos}.
\newblock {Holographic thermalization, quasinormal modes and superradiance in
  Kerr-AdS}.
\newblock {\em Journal of High Energy Physics}, 4:183, April 2014.

\bibitem{Niehoff16}
B.~E. {Niehoff}, J.~E. {Santos}, and B.~{Way}.
\newblock {Towards a violation of cosmic censorship}.
\newblock {\em Classical and Quantum Gravity}, 33(18):185012, September 2016.

\bibitem{Dias11}
{\'O}.~J.~C. {Dias}, G.~T. {Horowitz}, and J.~E. {Santos}.
\newblock {Black holes with only one Killing field}.
\newblock {\em Journal of High Energy Physics}, 7:115, July 2011.

\bibitem{Emparan08}
R.~{Emparan} and H.~S. {Reall}.
\newblock {Black Holes in Higher Dimensions}.
\newblock {\em Living Reviews in Relativity}, 11:6, September 2008.

\bibitem{Herdeiro16a}
C.~{Herdeiro} and E.~{Radu}.
\newblock {Einstein-Maxwell-Anti-de-Sitter spinning solitons}.
\newblock {\em Physics Letters B}, 757:268--274, June 2016.

\bibitem{Herdeiro16c}
C.~A.~R. {Herdeiro} and E.~{Radu}.
\newblock {Static Einstein-Maxwell Black Holes with No Spatial Isometries in
  AdS Space}.
\newblock {\em Physical Review Letters}, 117(22):221102, November 2016.

\bibitem{Duarte16}
M.~{Duarte} and R.~{Brito}.
\newblock {Asymptotically anti-de Sitter Proca stars}.
\newblock {\em \prd}, 94(6):064055, September 2016.

\bibitem{Holzegel13b}
Gustav Holzegel and Jacques Smulevici.
\newblock Decay properties of klein-gordon fields on kerr-ads spacetimes.
\newblock {\em Communications on Pure and Applied Mathematics},
  66(11):1751--1802, 2013.

\bibitem{Bantilan17}
H.~{Bantilan}, P.~{Figueras}, M.~{Kunesch}, and P.~{Romatschke}.
\newblock {Non-Spherically Symmetric Collapse in Asymptotically AdS Spacetimes,
  arXiv:1706:04199}.
\newblock {\em ArXiv e-prints}, June 2017.

\bibitem{Choptuik17}
M.~W. {Choptuik}, O.~J.~C. {Dias}, J.~E. {Santos}, and B.~{Way}.
\newblock {Collapse and Nonlinear Instability of AdS with Angular Momenta
  arXiv:1706.06101}.
\newblock June 2017.

\bibitem{Bekenstein73}
J.~D. {Bekenstein}.
\newblock {Black Holes and Entropy}.
\newblock {\em \prd}, 7:2333--2346, April 1973.

\bibitem{Hawking75}
S.~W. {Hawking}.
\newblock {Particle creation by black holes}.
\newblock {\em Communications in Mathematical Physics}, 43:199--220, August
  1975.

\bibitem{Hawking82}
S.~W. {Hawking} and D.~N. {Page}.
\newblock {Thermodynamics of black holes in anti-de Sitter space}.
\newblock {\em Communications in Mathematical Physics}, 87:577--588, December
  1982.

\bibitem{Gring12}
M.~{Gring}, M.~{Kuhnert}, T.~{Langen}, T.~{Kitagawa}, B.~{Rauer},
  M.~{Schreitl}, I.~{Mazets}, D.~A. {Smith}, E.~{Demler}, and
  J.~{Schmiedmayer}.
\newblock {Relaxation and Prethermalization in an Isolated Quantum System}.
\newblock {\em Science}, 337:1318, September 2012.

\bibitem{Trotzky12}
S.~{Trotzky}, Y.-A. {Chen}, A.~{Flesch}, I.~P. {McCulloch},
  U.~{Schollw{\"o}ck}, J.~{Eisert}, and I.~{Bloch}.
\newblock {Probing the relaxation towards equilibrium in an isolated strongly
  correlated one-dimensional Bose gas}.
\newblock {\em Nature Physics}, 8:325--330, April 2012.

\bibitem{Kinoshita06}
T.~{Kinoshita}, T.~{Wenger}, and D.~S. {Weiss}.
\newblock {A quantum Newton's cradle}.
\newblock {\em \nat}, 440:900--903, April 2006.

\end{thebibliography}

\end{document}